\documentclass[aps,prd,onecolumn,nofootinbib,preprintnumbers,superscriptaddress,longbibliography,nobibnotes]{revtex4-1}
\usepackage{style}

\begin{document}

\preprint{FERMILAB-PUB-23-779-SQMS-T}
\preprint{SLAC-PUB-17758}
\vspace*{1mm}

\title{Physical Signatures of Fermion-Coupled Axion Dark Matter}

\author{Asher Berlin\,\orcidlink{0000-0002-1156-1482}}
\email{aberlin@fnal.gov}
\affiliation{Theoretical Physics Division, Fermi National Accelerator Laboratory, Batavia, IL 60510, USA}
\affiliation{Superconducting Quantum Materials and Systems Center (SQMS), Fermi National Accelerator Laboratory, Batavia, IL 60510, USA}
\author{Alexander J. Millar\,\orcidlink{0000-0003-3526-0526}}
\email{amillar@fnal.gov}
\affiliation{Theoretical Physics Division, Fermi National Accelerator Laboratory, Batavia, IL 60510, USA}
\affiliation{Superconducting Quantum Materials and Systems Center (SQMS), Fermi National Accelerator Laboratory, Batavia, IL 60510, USA}
\author{Tanner Trickle\,\orcidlink{0000-0003-1371-4988}}
\email{ttrickle@fnal.gov}
\affiliation{Theoretical Physics Division, Fermi National Accelerator Laboratory, Batavia, IL 60510, USA}
\author{Kevin Zhou\,\orcidlink{0000-0002-9810-3977}}
\email{knzhou@stanford.edu}
\affiliation{SLAC National Accelerator Laboratory, 2575 Sand Hill Road, Menlo Park, CA 94025, USA}

\begin{abstract}
\noindent
In the presence of axion dark matter, fermion spins experience an ``axion wind'' torque and an ``axioelectric'' force. We investigate new experimental probes of these effects and find that magnetized analogs of multilayer dielectric haloscopes can explore orders of magnitude of new parameter space for the axion-electron coupling. We also revisit the calculation of axion absorption into in-medium excitations, showing that axioelectric absorption is screened in spin-polarized targets, and axion wind absorption can be characterized in terms of a magnetic energy loss function. Finally, our detailed theoretical treatment allows us to critically examine recent claims in the literature. We find that axioelectric corrections to electronic energy levels are smaller than previously estimated and that the purported electron electric dipole moment due to a constant axion field is entirely spurious.
\end{abstract}

\maketitle

{
\hypersetup{linkcolor=black}
\tableofcontents
}

\vspace{5mm}
\begin{center}
\textit{Conventions and Notation}
\end{center}

We use a mostly-negative spacetime metric and natural units, $\hbar = c = k_B = 1$, with rationalized Heaviside--Lorentz units for electromagnetic fields (i.e., SI units with $\eps_0 = \mu_0 = 1$). Instead of the chiral representation, we use the Dirac representation for the gamma matrices, 
\be
\g^0 = \begin{pmatrix} 1 & 0\\0 & -1 \end{pmatrix}
~~,~~
\g^i = \begin{pmatrix} 0  & \sigma^i \\ - \sigma^i & 0\end{pmatrix}
~~,~~ \g^5 = \begin{pmatrix}0 & 1 \\ 1 & 0 \end{pmatrix}
~,
\ee
where the $\sigma^i$ are the usual Pauli matrices. All states are normalized nonrelativistically, and all operators are in the Schr\"{o}dinger picture unless specified otherwise. In \Secs{EMsig}{absorb}, we work with complex axion and electromagnetic fields which oscillate with positive frequency, proportional to $e^{-i \w t}$, such that only the real part is physically meaningful. \timec

\newpage
\section{Introduction}
\label{sec:intro}

Axions are among the most well-motivated extensions to the Standard Model, generically arising as pseudo-Goldstone remnants of new approximate global symmetries broken at some high scale and in theories involving compactified extra dimensions~\cite{Svrcek:2006yi,Arvanitaki:2009fg,Cicoli:2012sz}. They are also motivated from a bottom-up perspective, as their existence could explain both the microscopic origin of dark matter~\cite{Abbott:1982af,Preskill:1982cy,Dine:1982ah} and the absence of CP violation in the strong interactions~\cite{Peccei:1977hh,Wilczek:1977pj,Weinberg:1977ma,Dine:1982ah}.

If axions account for the local dark matter density $\rhodm \simeq 0.4 \ \GeV/\cm^3$, their occupancy per quantum mode is large for axion masses $m_a \lesssim \text{few} \times 10 \ \eV$. In this case, the axion behaves as a nonrelativistic classical field, oscillating with an angular frequency set by its mass, $a(t) \simeq (\sqrt{2 \rhodm} / m_a) \, \cos{m_a  t}$. This behavior is coherent over macroscopic timescales $(m_a \vdm^2)^{-1} \sim 1 \ \mu \text{s} \times (\meV / m_a)$ and is uniform over length scales $(m_a \vdm)^{-1} \sim 10 \ \text{cm} \times (\meV / m_a)$, where $\vdm \sim 10^{-3}$ is the characteristic velocity of dark matter in the Galaxy. Axions generically couple to Standard Model currents via shift-symmetric higher-dimensional operators suppressed by a symmetry-breaking scale $\Lambda$, i.e., $\Lag \sim (\partial_\mu a) \, J^\mu_\text{SM} / \Lambda$. Most ongoing and proposed experiments search for coherent signals arising from the axion's coupling to photons via the Chern--Simons current~\cite{Irastorza:2018dyq,Sikivie:2020zpn,Semertzidis:2021rxs,Adams:2022pbo}.

Comparatively much less attention has been paid to the axion's coupling to color neutral fermions $f$, 
\be
\label{eq:IntroLag}
\Lag \supset g_{af} \, (\del_\mu a) \, \Psibar \g^\mu \g^5 \Psi
~.
\ee
Concretely, $f$ can be an electron or nucleon, with mass $\mf$ and charge $q_f$, $\Psi$ is its corresponding Dirac field, and $g_{af}$ is a dimensionful coupling inversely proportional to the symmetry-breaking scale.\footnote{Some works quantify the coupling in terms of $G_{af} = 2 g_{af}$ or use a dimensionless coupling $\tilde{g}_{af} = 2 \mf \, g_{af}$.} Independent of the axion's Galactic abundance, such interactions induce spin-dependent ``dipole--dipole'' forces between fermions~\cite{Moody:1984ba,Dobrescu:2006au}, which can be searched for by experiments involving electrons~\cite{Leslie:2014mua,Terrano:2015sna,Ficek:2016qwp,Ji:2016tmv,Wang:2022bpw}, nucleons~\cite{Adelberger:2006dh,Vasilakis:2008yn,Ledbetter:2012xd,Mostepanenko:2020lqe}, or both~\cite{Almasi:2018cob,Wang:2022wgs}. Stronger probes are possible in the presence of axion dark matter. For low-energy experiments, its physical effects are most directly seen in terms of the nonrelativistic single-particle Hamiltonian, 
\be
\label{eq:IntroHam}
H \supset - g_{af} \, (\grad a) \cdot \sigmav - \frac{g_{af}}{\mf} \, \dot{a} \, \sigmav \cdot \Piv 
~,
\ee
where $\Piv \equiv \p - q_f \, \A$ is the fermion's mechanical momentum, defined in terms of the canonical momentum $\p = -i \grad$ and electromagnetic vector potential $\A$. The first and second terms of \Eq{IntroHam} are called the ``axion wind'' and ``axioelectric'' terms, respectively, and give rise to a variety of spin-dependent effects.

The axion wind causes spins to precess about the axion gradient, like the effect of an effective magnetic field $\B_\eff$ on a magnetic dipole moment. Various experimental techniques have been developed and proposed to search for this anomalous torque. For instance, spin-polarized torsion pendulums~\cite{Terrano:2019clh,Graham:2017ivz} or precision magnetometers, such as comagnetometers and nuclear magnetic resonance setups~\cite{Graham:2013gfa,Graham:2017ivz,Garcon:2017ixh,JacksonKimball:2017elr,Abel:2017rtm,Garcon:2019inh,Wu:2019exd,Alonso:2018dxy,Bloch:2019lcy,Smorra:2019qfx,Jiang:2021dby,Aybas:2021nvn,Bloch:2021vnn,Bloch:2022kjm,Lee:2022vvb,Abel:2022vfg,Wei:2023rzs,Xu:2023vfn,Rosenzweig:2023kgb,huang2023axionlike}, can search for axions of mass $m_a \ll \mu \eV$. Other recent proposals involve searching for spin precession in superfluid helium~\cite{Gao:2022nuq,Chigusa:2023szl,Foster:2023bxl}, nitrogen vacancy centers~\cite{Chigusa:2023hms}, and storage rings~\cite{Graham:2020kai,Janish:2020knz,Nikolaev:2022wyi,Brandenstein:2022eif,JEDI:2022hxa}. At higher axion masses, axion absorption can induce spin-flip transitions in atoms~\cite{Sikivie:2014lha,Santamaria:2015gro,Braggio:2017oyt,Vergados:2018qdb} or excite in-medium magnons~\cite{Mitridate:2020kly,Trickle:2019ovy}. In the $\mu \eV - \meV$ mass range, several experiments~\cite{Barbieri:1985cp,Caspers:1989ix,Kakhidze:1990in,Ruoso:2015ytk,Barbieri:2016vwg,Crescini:2018qrz,QUAX:2020adt,Flower:2018qgb,Ikeda:2021mlv,Chigusa:2020gfs,Chigusa:2023hmz} search for magnon absorption by placing a magnetic sample in a cavity, thereby mixing the magnon with a cavity mode so that it can be read out as a photon. These efforts, which we term ``ferromagnetic haloscopes,'' leverage existing expertise in cavity magnonics. However, other methods to probe the axion-electron coupling in this mass range have been unexplored. 

The axioelectric term acts like an effective electric field $\E_\eff$ directed along the spin and hence produces a time-varying force. In analogy to the photoelectric effect, this force can ionize atoms by the ``axioelectric effect''~\cite{Dimopoulos:1985tm,Pospelov:2008jk,Dzuba:2010cw,Derevianko:2010kz}, which is used to search for, e.g., the absorption of high-energy solar axions~\cite{Avignone:1986vm,Arisaka:2012pb,Abdelhameed:2020hys,Gavrilyuk:2022yli,XENON:2022ltv}. More recent works have considered absorption of less energetic dark matter axions of mass $m_a \lesssim 10 \ \eV$, which can produce electronic excitations in molecules~\cite{Arvanitaki:2017nhi} and solids~\cite{Hochberg:2016ajh,Hochberg:2016sqx,Bloch:2016sjj,Mitridate:2021ctr,Chen:2022pyd} or single phonons in spin-polarized materials~\cite{Mitridate:2023izi}.

The physical effects of the axioelectric term in the ultralight regime ($m_a \ll 1 \ \eV$) are much less well-understood. While the ability for this oscillating force to generate currents in spin-polarized targets was briefly considered in Refs.~\cite{Krauss:1985ub,Slonczewski:1985oco,Smith:1988kw}, the prospects for detecting such effects were never carefully analyzed. Furthermore, some recent studies have investigated axioelectric-induced modifications to electronic energy levels~\cite{Arvanitaki:2021wjk,Roising:2021lpv,Balatsky:2023mdm}, but the sizes of these shifts were only very roughly estimated. Finally, there have been many recent independent claims that the axion-electron coupling generates an electron electric dipole moment (EDM)~\cite{Alexander:2017zoh,Chu:2019iry,Wang:2021dfj,Smith:2023htu,DiLuzio:2023ifp} proportional to the axion field value. This would seem to violate the shift symmetry of the axion field and, if true, would imply sensitivity up to twenty orders of magnitude stronger than existing astrophysical bounds. Thus, the physics of the axioelectric term is currently far from clear.

\begin{figure*}
\includegraphics[width=\textwidth]{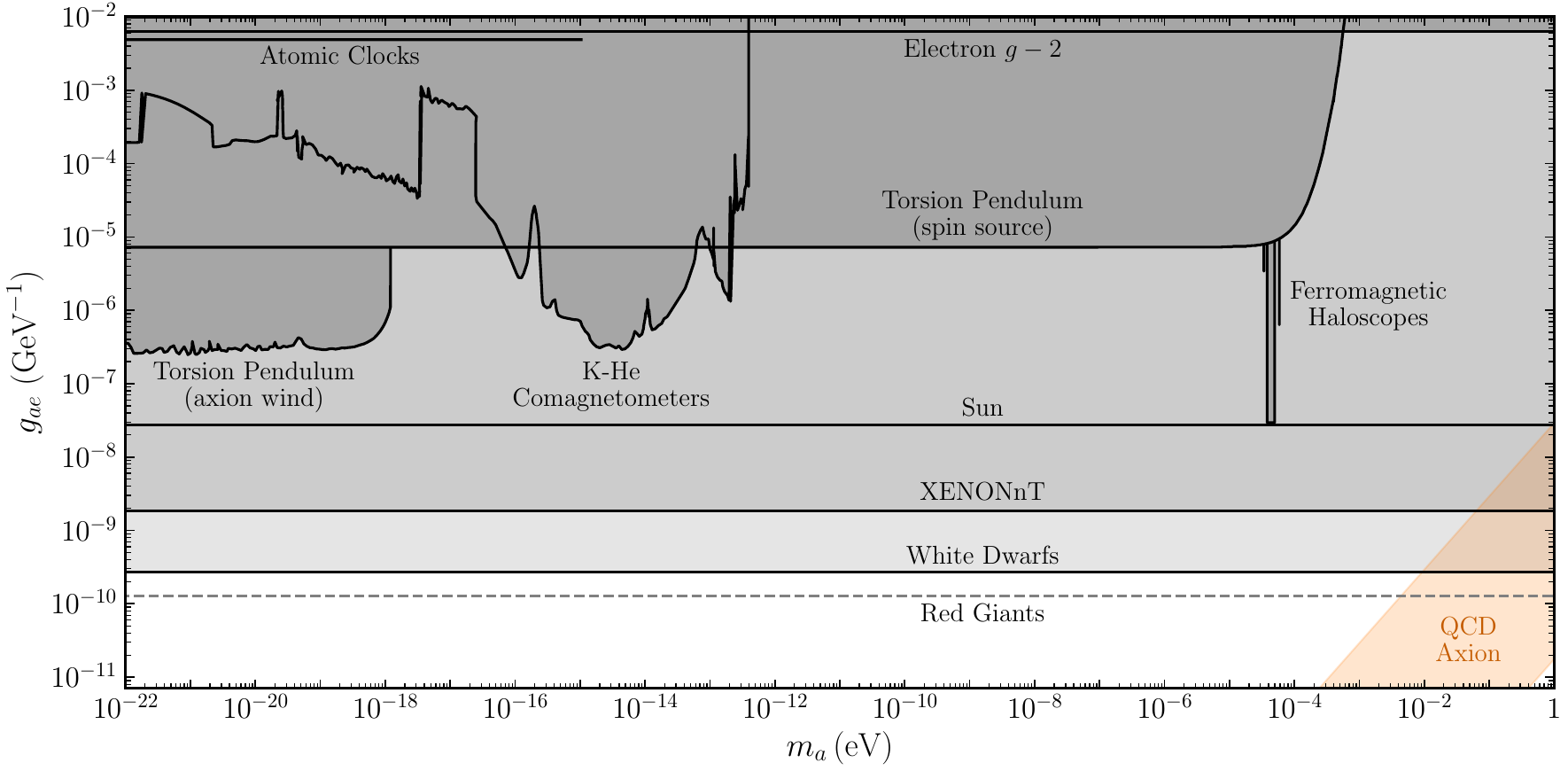} 
\caption{Existing constraints on the axion-electron coupling. Laboratory constraints are from torsion pendulums~\cite{Terrano:2019clh,Terrano:2015sna}, atomic clocks~\cite{Alonso:2018dxy}, comagnetometers~\cite{Bloch:2019lcy}, ferromagnetic haloscopes~\cite{Crescini:2018qrz,QUAX:2020adt,Flower:2018qgb,Ikeda:2021mlv}, and electron $g-2$ measurements~\cite{Yan:2019dar,Bauer:2021mvw} (using the $2\sigma$ uncertainty of the latest measurement~\cite{Fan:2022eto}).  Astrophysical constraints are from XENONnT solar axion searches~\cite{XENON:2022ltv} and from considerations of additional energy loss from the Sun~\cite{Gondolo:2008dd}, white dwarfs~\cite{MillerBertolami:2014rka}, and red giants~\cite{Capozzi:2020cbu}. However, the red giant bound may be significantly weakened when uncertainties in stellar parameters are accounted for~\cite{Dennis:2023kfe}. Furthermore, all of these astrophysical constraints are relaxed in axion models with environment-dependent couplings~\cite{DeRocco:2020xdt}. Bounds derived from supernova 1987A~\cite{Lucente:2021hbp,Ferreira:2022xlw} and Big Bang nucleosynthesis~\cite{Ferreira:2022xlw,Ghosh:2020vti} are about two orders of magnitude weaker than the solar bound and not shown here for clarity. The orange band corresponds to QCD axion models, reviewed in, e.g., Ref.~\cite{Irastorza:2018dyq}.
}
\label{fig:existing} 
\end{figure*}

\begin{figure*}
\includegraphics[width=\textwidth]{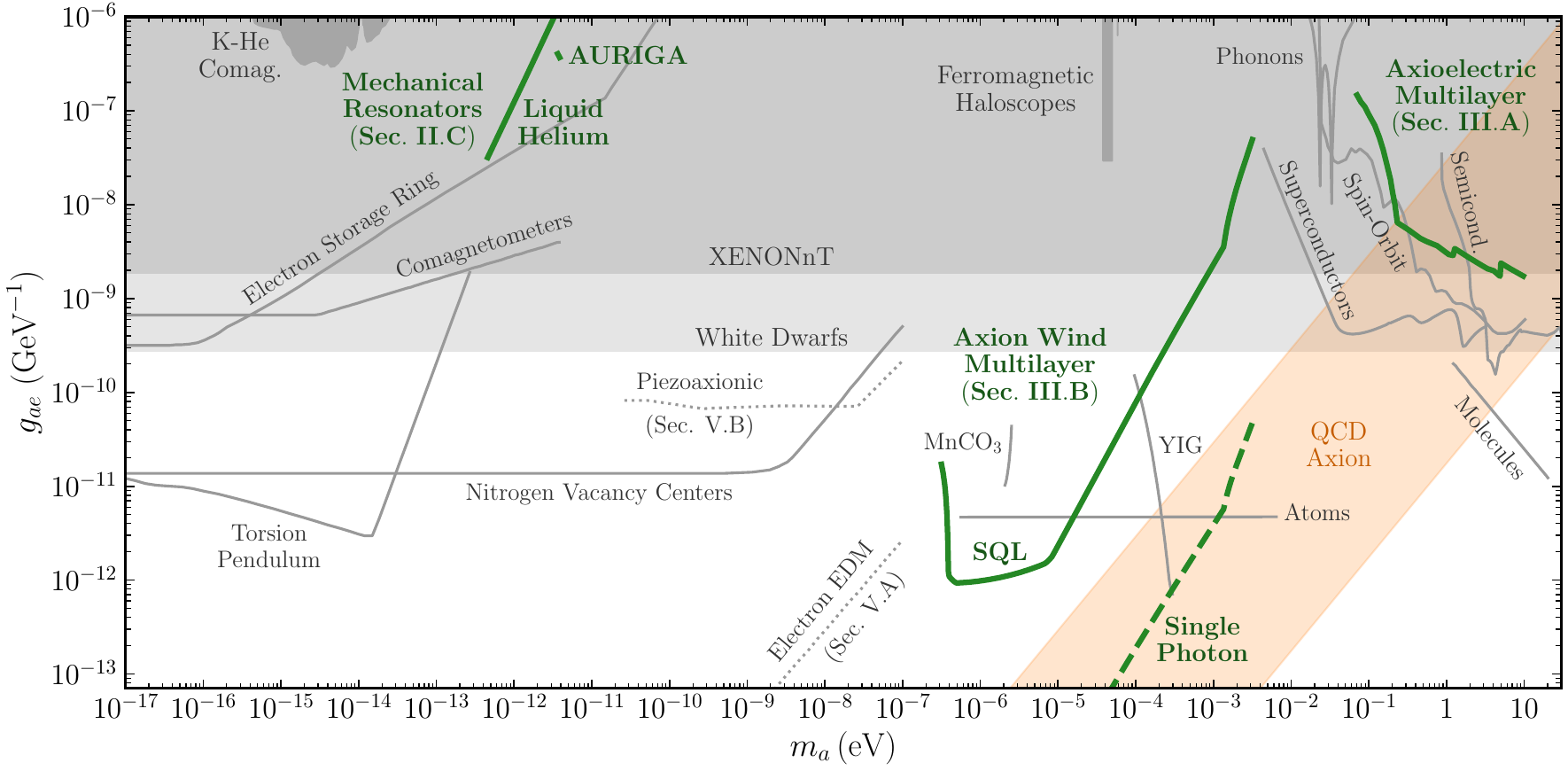} 
\caption{Proposed searches for the axion-electron coupling, with existing limits (shaded gray) as in \Fig{existing}. The new directions we explore are shown in solid green, and discussed in the corresponding sections. Dotted gray projections~\cite{Arvanitaki:2021wjk,Chu:2019iry} should be revised, as discussed in \Sec{EDMandDeltaE}. As solid gray lines, we show projections from torsion pendulums~\cite{Graham:2017ivz}, comagnetometers~\cite{Bloch:2019lcy}, an electron storage ring~\cite{Brandenstein:2022eif}, nitrogen vacancy centers~\cite{Chigusa:2023hms}, ferromagnetic haloscopes using $\mathrm{MnCO}_3$~\cite{Chigusa:2023hmz} and YIG~\cite{Chigusa:2020gfs}, and absorption into electronic excitations in atoms~\cite{Sikivie:2014lha}, superconductors~\cite{Hochberg:2016ajh}, spin-orbit coupled materials~\cite{Chen:2022pyd}, semiconductors~\cite{Hochberg:2016sqx}, and molecules~\cite{Arvanitaki:2017nhi}, as well as into phonon excitations in $\text{FeBr}_2$~\cite{Mitridate:2023izi}. For each proposal, we show the most optimistic projection, though the difficulty of experimentally realizing each one varies significantly.
}
\label{fig:future} 
\end{figure*}

In this work, we provide a firm foundation for the study of axion-fermion couplings, with an emphasis on new probes of the axion-electron coupling $g_{ae}$. Existing constraints on this coupling are reviewed in \Fig{existing}, while the projected sensitivities of future experiments are shown in \Fig{future}. We begin in \Sec{LowHam} by reviewing the derivation of the nonrelativistic Hamiltonian of \Eq{IntroHam}. In \Sec{HeisEOM}, we compute the associated classical torques and forces in an axion background and show how the leading effects can be expressed in terms of effective spin-coupled electromagnetic fields $\E_\eff$ and $\B_\eff$. In \Sec{Mechsig}, we consider the use of spin-polarized mechanical resonators to detect these effects. 

In \Sec{EMsig}, we show how $\E_\eff$ and $\B_\eff$ give rise to polarization and magnetization currents, respectively, in electron spin-polarized (magnetic) media. To detect these currents, we consider multilayer setups like those used by dielectric haloscopes to search for the axion-photon coupling at microwave~\cite{Caldwell:2016dcw,Millar:2016cjp,MADMAX:2019pub} and optical~\cite{Baryakhtar:2018doz,Chiles:2021gxk,Manenti:2021whp,DeMiguel:2023nmz} frequencies, where the signal appears as emitted electromagnetic radiation. As shown in \Sec{LAMPOST}, such a system can reach QCD axion sensitivity at optical frequencies, where the polarization current dominates. At microwave frequencies, the magnetization current instead dominates. In this case, a multilayer setup can explore orders of magnitude beyond astrophysical bounds, as shown in \Sec{MADMAX}. 

In \Sec{absorb}, we revisit axion absorption into in-medium excitations. For dark photons and axions, such results are often determined by the ``energy loss function'' $\Im(-1/\eps)$~\cite{Knapen:2021run,Hochberg:2021pkt,Knapen:2021bwg,Boyd:2022tcn,Berlin:2023ppd}, which has recently been applied to dark photons~\cite{Knapen:2021bwg} and photon-coupled axions~\cite{Berlin:2023ppd}. In \Sec{magnon}, we use a classical argument to show that in spin-polarized media, the absorption rate into magnons via the axion wind term is determined by $\Im(- 1/\mu)$, the magnetic analogue of the energy loss function. As for the axioelectric term, previous calculations have shown that the absorption rate into electronic excitations scales with $\Im(\eps)$ in unpolarized targets. In \Sec{absorption_into_electronic_excitations}, we show that in spin-polarized targets, it is instead proportional to the usual energy loss function $\Im(-1/\eps)$ and is thus generically screened.

In \Sec{EDMandDeltaE}, we critically examine recent claims about the physical effects of the axioelectric term. We show that any apparent EDM proportional to the axion field value is spurious, with no corresponding observable effect, and that energy level shifts from the axioelectric term are smaller than previously estimated. We conclude in \Sec{conclusion} by discussing directions for future investigation. The appendices are referred to throughout the text.

\section{The Nonrelativistic Limit}
\label{sec:the_NR_limit}

In this section, we review the physical effects of fermion-coupled axion dark matter. We begin by deriving the nonrelativistic Hamiltonian in \Sec{LowHam}. It is then used in \Sec{HeisEOM} to compute the torques and forces on a fermion, showing that the leading effects are simply described by effective spin-coupled magnetic and electric fields $\B_\eff$ and $\E_\eff$, respectively. In \Sec{Mechsig}, we consider the sensitivity of spin-polarized mechanical resonators to these fields.

\subsection{Deriving the Nonrelativistic Hamiltonian}
\label{sec:LowHam}

We first motivate the nonrelativistic Hamiltonian in \Eq{IntroHam} by considering the classical limit of the axial vector current. As shown in \App{bilinears}, in a classical single-particle state, the spatial integral of its expectation value is essentially the fermion's spin four-vector $s^\mu$, equal to $(0, \hat{\v{s}})^\mu$ in its rest frame, where $\shat$ is the unit vector aligned with its spin. Thus, to first order in the fermion's velocity $\vv$, we have
\be
\label{eq:classicalaxial}
\int d^3 \xv ~ \braket{\Psibar \g^\mu \g^5 \Psi} \simeq (\vv \cdot \shat ~, \, \shat)^\mu
~,
\ee
If we approximate $\del_\mu a$ as spatially uniform, we can use \Eq{classicalaxial} to evaluate the spatial integral of \Eq{IntroLag}, giving the particle Lagrangian
\be
\label{eq:IntroLagInt}
L \supset g_{af} \, (\grad a) \cdot \shat + g_{af} \, \dot{a} \, \vv \cdot \shat
~.
\ee
The nonrelativistic Hamiltonian is given by $H = \vv \cdot \p - L$, where the canonical momentum is
\be \label{eq:canonical_def}
\p = \partial L / \partial \vv = \mf \, \vv + q_f \, \A + g_{af} \, \dot{a} \, \shat
~.
\ee
Identifying $\p \to -i \grad$ and $\shat \to \sigmav$ in the Hamiltonian recovers \Eq{IntroHam}. The first term in \Eq{IntroLagInt} couples to the spin like an effective magnetic field $\mathbf{B}_\text{eff} = (g_{af}/\mu_f) \,  (\nabla a)$ with $\mu_f = q_f / 2 \mf$, while the second is of the form $L \supset q_f \, \A_{\eff} \cdot \vv$ with an effective vector potential $\A_\eff = (g_{af}/q_f) \, \dot{a} \, \shat$. The axion field thus exerts a torque on spins and a spin-dependent force,
\begin{align}
    \bm{\tau} & = \mu_f \, ( \shat \times \mathbf{B}_\text{eff} ) = \, g_{af} \, (\shat \times \nabla a) \\
    \v{F} & = q_f \, \E_\eff = - g_{af} \, \frac{d}{dt} \big( \dot{a} \, \shat \big)     \label{eq:IntroEeff}
    ~,
\end{align}
where $\E_\eff = - d\A_\eff / dt$ is an effective spin-coupled electric field. Though this derivation is merely heuristic, it does capture the axion's leading physical effects. 

In order to systematically extract the full set of physical effects, we take the nonrelativistic limit of the fermion's equation of motion, using the same procedure that is used to derive the Pauli Hamiltonian from the Dirac equation. Starting from \Eq{IntroLag}, the equation of motion for $\Psi$ is
\be
\label{eq:EOM1}
\big( i \slashed{\del} - \mf - q_f \, \slashed{A} + g_{af} \, (\slashed{\del} a) \, \g^5 \big) \, \Psi = 0
~.
\ee
In relativistic quantum mechanics, $\Psi$ is the four-component wavefunction of a state with a single particle or antiparticle, and \Eq{EOM1} governs the evolution of the wavefunction. Specifically, the positive-frequency solutions of \Eq{EOM1} represent the wavefunctions of particles, while negative-frequency solutions represent antiparticles. To reduce to the low-energy nonrelativistic theory, we ``integrate out'' the antiparticle component using the Pauli elimination method. To make this separation manifest, we divide $\Psi$ into upper and lower two-component wavefunctions,
\be
\label{eq:4to2}
\Psi = e^{- i \mf \, t} \, \begin{pmatrix} \psi \\ \psibar \end{pmatrix}
~,
\ee
where the rapid time-dependence of the positive-frequency solution has been factored out. The wavefunctions $\psi$ and $\psibar$ dominate for nonrelativistic particle and antiparticle states, respectively. In terms of $\psi$ and $\psibar$, \Eq{EOM1} is
\begin{align}
 \label{eq:psi_eom} 
\big( i \dt - q_f \, \phi + g_{af} \, (\grad a) \cdot \sigmav \big) \, \psi &= \big( \Piv \cdot \sigmav - g_{af} \, \dot{a} \big) \, \psibar
\\
\label{eq:psibar_eom}
\big(2 \mf + i \dt - q_f \, \phi + g_{af} \, (\grad a) \cdot \sigmav \big) \,  \psibar &= \big( \Piv \cdot \sigmav - g_{af} \, \dot{a} \big) \, \psi
~.
\end{align}
As anticipated, $\psibar$ is suppressed by $\sim 1/\mf$ compared to $\psi$ for nonrelativistic particle states. Thus, to economically describe a nonrelativistic particle, we can solve \Eq{psibar_eom} to leading order in the nonrelativistic expansion,
\be
\label{eq:psibarEOM}
\psibar \simeq  \frac{1}{2 \mf} \big( \Piv \cdot \sigmav - g_{af} \, \dot{a} \big) \, \psi
~,
\ee
and then use this to eliminate $\psibar$ from the equation of motion for $\psi$ in \Eq{psi_eom}. The result is
\be
\label{eq:penultimate_step}
\Big(i \dt - q_f \, \phi + g_{af} \, (\grad a) \cdot \sigmav \Big) \, \psi \simeq \Big( \frac{1}{2 \mf} \, |\Piv \cdot \sigmav|^2 - \frac{g_{af} \, \dot{a}}{\mf} \, \Piv \cdot \sigmav + \frac{i g_{af}}{2 \mf} \, (\grad \dot{a}) \cdot \sigmav \Big) \, \psi
~,
\ee
to leading order in the coupling $g_{af}$. The nonrelativistic Hamiltonian can then be identified by $i \dt \psi = H \psi$, which gives
\be 
\label{eq:HamDeriv}
H \simeq \frac{\Piv^2}{2 \mf} + q_f \, \phi - \frac{q_f}{2 \mf} \, \B \cdot \sigmav 
- g_{af} \, (\grad a) \cdot \sigmav - \frac{g_{af}}{2 \mf} \, \{ \dot{a} , \Piv \cdot \sigmav\} 
~,
\ee
where we used the identity $|\Piv \cdot \sigmav|^2 = \Piv^2 - q_f \, \B \cdot \sigmav$. The first three terms in \Eq{HamDeriv} form the usual Pauli Hamiltonian, and the axion-dependent terms are in agreement with Ref.~\cite{Smith:2023htu} (which instead used the Foldy--Wouthuysen method~\cite{Foldy:1949wa} to decouple the antiparticle component). Expanding the anticommutator in the final term yields $\{ \dot{a} , \Piv \cdot \sigmav\} = 2 \dot{a} \, \Piv \cdot \sigmav - i (\grad \dot{a}) \cdot \sigmav$, where the first term corresponds to the axioelectric term in \Eq{IntroHam}. Note that the second term is not determined by our heuristic classical argument; however, it is less phenomenologically interesting since it is qualitatively similar to the axion wind, but subdominant for $m_a \ll \mf$.

\subsection{Torques and Forces}
\label{sec:HeisEOM}

The Hamiltonian of \Eq{HamDeriv} implies that the axion field imparts spin torques and forces on fermions. To show this, we will use Ehrenfest's theorem, $d \langle \mathcal{O} \rangle / dt = \expect{\del_t \mathcal{O}} + i \expect{\,  [H , \mathcal{O}] \, }$, which governs the time evolution of the expectation value of any observable $\mathcal{O}$. For example, it implies that the velocity operator $\vv \equiv i [H, \xv]$ satisfies $\expect{\vv} = d \expect{\xv}/dt$. In the presence of an axion background, this is
\be
\label{eq:vEOM}
\vv = \frac{1}{\mf} \, (\Piv - g_{af} \, \dot{a} \, \sigmav)
~,
\ee
which is just the quantum analogue of \Eq{canonical_def}. Next, the expectation value of the spin operator $\v{S} = \sigmav/2$ evolves as 
\be
\label{eq:spinEOM}
\frac{d}{dt}\expect{\v{S}} = \expect{\, 2 \, \mu_f \, \v{S} \times \B + 2 \, g_{af} \, \v{S} \times ( \grad a +  \{\dot{a}, \vv\}/2 ) \, }
~.
\ee
Discarding subdominant contributions proportional to $\nabla \dot{a}$ or $g_{af}^2$, we find that the axion field imparts the same spin torque as an effective magnetic field
\be
\label{eq:Beff}
\B_\eff \simeq \frac{g_{af}}{\mu_f} \, (\grad a + \dot{a} \expect{\vv}) ~.
\ee
Above, the first term proportional to $\grad a$ is the usual axion wind effect, while the additional term proportional to $\dot{a}$ follows from Galilean invariance; it represents the additional contribution to the axion gradient in the frame of a moving fermion. Note that this new term is subdominant when the fermion velocity is much smaller than the dark matter velocity, which applies to most laboratory experiments. 

Evaluating the force from the axion field is somewhat more involved, as it requires taking the time derivative of \Eq{vEOM}. Using \Eq{spinEOM} to simplify the result, we find 
\begin{align}
\label{eq:forceEOM}
\F \equiv \mf  \, \frac{d}{dt}\expect{\vv} &\simeq \Big\langle q_f \, \E + \frac{q_f}{2} \, (\vv \times \B - \B \times \vv) + \mu_f \, \grad (\sigmav \cdot \B) \Big\rangle
\nl
& - g_{af} \, \frac{d}{dt} \, \big\langle \dot{a} \, \sigmav \big\rangle + g_{af} \, \Big\langle \grad(\sigmav \cdot \grad a) + \frac{1}{2} \{\sigmav \cdot \vv , \, (\grad \dot{a})\} - \frac{1}{2} \, (\vv \cdot \grad \dot{a} + \grad \dot{a} \cdot \vv) \, \sigmav \Big\rangle
~,
\end{align}
where the first line is the usual electromagnetic force on a minimally-coupled spin-$1/2$ particle, and the second line contains axion-induced effects. Since the gradient of an axion dark matter field is suppressed by its small velocity, the first term in the second line dominates. It is equivalent to an effective electric field aligned with the spin,
\be
\label{eq:Eeff}
\E_\eff \simeq - \frac{g_{af}}{q_f} \, \frac{d}{dt} \left( \dot{a} \expect{ \sigmav } \right)
~,
\ee
which is consistent with our classical result in \Eq{IntroEeff}, since $\expect{\sigmav} = \hat{\v{s}}$, and agrees with Ref.~\cite{Slonczewski:1985oco}. 

Since $\E_\eff$ and $\B_{\eff}$ can cause charges to move and dipole moments to precess, respectively, they indirectly give rise to electromagnetic signals, which we will consider in \Sec{EMsig}. However, first we consider their most direct effects, which are oscillating mechanical forces and torques in spin-polarized materials.

\subsection{Mechanical Signals}
\label{sec:Mechsig}

From \Eq{Eeff}, we see that when a material is electron or nuclear spin-polarized and its spins precess at angular frequency $\w_\text{spin}$, then its spins experience an oscillating axioelectric force with angular frequency $\w_{\text{sig}} \simeq m_a \pm \w_{\text{spin}}$, whose magnitude is proportional to $\w_{\text{sig}}$. For concreteness, we focus on electron-coupled axions and static electron spins, $\w_{\text{spin}} = 0$. In an insulating material, where the electrons are not free to move relative to the nuclei, forces on the electrons cause the material to accelerate as a whole. Thus, two nearby test masses of opposite electron spin polarization will experience an oscillating relative acceleration of magnitude
\be
\label{eq:axion_accel}
\Delta a_{ae} \simeq \frac{2 f_s \, g_{ae} \, \w_\sig \, \sqrt{\rhodm}}{m_N}
~,
\ee
where $m_N$ is the mass per atom and $f_s$ is the number of polarized electron spins per atom. 

At ultralow frequencies, $\w_\sig \lesssim 10 \ \Hz$, the axioelectric acceleration is very difficult to detect because it is suppressed by $\w_{\text{sig}}/m_N$. Using Refs.~\cite{Graham:2015ifn,Pierce:2018xmy,MAGIS-100:2021etm}, we have recast the sensitivity of future torsion pendulums, atom interferometers, and gravitational wave detectors, assuming they can be modified to have $f_s \simeq 1$. In each case we find that the ideal sensitivity is well below that of existing astrophysical bounds and too weak to appear in \Fig{future}.

Oscillating accelerations at $\kHz - \GHz$ frequencies can be detected with mechanical resonators. To analyze this case, we recast the results of Refs.~\cite{Arvanitaki:2015iga,Branca:2016rez,Manley:2019vxy}, which consider how dilaton dark matter modifies the equilibrium length of solid objects, leading to a differential acceleration across the object. Each of these setups can in principle be converted to a probe of the axioelectric force by, e.g., spin polarizing its two halves in opposite directions. Equating the resulting accelerations (under the assumption that $f_s \simeq 1$ can be realized and that the relevant $\order{1}$ mechanical form factors are comparable) yields the curves labeled ``Liquid Helium'' and ``AURIGA'' in \Fig{future}. Though they are more promising, they are still much weaker than existing astrophysical bounds. 

The axion wind's spin torque can lead to stronger mechanical effects. In a material where the orientation of electron spins is fixed relative to the atomic lattice, such as a hard ferromagnet, spin torques are converted to torques on the lattice. For an object of length scale $L$, this corresponds to a characteristic linear acceleration
\be
\label{eq:wind_accel}
\Delta a_{\text{wind}} \sim \frac{f_s \, g_{ae} \, \nabla a}{m_N L} \sim \frac{\vdm}{m_a L} \, \Delta a_{ae}
~.
\ee
The axion wind's mechanical effect was already considered for torsion pendulums in Ref.~\cite{Graham:2017ivz}, but, in principle, it could also excite ``toroidal'' (shearing) modes in mechanical resonators. For the resonators considered in Ref.~\cite{Manley:2019vxy}, $\Delta a_{\text{wind}}$ is enhanced over $\Delta a_{ae}$ by several orders of magnitude. However, a detailed analysis of this signature is beyond the scope of this work. 

\section{Electromagnetic Signals}
\label{sec:EMsig}

\begin{figure*}[t]
\includegraphics[width=12cm]{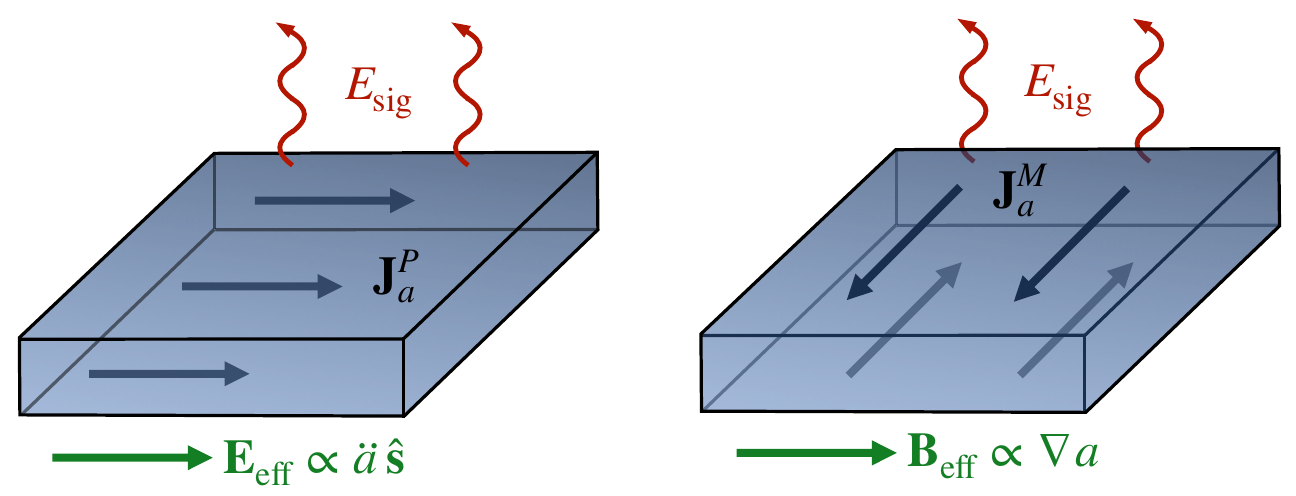} 
\caption{Schematic representation of the axion-induced electromagnetic signals considered in \Sec{EMsig}. \textbf{Left}: The axioelectric term's effective electric field $\E_\eff$ induces a bulk polarization current $\jv_a^P$ in a spin-polarized material. \textbf{Right}: The axion wind term's effective magnetic field $\Bv_\eff$ induces a surface magnetization current $\jv_a^M$.}
\label{fig:cartoon}
\end{figure*}

Unlike the axion-photon coupling, the axion-fermion coupling does not directly source electromagnetic fields. This is immediately apparent from both of the Lagrangian forms of the interaction, either in terms of relativistic fields (\Eq{IntroLag}) or nonrelativistic particles (\Eq{IntroLagInt}), which are independent of the vector potential. Although the Hamiltonian form of the axioelectric term $H \supset - (g_{af} / \mf) \, \dot{a} \, (\p - q_f \, \A) \cdot \sigmav$ does naively appear to contain a direct coupling between the axion field and the vector potential, it is straightforward to verify that the electromagnetic Heisenberg equations of motion (calculated using \Eq{HamDeriv}) do not involve axion-dependent source terms. 

Regardless, the effective fields $\E_\eff$ and $\B_{\eff}$ indirectly source real electromagnetic fields through their effect on the motion of charges. For concreteness, we focus here on the axion-electron coupling $g_{a e}$. In this case, the axioelectric field $\E_\eff$ drives polarization currents in spin-ordered material, while the axion wind field $\B_\eff$ induces magnetization currents on the boundary of any material with nonzero magnetic susceptibility. These currents in turn source electromagnetic radiation, as depicted in \Fig{cartoon}. Thus, dark matter experiments employing photon readout of dielectric stacks~\cite{Caldwell:2016dcw,Millar:2016cjp,Baryakhtar:2018doz,MADMAX:2019pub,Manenti:2021whp,Chiles:2021gxk} can be sensitive to the axion-electron coupling if they are modified to use appropriate materials.

To begin making this more precise, consider a linear medium with permittivity $\eps$ and permeability $\mu$. Since the effective fields act on electrons but do \textit{not} satisfy Maxwell's equations, we find it conceptually useful to split the induced polarization and magnetization into a standard and axion-induced part,
\begin{align}
\label{eq:PandM}
\Pv &= \Pv_0 + \Pv_a = (\eps - 1) \, \E + (\epsa - 1) \, \E_{\eff} \\
\Mv &= \Mv_0 + \Mv_a = (1 - \mu^{-1}) \, \B + (1 - \mu^{-1}) \, \B_{\eff}~. \label{eq:mag_def}
\end{align}
Here, $\Pv_0$ and $\Mv_0$ are induced by electromagnetic fields exactly as in ordinary electrodynamics, while $\Pv_a$ and $\Mv_a$ are induced by the axion field. Note that in \Eq{PandM} we have defined $\epsa$, the permittivity due to the spin-polarized electrons in the sample (where $\epsa = 1$ for an unpolarized sample). This will be more properly defined later in \Eqs{epsa}{epsaPi}. We similarly decompose the in-medium current as $\jv = \jv_0 + \jv_a$, where $\jv_0 = \del_t \Pv_0 + \grad \times \Mv_0$, as usual, and $\jv_a = \del_t \Pv_a + \grad \times \Mv_a$. While the case of an insulating medium may be more familiar, these equations can also be applied to conductors, which can be described by a permittivity with \timec $\text{Im}(\eps) = \tsw{-}{} \sigma/\w$, where $\sigma$ is the conductivity and $\w$ is the angular frequency. Since $\jv_0$ includes both the usual free and bound currents, only the axion-induced current $\jv_a$ appears as a source in the in-medium Maxwell's equations. For instance, Amp\`ere's law reads
\be
\grad \times (\mu^{-1} \, \B) = \jv_a + \eps \, \del_t \E
~.
\ee
Combining this with Faraday's law, which is unchanged, yields the inhomogeneous wave equation
\be
\label{eq:Efield1}
\grad \times \grad \times \E + n^2 \, \del_t^2 \E = - \mu \, \del_t \jv_a
~,
\ee
where $n = \sqrt{\eps \mu}$ is the refractive index. This is the key result we will need in \Secs{LAMPOST}{MADMAX}.

Explicitly, the axion-induced current contains two pieces, 
\be
\label{eq:axioncurrent}
\jv_a = \jv_a^P + \jv_a^M = (\epsa - 1) \, \del_t \E_{\eff} + \grad \times \big( (1 - \mu^{-1}) \, \B_{\eff} \big)
~,
\ee
each of which can produce electromagnetic signals. The axion-induced polarization current $\jv_a^P$ only exists in spin-polarized materials, such as ferromagnets, or paramagnets in an external magnetic field. The magnetization current $\jv_a^M$ exists on the surface of any finite material with $\mu \neq 1$, though it tends to be most significant in spin-polarized materials. From \Eqs{Beff}{Eeff} we see that for static spins ($d \langle \sigmav \rangle / d t = 0$), $\E_{\eff}$ depends on a second derivative of the axion field while $\B_{\eff}$ depends on only a first derivative. We therefore expect the axioelectric induced current $\jv_a^P$ to dominate for large axion masses and the axion wind induced current $\jv_a^M$ to dominate at smaller masses. In the following subsections, we consider these two currents in turn.

\subsection{Axioelectric Polarization Currents}
\label{sec:LAMPOST}

Let us first consider the polarization current $\jv_a^P$ resulting from the axioelectric field $\E_{\eff}$, by specializing to a monochromatic spatially uniform axion dark matter field $a \propto e^{\tsw{}{-} i m_a t}$ \timec. To build intuition, we begin with the example of an infinite uniformly spin-polarized medium. In this case, we can ignore the curl in \Eq{Efield1}, which then gives
\be 
\label{eq:Einfinite}
\E = \frac{1 - \epsa}{\eps} \, \E_{\eff}
~.
\ee
The total current is then
\be 
\label{eq:totj}
\jv = (\eps - 1) \, \del_t \E + (\epsa - 1) \, \del_t \E_{\eff} = \frac{\epsa - 1}{\eps} \, \del_t \E_{\eff} = \frac{\jv_a^P}{\eps}
~.
\ee
This equation encompasses all of the results of Ref.~\cite{Slonczewski:1985oco}; in particular, it shows that for large $\eps$, which occurs in conductors with $\sigma \gg m_a$, the axion-induced current is significantly screened. On the other hand, it was not previously realized that the current can be resonantly enhanced when $\eps$ approaches zero, which occurs when the axion mass matches that of a quasiparticle that mixes with the photon, such as a plasmon or phonon~\cite{Marsh:2018dlj,Lawson:2019brd,ALPHA:2022rxj,Mitridate:2020kly,Balafendiev:2022wua,Schutte-Engel:2021bqm,Marsh:2022fmo,Berlin:2023ppd}. However, the frequencies of such resonances are often not easily tunable. 

In any case, dielectric haloscope and dish antenna experiments do not measure the current inside a medium, but rather the propagating radiation produced outside of it. The simplest setup which produces such radiation is an infinite slab of material of thickness $d$ placed in vacuum, carrying a uniform spin polarization lying in the slab's plane, as shown in the left panel of \Fig{cartoon}. Within the slab, the axion induces an electric field as in \Eq{Einfinite}. However, if this was the total electric field, then the component of the electric field tangential to the plane $\E_\parallel$ would be discontinuous at the slab's boundaries. Instead, the continuity of $\E_\parallel$ is restored by including plane wave solutions of \Eq{Efield1}, which propagate outward from the slab; such fields also exist within the slab, though they are exponentially damped on the scale of the skin depth $1/ (\text{Im}(n) \, m_a)$. In other words, the slab's finite thickness breaks translational invariance, providing the momentum mismatch required to generate photons.

As shown in \App{appendixlayer}, enforcing continuity of $\E_\parallel$ and $\B_\parallel/\mu$ yields a signal amplitude \timec
\be
\label{eq:ElecSlab}
E_\sig = \bigg| \frac{J_a^P / m_a}{\eps \tsw{-}{+} i \, n \, \cot{(n \, m_a \, d / 2)}} \bigg|
\ee
for the outgoing radiation field. For real $\eps$ and $\mu$, the amplitude is maximized for $d \simeq \pi / (n \, m_a)$, in which case $E_\sig = J_a^P / (m_a \, \eps)$. Note that photon-coupled axions or kinetically-mixed dark photons also produce such signals, but through different physical mechanisms; for photon-coupled axions the fields are sourced by effective currents which exist wherever a background magnetic field $B_0$ is present, while for electron-coupled axions they are from real currents, which exist only within the slab. Despite these differences, the boundary conditions in these various model-examples have the same form, such that the expression in \Eq{ElecSlab} holds in all three cases~\cite{Millar:2016cjp,Baryakhtar:2018doz} if one formally substitutes
\be 
\label{eq:PhotonSlab}
J_a^P \to \sqrt{\rhodm} \times 
\begin{cases} (\epsa - 1) \, g_{ae} \, m_a^2 / e & \text{(electron-coupled axion)} 
\\ (\eps - 1)\, g_{a \g \g} \, B_0 & \text{(photon-coupled axion)} 
\\ (\eps - 1) \, \kappa \, m_{A'} & \text{(kinetically-mixed dark photon)}
~.
\end{cases}
\ee
In the second line, $g_{a \g \g}$ is the axion-photon coupling, and in the third line, $\kappa$ and $m_{A'}$ are the kinetic mixing parameter and mass of the dark photon field, which we assume is polarized tangential to the slab (although a real experiment would have to average over its varying orientation~\cite{Horns:2012jf,Caputo:2021eaa}). 

To compare these signals, we must estimate $\epsa$, the permittivity due to the spin-polarized electrons. Consider a material in which the electrons in the outermost partially-filled shell are completely spin-polarized. Those spin-polarized electrons contribute to $\epsa$, and at low frequencies $\w \lesssim 10 \ \eV$, below the characteristic scale of electronic excitations, they are also the ones that primarily contribute to $\eps$. Therefore, in general we expect $\epsa \sim \eps$ up to an $\order{1}$ factor. This argument is discussed in more detail in \Sec{abs_classical}, and for this initial study we simply take $\epsa = \eps$.

Under this assumption, \Eq{PhotonSlab} allows us to map between couplings which yield signals of equivalent strength,
\be 
\label{eq:ElecMapping}
g_{ae} ~\leftrightarrow~ g_{a \g \g} \, (e \, B_0 / m_a^2) ~\leftrightarrow~ \kappa \, (e / m_a)
~.
\ee
As discussed in \App{appendixlayer}, this mapping is general, holding for an arbitrary series of spin-polarized layers. It can thus be used to recast the sensitivity of dielectric haloscopes, provided they are modified to incorporate spin-polarized media. The strongest sensitivity for this type of experiment is achieved at large axion masses, at which the optical dielectric haloscopes LAMPOST and MuDHI operate. Since LAMPOST has published projections~\cite{Baryakhtar:2018doz}, we recast them in \Fig{future} to yield the curve labeled ``Axioelectric Multilayer,'' which shows that such a setup could have sensitivity comparable to existing solar limits, corresponding to couplings motivated by the QCD axion.

\subsection{Axion Wind Magnetization Currents}
\label{sec:MADMAX}

Next, we turn to the magnetization current $\jv_a^M$ of \Eq{axioncurrent}. This term results from the axion wind effective magnetic field $\B_{\eff} \propto \grad a$, which is nonzero due to the solar system's motion through the dark matter halo, i.e., $\grad a \sim m_a \, \vv_{_\text{DM}} \, a$ with $\vdm \sim 10^{-3}$ the velocity of the dark matter wind. First, note that $\jv_a^M$ vanishes in an infinite medium with a uniform and scalar permeability $\mu$, since from \Eqs{Beff}{axioncurrent} we have in this case that $\jv_a^M \propto \grad \times \B_{\eff} \propto \grad \times \grad a = 0$. Thus, for scalar $\mu$, the simplest nontrivial situation is a planar slab of material of finite thickness, where the discontinuity of the magnetization at the slab's boundaries leads to surface magnetization currents, shown on the right in \Fig{cartoon}. 

Such a signal arises in any finite medium with nonzero magnetic susceptibility $\x_m = \mu - 1$, but in practice $\x_m$ is small unless the medium is magnetically ordered, in which case $\x_m$ is a nontrivial tensor. Furthermore, since the axion field oscillates in time and the medium can be placed in a tunable magnetic field, we must also consider how $\x_m$ depends on frequency and external field. 

\subsubsection{Magnetization Dynamics}
\label{sec:magdyn}

The form of $\x_m$ can be derived from the well-understood theory of classical magnetization dynamics. Here we review the relevant results, following standard introductory treatments~\cite{lan84,chikazumi1997physics,coey2010magnetism,pozar2011microwave}. The starting point is the Landau--Lifshitz equation, which governs the time evolution of a medium's magnetization density $\Mv$,
\be 
\label{eq:LL_equation}
\frac{d\Mv}{dt} = \gamma \, \Mv \times \Hv + \frac{\alpha \, \gamma}{|\Mv|} \, \Mv \times (\Mv \times \Hv)
~,
\ee
where $\Hv = \B - \Mv$ is the auxiliary magnetic field, $\alpha$ is a small dimensionless damping parameter due to internal losses, and $\gamma$ is the in-medium gyromagnetic ratio, approximately equal to the electron's gyromagnetic ratio $\gamma_e \simeq -e/m_e\, $. 

For concreteness, consider a medium prepared with high magnetization density $\Mv_0 = M_0 \, \zhat$, aligned with a large applied magnetic field $\Bv_0 = B_0 \, \zhat = \Hv_0 + \Mv_0$. We are interested in the magnetization's response to the small axion effective magnetic field $B_\eff \ll B_0$, which contributes small corrections $\Mv = \Mv_0 + \Delta \Mv$ and $\Hv = \Hv_0 + \Delta \Hv$. Defining $\Delta \Mv = \x_m \, \Delta \Hv$, we can solve \Eq{LL_equation} for $\x_m$ in frequency space by linearizing in the small components. This yields 
\be
\label{eq:Polder1}
\x_m \simeq - \, \frac{\w_M}{\w^2 - (1 + \alpha^2) \, \w_H^2 + 2 i \alpha \, \w \, \w_H} \begin{pmatrix} (1 + \alpha^2) \, \w_H - i \alpha \, \w & - i \w & 0 \\ i \w & (1 + \alpha^2) \, \w_H - i \alpha \, \w & 0 \\ 0 & 0 & 0 \end{pmatrix}
~,
\ee
which in the absence of damping ($\alpha = 0$) is known as the Polder tensor. In the presence of damping, the width of the resonant response is controlled by the magnetic quality factor $Q \equiv 1 / (2 \alpha)$. Here, we have introduced the angular frequencies $\w_M \equiv |\gamma| \, M_0$ and $\w_H \equiv |\gamma| \, H_0$, and we also define $\w_B \equiv |\gamma| B_0 = \w_H + \w_M$. Note that the maximum realistic value of the external field, $B_0 = 10 \ \text{T}$, corresponds to $\w_B \sim 1 \  \meV$, such that for larger frequencies the susceptibility is necessarily strongly suppressed. As a result, $\mu \simeq 1$ at optical frequencies, which is why we were able to neglect the tensorial nature of $\mu$ in \Sec{LAMPOST}. 

It is convenient to diagonalize \Eq{Polder1} by describing the transverse components with circular polarizations, defined as $M_\pm \equiv (M_x \pm i M_y)/\sqrt{2}$, $H_\pm \equiv (H_x \pm i H_y)/\sqrt{2}$, and similarly for the other fields. In this basis, the magnetic susceptibility is $\x_m \simeq \text{diag}(\x_+, \x_-, 0)$, where the diagonal elements are given by
\be
\label{eq:xpm}
\x_\pm = \frac{\pm \w_M + i \w_M/2Q}{\w \pm \w_H + i \w_H / 2 Q}
~.
\ee
Note that for positive frequencies, only $\x_-$ can be resonantly enhanced. In other words, if the material is driven by a positive frequency linearly polarized magnetic field (which contains equal magnitude plus and minus circular polarization components), then on resonance its magnetization preferentially rotates in one direction. This predominantly leads to clockwise circularly polarized radiation propagating along the direction of $\v{B}_0$, and counterclockwise circularly polarized light propagating in the opposite direction.

\subsubsection{Form Factor for a Single Slab}
\label{subsubsec:form_factor_for_a_single_slab}

Let us now return to the case of a single planar slab of finite thickness $d$, extending infinitely along the $xy$-plane. For simplicity, we suppose that $\grad a$ is uniform and also lies in the $xy$-plane. The slab is placed inside a strong external magnetic field $\Bv_0 = B_0 \, \zhat$, which fully magnetizes it along the $z$-direction, normal to the slab's surface. Note that since $\Bv \cdot \zhat$ is continuous across the slab's boundaries, the $z$-component of the magnetic field inside the slab is also equal to $B_0$. We can compute the outgoing radiation produced by a single slab using the same method as in~\Sec{LAMPOST}, provided we work in a basis of circular polarizations and account for the discontinuity of $(\mu^{-1} \, \B) \times \zhat$ at the slab boundaries due to the axion-induced magnetization surface currents. As shown in \App{appendixlayer}, the resulting amplitudes of the outgoing radiation components have magnitude $|E_\sig^\pm| = \FF_\pm \, B_{\eff} / \sqrt{2} \, $, where \timec
\be
\label{eq:MagSlab}
\FF_\pm = \bigg|\frac{\x_\pm}{\mu_\pm \tsw{-}{+} i n_\pm \, \cot{(n_\pm \, m_a \, d / 2)}} \bigg|
~,
\ee
is a dimensionless form factor and $n_\pm = \sqrt{\eps \mu_\pm}$ is the polarization-dependent refractive index. Comparing \Eqs{ElecSlab}{MagSlab}, we see that if $\FF_\pm$ is $\order{1}$, then the radiation amplitude due to the axion wind is larger than that due to the axioelectric effect for all $m_a \lesssim m_e \, \vdm \sim \keV$. However, a multilayer experiment would become impractically large for $m_a \ll 1 \ \mu\eV$ and, as mentioned above, the magnetic susceptibility is suppressed for $m_a \gtrsim 1 \ \meV$. Thus, the approach described in this section is most useful for targeting axions at microwave frequencies, in a setup analogous to the MADMAX dielectric haloscope that we refer to as a ``magnetized multilayer"~\cite{Caldwell:2016dcw,Millar:2016cjp,MADMAX:2019pub}.

\begin{figure*}[t]
\includegraphics[width=9cm]{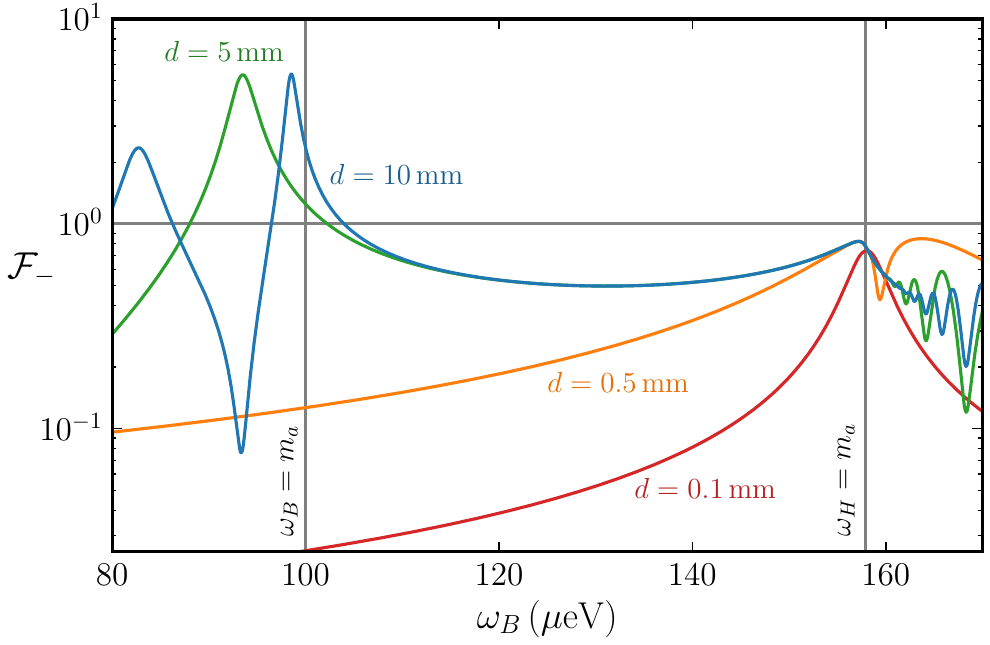} 
\caption{The form factor $\mathcal{F}_-$ (\Eq{MagSlab}) for slabs of varying thickness $d$ as a function of $\w_B = |\gamma| B_0$, for a fixed axion mass of $m_a = 100 \ \mu \eV$ and the material properties listed in Table~\ref{tab:noise}. While the form factor always has a peak near $\w_H = m_a$, it has a higher peak near $\w_B = m_a$ for thick slabs.}
\label{fig:form} 
\end{figure*}

The external field strength $B_0$ can easily be tuned in the laboratory; it affects $\w_B$ and $\w_H$, and thereby the susceptibility $\x_\pm$ and related quantities. We now consider how it should be chosen to maximize the form factor $\mathcal{F}_\pm$ for a given axion mass. For concreteness, we focus on the ``minus'' polarization, which has permeability
\be \label{eq:mu_expr}
\mu_- = \frac{m_a - \w_B + i \w_B/2Q}{m_a - \w_H + i \w_H/2Q}
~.
\ee
\Fig{form} shows the value of $\FF_-$ as a function of $\w_B$, for an axion mass of $m_a = 100 \ \mu \eV$ and slabs of various thickness $d$. For thin slabs, the form factor approaches a peak value of $\FF_- \simeq 1$ when $\w_H = m_a$. On the other hand, for much thicker slabs, there is a second, parametrically higher peak when $\w_B$ is close to but slightly below $m_a$. 

These results can be understood by carefully considering limiting cases of \Eq{MagSlab} for fixed $m_a$. First, in the thin slab regime, $|n_- m_a d| \ll 1$, the cotangent in \Eq{MagSlab} is large and can be expanded as $\cot{x} \simeq 1/x$. The form factor has a resonant peak about $\w_H = m_a$, where the numerator $\x_- \simeq \mu_-$ becomes large. Near this peak, we have
\be 
\label{eq:simple_form}
\FF_- \simeq \bigg|\frac{\w_M \, m_a \, d / 2}{m_a - \w_H + i \w_H/2 Q_\eff}\bigg|
~,
\ee
where we took $Q \gg 1$ and defined an effective quality factor
\be 
\label{eq:q_eff}
\frac{1}{Q_\eff} = \frac{1}{Q} + \w_M \, d
~.
\ee
These results are physically sensible. First, the resonant frequency $\w_H$ is the usual Kittel frequency (i.e., the lowest, zero-momentum magnon frequency) for a thin slab in an orthogonal magnetic field. Next, the second term in \Eq{q_eff} describes radiation damping due to the emission of electromagnetic waves. This damping ensures $\FF_- \leq 1$ on resonance, regardless of the value of $Q$, while increasing $d$ simply broadens the width of the resonant response. Thus, for thin slabs in the regime $Q \, \w_M\, d \gtrsim 1$, increasing the slab thickness can increase the scanning rate of an experiment, but not the peak signal power.

Radiation damping is a familiar concern in ferromagnetic resonance studies~\cite{sanders1974radiation,wende1976radiation,schoen2015radiative,weymann2021magnetic} and is one of the main reasons that ferromagnetic haloscope experiments enclose their spin-polarized sample in a microwave cavity. Note, however, that \Eqs{simple_form}{q_eff} apply only to thin slabs; qualitatively different behavior can occur in the thick slab limit $|n_- m_a d| \gtrsim 1$. In this case the cotangent $
\mucot$ is not necessarily large, but is instead generically $\order{1}$. From \Eq{mu_expr}, when the quality factor is high and the axion mass matches the Kittel frequency of an infinite medium, $m_a = \w_B$, the permeability is small and approximately imaginary, $\mu_- \simeq i m_a / (2 \, Q \, \w_M)$, so that $|\mu_-| \ll |n_-|$. Then
\be \label{eq:Fm_eq}
\mathcal{F}_- \simeq \frac{1}{|\mu_- + i n_-
\mucot|} \simeq \frac{1}{|
\mucot|} \, \sqrt{\frac{2 \, Q \, \w_M}{m_a \, \eps}}
~,
\ee
so that the form factor can be much greater than one if $Q$ is sufficiently large. This enhancement is possible because for thick slabs, radiation is not merely a source of energy loss. Instead, it couples the magnon and photon degrees of freedom within the slab, forming a propagating hybrid ``magnon-polariton'' mode. 

As shown in \Fig{form}, we find numerically that the highest form factors occur when $\w_B$ is shifted slightly below the axion mass, $\w_B = m_a - \Delta \w$ with $\Delta \w > 0$, even though this increases $|\mu_-|$. This result also has a simple physical interpretation. For concreteness, suppose that $\Delta \w$ is nonzero and $Q$ is very high, which implies that $\mu_-$ is small and approximately real. In this case, the slab acts as an effective cavity due to the discontinuity of $n_-$ at its boundaries. When $n_- m_a d = \pi$, the in-medium wavelength of the magnon-polariton mode matches the slab's thickness, so that $\mucot$ vanishes, leading to a greatly enhanced form factor. This phenomenon is familiar from other axion searches using quasiparticle resonances~\cite{Lawson:2019brd,Schutte-Engel:2021bqm,Balafendiev:2022wua,Marsh:2022fmo,ALPHA:2022rxj} and remains approximately true even at finite $Q$. 

\subsubsection{Signal-to-Noise Ratio for a Multilayer Setup}

Now that we have discussed the outgoing radiation from a single magnetized layer, we turn to the signal for many layers. Optimizing the response of a general multilayer experiment is analytically intractable. Therefore, for concreteness, we focus on a ``transparent mode'' setup, where $N$ slabs, each of area $A$ and thickness $d = \pi / (\text{Re}(n_-) \, m_a)$, are separated by vacuum gaps of thickness $\pi/m_a$~\cite{Jaeckel:2013eha,Millar:2016cjp}. For such multilayers, an emitted electromagnetic wave accumulates a phase of $2 \pi$ upon traveling from one slab to the next, so that the total signal amplitude emitted from the $+ z$ side of the stack is ideally $N E_\sig^-$.\footnote{Radiation of the ``plus'' circular polarization is also emitted from this side, but its amplitude is negligible, since generically $\FF_+$ will not be resonantly enhanced and its contributions from the different slabs will not interfere constructively. Thus, from this point onward we will only consider the ``minus'' polarization.} In this case, the time-averaged signal power emitted from each end of the stack is therefore 
\be \label{eq:P_sig_1}
P_\sig = \frac{1}{2} \, N^2 \, |E_\sig^-|^2 ~ A = \frac{1}{4} \, \FF_-^2 \, N^2 \, |B_\eff^2| \, A 
~.
\ee
As in the MADMAX experiment~\cite{Millar:2016cjp}, this power can be focused with a horn antenna onto a pickup circuit, which is coupled to an amplifier. We demand that the total thickness $L = N d$ of the slabs be no larger than the characteristic screening length $1/(\text{Im}(n_-) \, m_a)$. This implies that the maximum number of layers is 
\be 
\label{eq:N_eq}
N = \frac{\text{Re}(n_-)}{\pi \, \text{Im}(n_-)}
~.
\ee
Finally, the signal-to-noise ratio is given by the Dicke radiometer equation~\cite{dicke1946measurement},
\be
\text{SNR} = \frac{P_\sig}{T_n} \sqrt{\frac{t_{\text{int}}}{\Delta \nu_a}} \label{eq:Dicke_radiometer_equation}
~,
\ee
where $T_n$ is the noise temperature, $t_{\text{int}}$ is the integration time, and $\Delta \nu_a \simeq m_a / (2 \pi Q_a)$ is the axion bandwidth, where $Q_a \sim 1/\vdm^2 \sim 10^6$ is the effective axion quality factor. The integration time is set by $t_{\text{int}} = (\Delta \w_s/m_a) \, t_e$, where $t_e$ is the total time to scan one $e$-fold in axion masses and $\Delta \w_s$ is the sensitivity bandwidth (i.e., the spread in axion masses for which the signal power is near its maximal value for a fixed multilayer geometry and applied field). 

To estimate the sensitivity bandwidth, we consider what happens when the axion mass is shifted by a small amount $\delta$ from the optimal value, such that $m_a = \w_B + \Delta \w + \delta$. When $\delta = 0$, radiation accumulates a phase of $2 \pi$ upon propagating from one slab to the next, for a total of $2\pi N$ through the entire stack. When $\delta \neq 0$, the wave frequency is shifted, which changes the phase by a fractional amount $\sim \delta / m_a$. Demanding that the total change in phase is less than $\order{1}$ (so that the radiation from each slab still constructively interferes) then yields the constraint $\delta \lesssim m_a/N$, as in standard dielectric haloscope experiments~\cite{Millar:2016cjp}. However, in our case there is another constraint: changing the axion mass also changes the wavelength of the radiation within the slab, by a fractional amount $\Delta (\text{Re}(n)) \, / \, \text{Re}(n) \simeq \delta / (2 \, \Delta \w)$, and this quantity must also be less than $\sim 1/N$. This yields the stronger condition $\delta \lesssim 2 \, \Delta \w / N$ and therefore fixes the sensitivity bandwidth to
\be
\Delta \w_s = 2 \, \Delta \w / N
~.
\ee
Note that because $P_\sig \propto N^2$ and $\Delta \w_s \propto 1/N$, the setup obeys the so-called ``area law,'' $\int d \w \, P_\sig (\w) \propto N$. This is a very general feature of axion dark matter experiments~\cite{Lasenby:2019hfz,Chaudhuri:2021xjd} that was first observed for dielectric haloscopes~\cite{Millar:2016cjp}.

Given the above discussion, one can determine the signal-to-noise ratio in terms of $\Delta \w = m_a - \w_B$ and fixed parameters. Since the applied magnetic field can be tuned experimentally, we numerically optimize the sensitivity with respect to $\Delta \w$. Qualitatively, moving away from $\w_B = m_a$ by increasing $\Delta \w$ increases $\text{Re}(\mu_-)$ (which decreases the thickness of each slab) and decreases $\Im(\mu_-)/\text{Re}(\mu_-)$ (which increases the maximum number of layers). On the other hand, the increasing separation between the Kittel frequency and the axion frequency eventually begins to suppress the form factor. In the absence of other constraints, the optimal value of $\Delta \w$ is set by a trade-off between these effects.

In \App{YIG}, we show analytically that for real $\eps$ and sufficiently high $Q$, the optimized signal power scales as 
\be
\label{eq:Psigopt}
P_\sig \sim \bigg(\frac{Q \, \w_M}{m_a} \bigg)^2 \, |B_\eff^2| \, A
~.
\ee
The enhancement with $\w_M$ is simply due to the fact that a higher magnetization improves the form factor. The unusual quadratic scaling with $Q$ is also simple to understand; a larger quality factor reduces $\Im(\mu_-)$, and thereby improves both the resonant enhancement in the form factor and increases the maximum possible number of layers. As a result, \Eq{Psigopt} can be reexpressed as $P_\sig \propto Q \, V$, where $V$ is the total volume of the experiment, as in other resonant setups. 

We caution that in our discussion below, we will focus on a material for which $Q$ is too small for the approximations used to derive \Eq{Psigopt} to accurately apply. We also note that \Eq{Psigopt} cannot be applied at sufficiently small axion masses, where the signal power will be further constrained by an upper limit on the total slab thickness. Regardless, \Eq{Psigopt} does display the correct qualitative dependence of the signal power on material properties. 

\subsubsection{Material Properties and Experimental Parameters}

\renewcommand{\arraystretch}{1.2}
\setlength{\tabcolsep}{12pt}
\begin{table*}
    \begin{center}
        \begin{tabular}{@{}llccc@{}} \toprule
        \textbf{Parameters} & Description & Variable & \multicolumn{2}{c}{Value} \\\toprule
        \textbf{Material} & Saturation magnetization & $M_S$ & \multicolumn{2}{c}{$0.5 \ \text{T}$} \\ 
        & Magnetic quality factor & $Q$ & \multicolumn{2}{c}{$10^2$} \\
        & Permittivity & $\varepsilon$ & \multicolumn{2}{c}{$15$} \\ \hline
        \textbf{Experimental} & Slab area & $A$ & \multicolumn{2}{c}{$1\ \text{m}^2$} \\ 
        & $e$-fold scanning time & $t_e$ & \multicolumn{2}{c}{$1\ \text{year}$} \\
        & Maximum number of layers & $N_\text{max}$ & \multicolumn{2}{c}{80} \\ 
        & Maximum total material thickness & $L_\text{max}$ & \multicolumn{2}{c}{$5\ \text{m}$} \\ 
        & Maximum applied $B$ field & $B_\text{max}$ & \multicolumn{2}{c}{$10\ \text{T}$} \\ \hline
        \textbf{Noise} & & & HEMT & SQL \\
        & Physical temperature & $T$ & $4 \ \mathrm{K}$ & $40 \ \mathrm{mK}$ \\ & Amplifier noise temperature & $T_\text{amp}$ & 
        $1 \ \mathrm{K} ~ \left(\dfrac{m_a}{2 \pi \times 4 \ \GHz}\right)$ & $m_a$ \\\bottomrule
        \end{tabular}
    \end{center}
    \caption{Material, experimental, and noise parameters assumed when optimizing the experimental setup (Fig.~\ref{fig:values}) and computing the sensitivity (Fig.~\ref{fig:projection_ferrite}) for a polycrystalline spinel ferrite multilayer. The dielectric loss tangent is negligible in these materials, $\tan \delta_\eps \lesssim 10^{-4}$, and so the permittivity can be approximated as real. For HEMT and SQL amplifiers, the physical temperatures correspond to cryostat and dilution fridge cooling, respectively. The total noise temperature used in \Eq{Dicke_radiometer_equation} is $T_n = T + T_\text{amp}$.}
    \label{tab:noise}
\end{table*}

The best materials are those with high saturation magnetization $M_S$, which sets the maximum possible value of $M_0$, and large magnetic quality factor $Q$. In addition, the permittivity $\eps$ must be approximately real, as a large imaginary component would lower the screening length. In particular, magnetic alloys cannot be used, as they have $\text{Im}(\eps) = \sigma/\w$ for a large conductivity $\sigma$. For such materials, the screening length (i.e., the skin depth) would be well below the thickness of a single slab at the microwave frequencies considered here. 

We therefore choose to focus on ferrites, which have negligible conductivity and are widely commercially available. The properties of these materials are well known, and discussed in detail in \App{YIG}. To date, ferromagnetic haloscope experiments have exclusively used single crystal yttrium iron garnet (YIG), as it has the highest known quality factor $Q \sim 10^4$. However, YIG crystals are extraordinarily difficult to grow~\cite{emori2021ferrimagnetic}, and currently YIG spheres and films can only be produced individually at $\sim 1 \ \text{mm}$ scales. By contrast, polycrystalline spinel ferrites are mass-produced and can be purchased at a per-kilogram cost over five orders of magnitude lower than YIG. Though they have a relatively low quality factor $Q \sim 10^2$, they possess a saturation magnetization $M_S$ twice as large as that of YIG. 

Thus, we will focus on polycrystalline spinel ferrites, since an experiment employing them can benefit tremendously by the increased detector volume. The benchmark values we assume are shown in Table~\ref{tab:noise} and discussed further in \App{YIG}. We assume the material is fully magnetized, so that $M_0 = M_S$. As for the other experimental parameters, we adopt a slab area $A = 1 \ \mathrm{m}^2$ and $e$-fold scanning time $t_e = 1 \ \mathrm{yr}$, similar to the MADMAX experiment. We allow a maximum external magnetic field of $B_\text{max} = 10 \ \text{T}$, require that the number of layers does not exceed $N_\text{max} = 80$, and cap the total thickness $L = N d$ of the slabs at $L_{\text{max}} = 5 \ \text{m}$.

\subsubsection{Experimental Sensitivity}
\label{sec:subsection_experimental_sensitivity}

\begin{figure*}[t]
\includegraphics[width=2.35in]{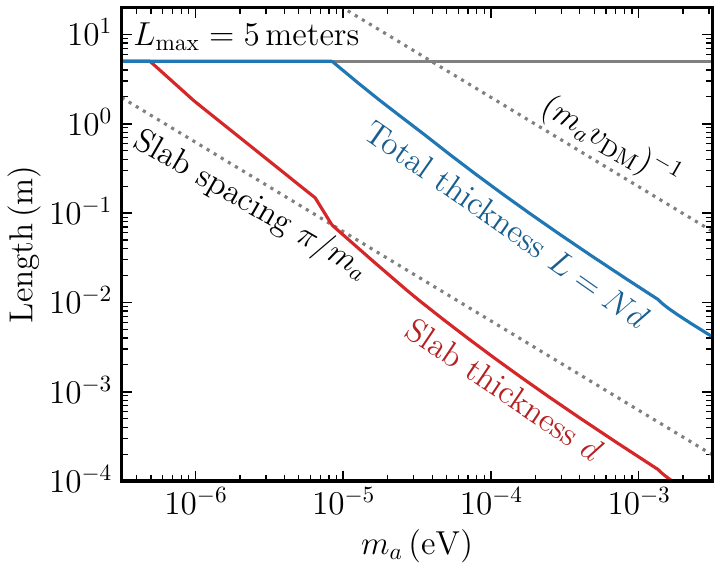} \hspace{-0.1in}
\includegraphics[width=2.35in]{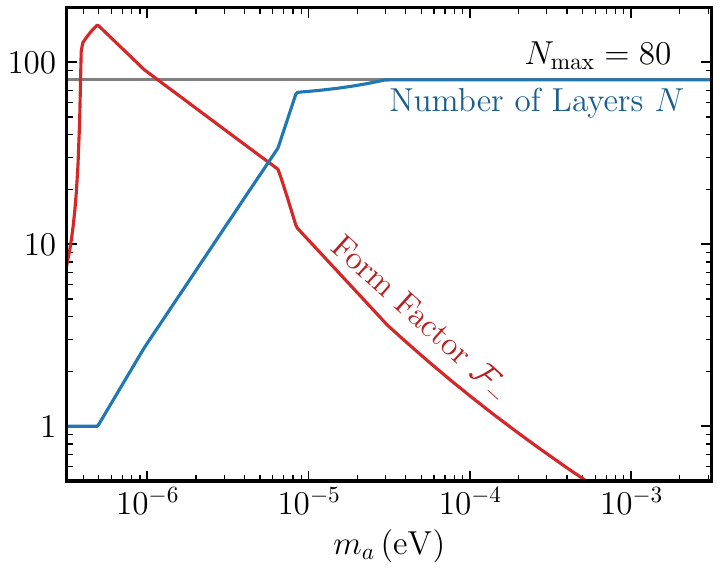} \hspace{-0.1in}
\includegraphics[width=2.35in]{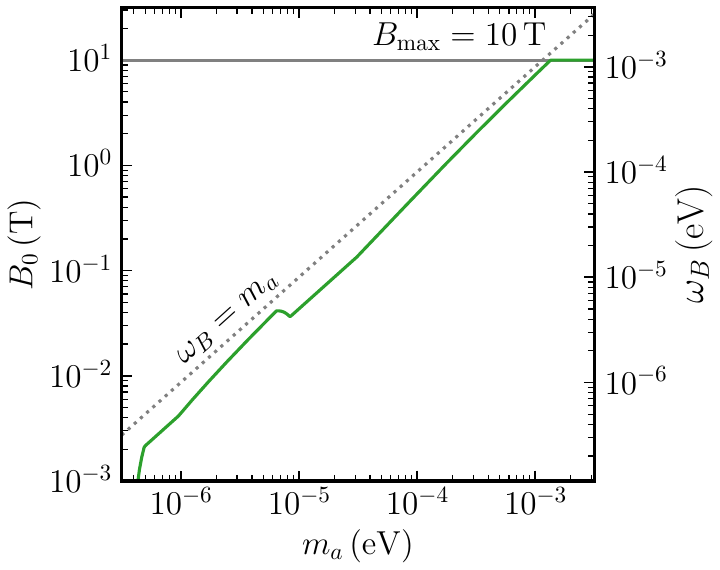} 
\caption{Numerically optimized values of the slab thickness and total thickness (left panel), form factor and number of layers (middle panel), and applied external magnetic field (right panel), as a function of the axion mass. The middle panel shows that the signal power, proportional to $\FF_-^2 N^2$, is predominantly enhanced by the form factor at small $m_a$ and the number of layers at large $m_a$. The other panels show that a large total volume is only required at small $m_a$, and a strong external field is only required at large $m_a$.}
\label{fig:values} 
\end{figure*}

For each axion mass, we numerically optimize the signal-to-noise ratio by varying $\w_B$, which in turn determines the form factor $\FF_-$, number of layers $N$, slab thickness $d$, and sensitivity bandwidth $\Delta \w_s$. The results of the optimization are shown in \Fig{values}, which shows how the setup qualitatively changes as the axion mass is varied.

At small axion masses, $m_a \lesssim 10^{-5} \ \eV$, the length of the setup is large, but it only requires a weak applied magnetic field. In this regime, the number of layers is suppressed due to the constraint $L \leq L_\text{max}$, and the sensitivity boost primarily comes from operating at small $\mu_-$, corresponding to a large form factor. Note that this implies the thickness of a single slab is much greater than the spacing between slabs, so that $L$ is approximately the total length of the multilayer setup. In this regime, the experiment most closely resembles axion searches using tunable quasiparticle resonances, such as TOORAD~\cite{Marsh:2018dlj,Schutte-Engel:2021bqm} and ALPHA~\cite{Lawson:2019brd,ALPHA:2022rxj,Balafendiev:2022wua}.

Increasing the axion mass increases the minimum value of $|\mu_-|$ and therefore decreases the maximum possible form factor. Thus, at large axion masses the sensitivity boost primarily comes from having a large number of layers and is limited by the constraint $N \leq N_\text{max}$. For polycrystalline ferrite, one can only use this many layers if the setup is operated well away from the resonance. Numerically, we find $\Delta \w \gtrsim \w_M$ in this regime, so that $n_-$ does not have sharp frequency dependence, and the experiment most closely resembles the microwave dielectric haloscope MADMAX. Also note that for $m_a \gtrsim 10^{-3} \ \eV$, our constraint $B_\text{max} = 10 \ \text{T}$ limits the size of $B_0$, preventing resonant enhancement of the form factor entirely. For axion masses near this upper limit, the required magnetic field is large, but the total length of the setup is small, i.e., $\lesssim 1 \ \text{cm}$. 

For the parameter values we have chosen, the analytic approximation for the signal power in \Eq{Psigopt} is only accurate in a narrow mass range centered near $m_a \sim 10^{-5} \ \eV$. For larger axion masses, we find numerically that the optimized signal power is an $\order{1}$ factor larger. In \App{YIG}, we also carry out the same computation for single crystal YIG, which we show has a sufficiently high $Q$ for the analytic results to work accurately. For YIG, the reach is slightly stronger, but the results are qualitatively very similar. 

\begin{figure*}[t]
\includegraphics[width=12cm]{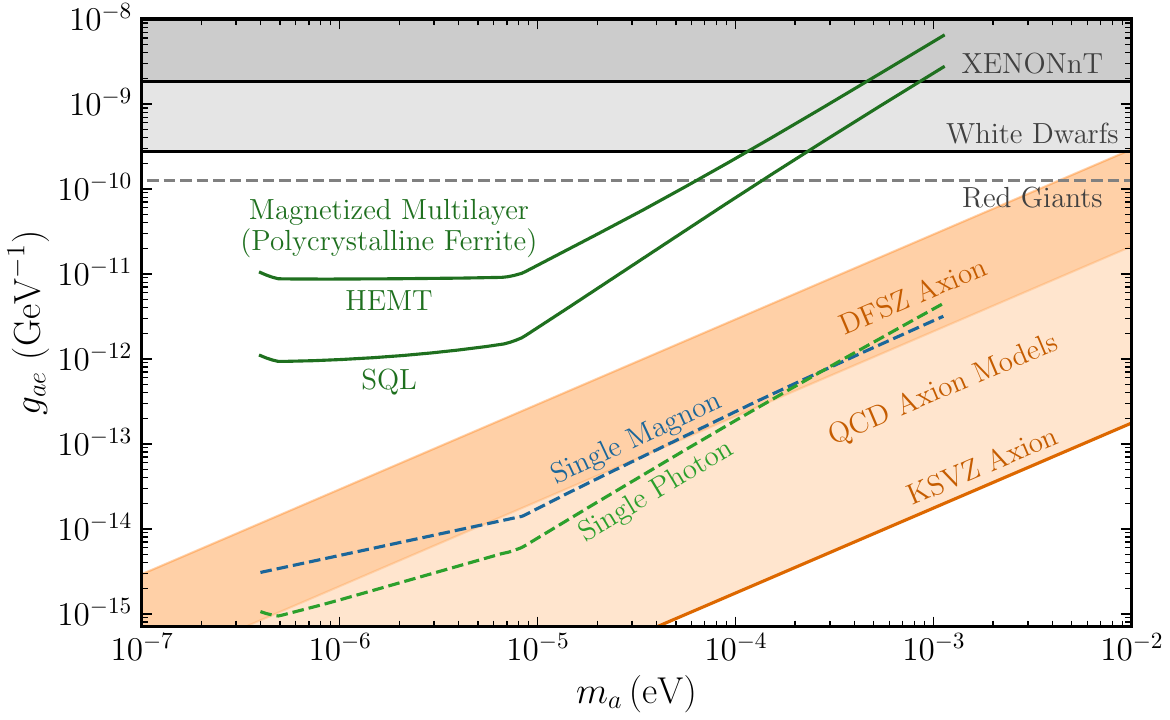} 
\caption{In solid green, we show the projected sensitivity at $\text{SNR} = 2$ to the axion-electron coupling of a magnetized multilayer experiment based on detecting the radiation generated by the axion wind induced magnetization current, with layers made of polycrystalline ferrite and either HEMT or SQL amplifiers. All parameters assumed for the projections are listed in Table~\ref{tab:noise}. We cut off these reach curves at large axion masses when the required applied magnetic field exceeds $B_{\text{max}} = 10 \ \text{T}$ and at small axion masses when the optimal thickness of a single layer exceeds $L_{\text{max}} = 5 \ \text{m}$. In dashed green and dashed blue, we show the maximum possible reach given noise-free detection of single photon and magnon quanta, respectively (see \Sec{subsection_experimental_sensitivity} for more details). Astrophysical bounds are as in \Fig{existing}, and the band for DFSZ axions and loop-induced couplings for KSVZ axions are summarized in, e.g., Ref.~\cite{Irastorza:2018dyq}.}
\label{fig:projection_ferrite}
\end{figure*}

The numeric values of $d$ and $L$ shown in the left-panel of \Fig{values} allow us to confirm that certain corrections to our results are indeed negligible. First, we have treated $\grad a$ as spatially uniform. This mathematical assumption corresponds to the physical requirement that radiation from different slabs interferes constructively, $L \ll (m_a \, \vdm)^{-1}$, which is indeed true here. Note that at large axion masses we have $\sqrt{A} \gtrsim (m_a \, \vdm)^{-1}$, but this is acceptable since there is no requirement that radiation from different parts of the same slab be emitted in phase. Second, we remind the reader that we have only considered surface magnetization currents. In the beginning of this subsection, we noted that for scalar $\mu$, the volume magnetization current vanishes. However, it is in general nonzero for the tensorial $\mu$ considered here, albeit suppressed by an additional gradient of the axion field. Thus, the relative effect of the volume magnetization current is $\sim d \, m_a \, \vdm \ll 1$, which is completely negligible. 

Finally, the total noise temperature can be decomposed as $T_n = T_{\text{th}} + T_{\text{amp}} + \Delta T$, where $T_{\text{th}}$ is due to thermal noise, $T_{\text{amp}}$ is the effective noise temperature of the amplifier, and $\Delta T$ is due to additional, reducible noise sources. By the fluctuation-dissipation theorem, thermal noise is sourced from any part of the system that dissipates energy. This includes thermal radiation sourced from the charges in the walls of the magnet and the thermally fluctuating spins in the multilayer, and Johnson--Nyquist noise sourced from the resistance in the pickup circuit. However, when the entire system is at the same physical temperature $T$, these contributions sum to $T_{\text{th}} = T$ independently of the details of the setup. As for the additional noise $\Delta T$, it receives contributions from, e.g., nonthermalized magnetic impurities, vibrations of the multilayer or magnet, and Barkhausen noise due to the relaxation of domain walls. However, these effects fall off rapidly above $\sim \kHz$ frequencies, and we expect they are subdominant at the $\GHz$ to $\text{THz}$ frequencies and long integration times relevant in this work~\cite{Barkhausen1,wolf1978noise,Barkhausen2,Barkhausen3,Barkhausen4,Barkhausen5}. 
Thus, for this initial study we neglect $\Delta T$.

In \Fig{projection_ferrite}, we show the experimental sensitivity for various noise benchmarks, with corresponding noise parameters shown explicitly in Table~\ref{tab:noise}. For the line labeled ``HEMT," we have inferred an amplifier noise temperature $T_{\text{amp}}$ from manufacturer datasheets~\cite{lownoise}. We also show a ``standard quantum limit'' (SQL) benchmark with a lower physical temperature and a quantum limited amplifier. In this case quantum noise is important, and the fluctuation-dissipation theorem implies that it arises from a variety of sources, including quantum fluctuations in the pickup circuit voltage, and spin-projection noise arising from the uncertainty in the transverse component of the multilayer magnetization~\cite{Aybas:2021cdk}. The details of the setup determine which of these effects contributes more to the quantum noise seen by the amplifier, but the SQL itself always corresponds to taking $T_{\text{amp}} = m_a$, independent of these details. The detailed contributions of individual sources to the quantum noise do not affect this result, but may be important for detailed optimization of the detector, which we defer to future work.

For concreteness, we set a sensitivity threshold by taking $\text{SNR} = 2$, though we note that this does not have a definite statistical interpretation, as various $\order{1}$ factors have been dropped throughout our analysis. Finally, the dashed line labeled ``Single Photon" indicates a theoretical upper limit on sensitivity; it shows axion couplings for which $P_{\text{sig}} \, t_{\text{int}} = m_a$, corresponding to an average of a single photon emitted within the integration time~\cite{Ioannisian:2017srr}. 

To illustrate the effect of the multilayer geometry, we can compare the photon power absorption rate $P_{\text{sig}}$ in an optimized multilayer to the magnon power absorption rate $P_{\text{mag}}$ in a uniform medium. As shown in \Sec{abs_classical} and \App{Ohms}, $P_{\text{mag}}$ is maximized when $\w_B = m_a$. When normalized to the same volume $V$ of material, we have
\be
\frac{P_{\text{mag}}}{P_{\text{sig}}} = \frac{2 \, Q \, \w_M \, L}{(N \, \FF_-)^2} 
~.
\ee
For a setup employing direct readout of magnons, the sensitivity bandwidth is simply the ferromagnetic linewidth $\Delta \w_s \simeq m_a/Q$, which fixes the integration time as $t_{\text{int}}^{\text{mag}} = t_e / Q$. Setting $P_{\text{mag}} \, t_{\text{int}}^{\text{mag}} = m_a$ gives the line labeled as ``Single Magnon'' in \Fig{projection_ferrite}. This represents the strongest possible sensitivity of an experiment of equal volume but trivial bulk geometry. The fact that the single magnon sensitivity is weaker than the single photon sensitivity shows that the magnon-polariton mode in an optimized multilayer effectively couples more strongly to the axion field than an infinite-medium magnon mode. Moreover, since these magnon-polaritons propagate out of the multilayer in the form of electromagnetic radiation, it is much easier to detect them precisely. 

\subsubsection{Comments on Experimental Realization}

We conclude this section with some additional comments on the experimental realization of such a ``magnetized multilayer'' experiment. First, unlike the planned MADMAX experiment, the external magnetic field here can significantly affect the medium's properties, allowing strong sensitivity to a very wide range of axion masses. Due to this wide mass range, the experimental implementation is qualitatively different at each end. Referring to \Fig{values}, at small masses ($m_a \lesssim 10^{-5} \ \eV$) one requires a large amount of material, but only a few layers and a weak external field. At larger masses ($m_a \sim 10^{-3} \ \eV$), a strong external field is required, but the entire stack is only a few centimeters long. 

Unlike many axion experiments, our concept never requires a magnetic field that is simultaneously strong and large in volume, thus avoiding the need for expensive magnets; however, we do require the magnetic field to be highly uniform. Also note that while MADMAX requires tangential magnetic fields and thus a custom dipole magnet, we require a magnetic field normal to the slabs' surfaces. These can be produced by solenoidal magnets, which are more common and less expensive. In addition, while the signal in multilayer setups can be calibrated indirectly using reciprocity theorems~\cite{Egge:2022gfp,Egge:2023cos}, in our case an electron-coupled axion acts almost exactly like an oscillating transverse magnetic field, so the response can also be calibrated by simply applying a real transverse magnetic field.

Since the sensitivity bandwidth $\Delta \w_s$ is small, our setup must be tuned to scan across a substantial range of possible axion masses. As we have noted below \Eq{Dicke_radiometer_equation}, a magnetized multilayer setup has a somewhat smaller $\Delta \w_s$ compared to a standard dielectric haloscope since the permeability varies with frequency. However, this is compensated by the fact that there are two independent tuning mechanisms. For fine adjustments, one can alter the refractive index of the material by changing the applied magnetic field, which can be done quickly and precisely. For coarse adjustments, one may instead adjust the spacing between the slabs, as planned for MADMAX~\cite{Caldwell:2016dcw}. This two-stage strategy ensures that the amount of mechanical tuning required for our setup is never greater, even in the regime $Q \gg N$. 

There are a number of experimental details that we have neglected in this work, all of which could be addressed in a more detailed analysis, along the lines of Refs.~\cite{Knirck:2019eug,MADMAX:2021lxf}. For simplicity, we have only considered transparent mode setups. These setups are analytically tractable, and as discussed below \Eq{Fm_eq} correspond to near-optimal form factors. However, other multilayer geometries may be more flexible and effective in a real experiment; determining this will require a detailed numeric optimization. Furthermore, we have treated the system as essentially one-dimensional, neglecting finite size effects, and assumed the slab thickness and spacing is perfectly uniform. Finally, we have approximated $\grad a$ as spatially constant and have taken it to lie along the plane of the slabs. In reality, the direction of $\grad a$ is anti-aligned with the Galactic motion of the solar system and thus sweeps over various directions throughout a sidereal day, incurring an $\order{1}$ penalty to the average signal power. However, the signal also exhibits a predictable modulation, which can help disentangle it from background.  
It is also important to understand the material properties in detail. As discussed in \App{YIG}, magnetic losses at the relevant cryogenic, low-power conditions are somewhat uncertain and should be measured experimentally. Ideally, the permeability itself could be measured as a function of frequency and applied field, as it entirely determines the relevant response. We have also neglected the small magnetic anisotropy fields generated inside spinel ferrites, which slightly affect the relationship between $B_0$ and $\w_B$. Finally, other materials, such as hexagonal ferrites, could also yield good sensitivity at low cost. We defer investigation of these questions to future work. 

\section{Absorption Into In-Medium Excitations}
\label{sec:absorb}

Absorption of electron-coupled axion dark matter produces a variety of in-medium excitations. In analogy to the photoelectric effect, absorption through the axioelectric term gives rise to electronic excitations. This process has been well-studied for non-spin-ordered targets, such as noble liquids~\cite{Pospelov:2008jk,Derevianko:2010kz,Dzuba:2010cw,Arisaka:2012pb,Bloch:2016sjj}, semiconductors~\cite{Bloch:2016sjj,Hochberg:2016sqx,Mitridate:2021ctr,Trickle:2022fwt}, and spin-orbit coupled materials~\cite{Chen:2022pyd}, which target axions of mass $m_a \gtrsim 10  \ \eV$, $1 \ \eV$, and $1 \ \meV$, respectively.\footnote{For $m_a \gtrsim 1 \ \meV$, the axion-electron coupling also induces $a + e \rightarrow e + \text{phonon}$ transitions in non-spin-ordered superconductors~\cite{Hochberg:2016ajh,Mitridate:2021ctr}.} In spin-polarized targets, electron-coupled axion dark matter can generate a wider range of excitations, such as meV-scale phonons~\cite{Mitridate:2023izi} and magnons~\cite{Mitridate:2020kly,Chigusa:2020gfs}, electronic transitions between Zeeman-split levels~\cite{Sikivie:2014lha}, and ``nuclear magnons'' in materials with strong hyperfine interactions~\cite{Chigusa:2023hmz}.

In this section, we extend the results for two of these excitation channels. In \Sec{magnon}, we revisit the calculation of axion dark matter absorption into magnons. We derive the absorption rate using the constitutive relations of classical electromagnetism, and show that it is determined by a magnetic energy loss function, complementing previous quantum mechanical derivations~\cite{Chigusa:2020gfs,Mitridate:2020kly}. In \Sec{absorption_into_electronic_excitations}, we compute the axioelectric absorption rate in spin-polarized targets in two complementary ways, and show that the result is parametrically different than in non-spin-polarized targets. 

\subsection{Classical Estimate of Absorption into Magnons}
\label{sec:magnon}

Spin-ordered targets support collective spin excitations known as magnons. Similar to phonons, the energy scale of magnons is typically $\sim (1 - 100) \ \text{meV}$, making them useful in the search for scattering of sub-GeV dark matter~\cite{Trickle:2019ovy,Esposito:2022bnu}, especially in models preferentially coupling to the electron spin. These excitations have also been studied in the context of axion dark matter, since the wind coupling allows an axion to be absorbed into a magnon~\cite{Barbieri:1985cp,Kakhidze:1990in,Chigusa:2020gfs,Mitridate:2020kly}. Pioneering work focused on the absorption of an axion into the lowest magnon mode, sometimes referred to as the ``Kittel mode," which can be tuned with an applied magnetic field~\cite{Barbieri:1985cp,Kakhidze:1990in,Chigusa:2020gfs}. However, gapped magnon modes exist in any spin-ordered target with more than one magnetic ion in the unit cell and therefore can be used even in the absence of an external magnetic field.\footnote{While the existence of gapped magnon modes makes the axion absorption process kinematically viable, the usefulness of these modes is limited in simple magnets by selection rules. See Ref.~\cite{Mitridate:2020kly} for more details.} A general formalism to understand such axion interactions with spin-ordered targets, with and without an external magnetic field, was developed in Ref.~\cite{Mitridate:2020kly}. 

The axion absorption rate into magnons is typically computed (at least partially) quantum mechanically. In particular, the dynamics of the spins are assumed to be governed by a Heisenberg-like Hamiltonian. This Hamiltonian is diagonalized, which defines the magnon eigensystem, and then coupled to the axion wind. This approach works for any spin-ordered target, and the absorption rate depends on the model parameters of the Heisenberg-like Hamiltonian, which are usually determined from a first-principles calculation. While this provides a starting point to understand general dark matter interactions with spin, it introduces some uncertainty since the model parameters may be difficult to measure directly. Therefore, it is useful to understand if specific calculations may be written in terms of experimentally measurable properties, within the kinematic regime appropriate for the incoming dark matter. Such an approach has been used for axioelectric absorption, and more recently developed for light dark matter coupling via a kinetically-mixed dark photon~\cite{Hochberg:2021pkt,Knapen:2021run,Knapen:2021bwg,Boyd:2022tcn}, as well as for the absorption of electromagnetically-coupled axion dark matter in magnetized media~\cite{Berlin:2023ppd}. Both of these dark matter interaction rates have been related to the ``energy loss function" $\Im ( -1/\eps )$. Here, we show that an analogous classical derivation can be used to derive the axion absorption rate into magnons for a uniform bulk material in terms of a ``magnetic energy loss function" $\Im\left( -1/\mu \right)$, where $\mu$ is the target permeability. 

Generally, the axion dark matter absorption rate per unit target mass is determined by the imaginary component of the axion field's angular frequency $\w$~\cite{Dubovsky:2015cca,Hardy:2016kme,Berlin:2023ppd},
\begin{align}
\label{eq:classical_rate}
R \simeq \frac{\rhodm}{\rhoT} \, \frac{\Im\left( \tsw{}{-} \w^2 \right)}{m_a^2} 
~,
\end{align}
where $\rhoT$ is the mass density of the target. The axion frequency can be evaluated by solving the axion's classical equation of motion, 
\be
\label{eq:aEOM0_classical}
(\partial^2 + m_a^2) \, a = - g_{ae} \, \del_\mu \left( \Psibar \g^\mu \g^5 \Psi \right)
~.
\ee
Thus, the absorption rate is related to the imaginary component of the source term $g_{ae} \, \del_\mu \left( \Psibar \g^\mu \g^5 \Psi \right)$. Classically, this source term is the sum of the contributions from each individual electron in the target, which we define as its ``expectation value,"
\be
\label{eq:aEOM1_classical}
g_{ae} \, \expect{\del_\mu \left( \Psibar \g^\mu \g^5 \Psi \right)} 
\simeq  \frac{g_{ae}}{e} \, \, \sum_{i} \, \Big( (\dt \jv_i) \cdot \shat_i \Big) + \frac{g_{ae}}{\mu_B} \, \grad \cdot \Mv
~,
\ee
where the classical expectation value of the electron axial current $\Psibar \g^\mu \g^5 \Psi$ was evaluated using the single-particle classical mapping of \Eq{av_current}. In the first term of \Eq{aEOM1_classical}, the subscript $i$ indexes the individual electrons, such that $\jv_i \equiv e \, \vv_i / V$ is the single particle current density with $V$ the target volume, and $\shat_i$ is the direction of the electron spin. In the second term, $\mu_B$ is the Bohr magneton and $\Mv = \sum_i \Mv_i$ is the total magnetization density of the target.\footnote{Since $\Mv$ enters the axion absorption rate and the usual constitutive relationships of electromagnetism in the same way, we can perform a trivial sum rather than explicitly writing down the contributions from each electron in \Eq{aEOM1_classical}. However, this is not the case for the first term in \Eq{aEOM1_classical}, and care must be taken in understanding the individual electronic response.} The first term is dominant for electronic excitations via the axioelectric effect and will be considered in detail in \Sec{abs_classical}. The second term governs absorption via the axion wind and will be the focus here.

As stated in \Eqs{Beff}{mag_def}, the axion wind electron coupling produces an effective magnetic field $\B_{\eff}$ which induces a magnetization $\v{M}_a = (1 - \mu^{-1}) \, \B_\eff$. This magnetization contributes to the second term of \Eq{aEOM1_classical}, which becomes
\be
\label{eq:aEOM_wind}
\frac{g_{ae}}{\mu_B} \, \grad \cdot \Mv = - \Big( \frac{g_{ae} \, m_a \, \vdm}{\mu_B} \Big)^2 \, \big( 1 - \qhat \cdot \mu^{-1} \cdot \qhat \big) \, a
~,
\ee
where $\qhat$ is the unit vector aligned with the axion gradient. To determine the absorption rate $R$ in \Eq{classical_rate}, we substitute \Eq{aEOM_wind} into \Eq{aEOM0_classical}, Fourier transform, and evaluate the imaginary component of the axion frequency, which yields \timec
\be
\label{eq:classicalmagnon}
R \simeq  \Big( \frac{g_{ae} \, \vdm}{\mu_B} \Big)^2 ~ \frac{\rhodm}{ \rhoT} ~ \qhat \cdot \Im\big( \tsw{}{-}\mu^{-1} \, \big) \cdot \qhat  
~.
\ee
The last factor in \Eq{classicalmagnon}, related to the imaginary part of the permeability, is the magnetic analogue of the so-called ``energy loss function," previously identified within the context of dark matter scattering and absorption~\cite{Knapen:2021run,Hochberg:2021pkt,Knapen:2021bwg,Boyd:2022tcn,Berlin:2023ppd}. This implies a direct connection between $\mu$ and the magnon eigensystem derived previously in Ref.~\cite{Mitridate:2020kly}, which is worth exploring more generally. Along these lines, we thus anticipate that $\mu$ also dictates the dark matter-magnon scattering rate, originally derived in Ref.~\cite{Trickle:2019ovy}, exploration of which we leave to future work.

While, $\mu$ is experimentally measurable in principle, data is typically fit to the Landau--Lifshitz model of \Eq{Polder1} (see \App{YIG} for details). This model only accounts for the Kittel mode resonance. In particular, absorption into the Kittel mode is controlled by $1/\mu_-$, since $\qhat \cdot \mu^{-1} \cdot \qhat \simeq |\hat{\mathbf{e}}_- \cdot \qhat|^2 / \mu_-$ for $m_a \simeq \w_B$, where $\hat{\mathbf{e}}_-$ is the unit vector of the minus circular polarization. From \Eq{xpm}, the magnetic energy loss function for $\mu_-$ is given by
\be
\Im\bigg( \frac{\tsw{}{-} 1}{\mu_-}\bigg)  = \frac{m_a \, \w_M / 2 Q}{(m_a - \w_B)^2 + (\w_B / 2 Q)^2}
~.
\ee
When $m_a \simeq \w_B$ and $\hat{\v{q}}$ is perpendicular to the background magnetization, \Eq{classicalmagnon} reduces to
\be 
\label{eq:final_magnon_rate_IV}
R \simeq 2 \, (g_{ae} \, \vdm)^2 \, \frac{\rhodm}{\rhoT} \, \frac{Q \, n_{\text{spin}}}{m_a} 
~,
\ee
where we defined the spin density $n_{\text{spin}} = M_0 / \mu_B$. This parametrically matches the magnon absorption rate computed in Ref.~\cite{Mitridate:2020kly}. Another classical derivation of this result is presented in \App{Ohms}, which shows that it can be associated to the work done by rotating the magnetic dipoles in the material against the field $\B + \B_\eff$. While we have neglected boundary conditions throughout this section, it is also possible to include finite volume effects as was done in  Ref.~\cite{Chigusa:2023bga}.

\subsection{Absorption into Electronic Excitations}
\label{sec:absorption_into_electronic_excitations}

In this section, we show that the axioelectric absorption rate into electronic excitations is given by
\be
\label{eq:Rabs}
R \simeq \Big( \frac{g_{ae} \, m_a}{e} \Big)^2 ~ \frac{\rhodm}{ \rhoT} 
\times  
\begin{cases} \displaystyle 3 \, \Im\left[ \,\tsw{-}{} \eps(m_a) \,\right] & (\text{unpolarized target})\vspace{1em}
\\
\displaystyle \Im\left[ \frac{\tsw{}{-} 1}{\eps(m_a)} \right] & (\text{polarized target, spin splitting} \gg m_a) 
~,
\end{cases}
\ee
where $\eps(m_a)$ is the permittivity (i.e., dielectric function) evaluated at energy $\w = m_a$ and zero momentum-transfer, appropriate for nonrelativistic dark matter. The first line of \Eq{Rabs} has been derived previously in, e.g., Refs.~\cite{Pospelov:2008jk,Hochberg:2016sqx,Mitridate:2021ctr}. The second line is a new result and applies to completely spin-polarized targets with a large energy splitting between electron spin states. 

The key difference between the two cases in \Eq{Rabs} is that $R \propto 1/|\eps(m_a)|^2$ in polarized targets and thus is typically suppressed.\footnote{However, it can be enhanced when $|\eps(m_a)| \ll 1$, corresponding to $m_a$ close to an in-medium resonance.} This has a simple physical interpretation. The axioelectric force drives each electron along the direction of its spin. In a spin-polarized target, each electron is thus driven in the same direction, generating a coherent electromagnetic field which backreacts on the electrons, screening the axion's effect. By contrast, in an unpolarized target the electrons respond incoherently, and their motion does not produce any net electromagnetic effects. 

Though this screening reduces the signal rate, it can be useful for background rejection, since counting experiments are currently limited by large dark counts~\cite{Essig:2022dfa}. More specifically, the act of comparing polarized and unpolarized targets allows the dark count rate to be directly measured and separated from a potential signal. Ideally, a signal then only needs to overcome \textit{fluctuations} in the background rate to become detectable, allowing the sensitivity to increase with larger exposure (which is not the case when background systematics dominate).

Below we derive \Eq{Rabs} in two complementary ways. In \Sec{abs_classical}, we use the constitutive relationships of classical electrodynamics, as was done in \Sec{magnon}. In \Sec{elecabs}, we compute the absorption rate quantum mechanically in terms of self-energy diagrams using the formalism of Refs.~\cite{Hardy:2016kme,Mitridate:2021ctr,Chen:2022pyd,Krnjaic:2023nxe}. This more rigorous derivation produces correct $\order{1}$ factors, and allows us to generalize the second line of \Eq{Rabs} to arbitrary spin splitting in \Eq{Rgen}.

\subsubsection{Classical Derivation}
\label{sec:abs_classical}

The axioelectric absorption rate into electronic excitations is dominantly controlled by the first term in \Eq{aEOM1_classical}. In order to evaluate this term, we must determine the current density of each electron $\jv_i$. We do this by using the classical electronic equation of motion, $m_e \, \ddot{\xv}_i = \mathbf{F}_i$, where $\xv_i$ and $\mathbf{F}_i$ are the position of and total force acting on the $i^\text{th}$ electron, respectively. The internal forces are packaged into their contribution to the electric susceptibility $\x_{ei}$, defined such that in the presence of some external electric field $\E$, the equation of motion becomes $\jv_i = \x_{ei} \, \dt \E$. This form also makes it clear that $\x_{ei}$ are related to the dielectric function as $\sum_i \x_{ei} = \eps - 1$, since this gives the usual constitutive relation $\sum_i \jv_i  = (\eps - 1) \, \dt \E$. Note that in writing this, we have assumed that the target medium is sufficiently large (i.e., larger than the decay length in medium) so that boundary conditions can be neglected~\cite{Balafendiev:2022wua,ALPHA:2022rxj}.

In the presence of the axion field, we must also account for the effective electric of \Eq{Eeff} in the electron's equation of motion, such that $\jv_i = \x_{ei} \, \big( \dt \E + (\dt E_\eff) \, \shat_i \big)$. Here, the electric field $\E$ incorporates the backreaction from any coherent motion of charges induced by the axion field. From the long-wavelength limit of Amp\`ere's law (or alternatively \Eq{Efield1}), this field is $\dt \E \simeq - \jv_a^P / \eps$, where $\jv_a^P = (\dt E_\eff) \, \sum_i \x_{ei} \, \shat_i$ is the total polarization current induced by the axioelectric term, as in \Eq{axioncurrent}. As an aside, note that by comparing this form of $\jv_a^P$ to that given previously in \Eq{axioncurrent}, we arrive at a concrete expression for the spin-polarized contribution to the permittivity, denoted as $\epsa$ in \Sec{EMsig}. In particular, for a material with a net polarization along the $\shat$ direction, we have
\be
\label{eq:epsa}
\epsa \equiv 1 + \sum_i \x_{e i} \, \shat_i \cdot \shat \leq \eps
~,
\ee
so that $\epsa \simeq \eps$ for a fully spin-polarized medium. 

Now, using the result for the backreaction field $\E$ in the expression for $\jv_i$, we have that the single-particle current is
\be
\label{eq:single_electron_current_density}
\jv_i 
\simeq \x_{ei} \, (\dt E_\eff) \,  \Big( \shat_i - \frac{1}{\eps} \, \sum_j \x_{ej} \, \shat_j  \Big)
=
\x_{ei} \, (\dt E_\eff) \times
\begin{cases}
\shat_i & (\text{unpolarized target})
\\
\shat / \eps & (\text{polarized target})
~,
\end{cases}
\ee
where in the first equality, the first term is from the direct axion interaction with the $i^\text{th}$ electron, and the second term is due to the electric backreaction from the collective motion of many electrons. In the second equality, we used that if electron spins of the same $\x_{ei}$ are oppositely paired, as in an unpolarized target, then $\sum_i \x_{e i} \, \shat_i = 0$, and if the target is instead completely spin-polarized, then $\shat_i = \shat$ and $\sum_i \x_{e i} \, \shat_i = (\eps - 1) \, \shat$. From \Eq{single_electron_current_density}, we then have that the first term on the right-hand side of \Eq{aEOM1_classical} is
\begin{align}
\label{eq:Jdots}
\sum_i \, (\dt \jv_i) \cdot \shat_i & = (\eps - 1) \, \del_t^2 E_{\eff}
\times
\begin{cases}
1 & (\text{unpolarized target})
\\
1/\eps & (\text{polarized target})
~,
\end{cases}
\end{align}
where the second line is in agreement with \Eq{totj}. From this point, the absorption rate $R$ can again be computed using \Eq{classical_rate}. The quantity $\text{Im}(-\w^2)$ is determined by substituting \Eq{Jdots} in \Eq{aEOM1_classical}, and then using \Eq{aEOM0_classical}. This recovers the main result of \Eq{Rabs}, but without the factor of three for unpolarized targets. 

This mismatch occurs because the classical picture provides an incomplete description of quantum spins. More precisely, the other classical arguments in this work gave correct numeric factors because they were linear in quantum operators and therefore were guaranteed to match quantum results in expectation by the Ehrenfest theorem. By contrast, since $\jv_i \propto \shat_i$ for unpolarized targets, the classical treatment gives an answer proportional to $\shat_i \cdot \shat_i = 1$, while the analogous quantum mechanical treatment would give $\langle \sigmav_i \cdot \sigmav_i \rangle = 3$. For the polarized case, there is no such discrepancy because the mean spin polarization $\shat$ is inherently a classical vector and is treated as such in a fully quantum calculation. Finally, we note that an alternative classical derivation of the axioelectric absorption rate for a spin-polarized target is presented in \App{Ohms}, which shows that it can be associated to the work done by the force on the electrons, proportional to $\E + \E_\eff$.

\subsubsection{Matrix Element Calculation}
\label{sec:elecabs}

While the derivation in \Sec{abs_classical}, which is based on classical electrodynamics, elucidates some of the underlying physics, a quantum mechanical derivation makes direct contact with electronic states in the system and is therefore necessary for first-principles calculations. Moreover, it provides a general framework to understand any dark matter absorption rate into electronic excitations, which can then be simplified further by including assumptions about the target. The tradeoff is that this derivation is more technically involved than in \Sec{abs_classical}. Below, we will use the self-energy formalism recently applied to dark matter absorption in Refs.~\cite{Hardy:2016kme,Caputo:2020quz,Gelmini:2020xir,Gelmini:2020kcu,Mitridate:2021ctr,Chen:2022pyd,Krnjaic:2023nxe} and refer the reader to Refs.~\cite{Raffelt:1996wa,Mitridate:2021ctr} for an introduction. We will adopt absorption kinematics throughout, i.e., that the dark matter is nonrelativistic with $\grad a \ll \dot a$ (or, in other words, that the momentum $k$ satisfies $k\ll \w$). 

To calculate the absorption rate $R$, we note that the presence of interactions mixes the free axion and photon dispersion relations. In the language of self-energies, the mixed dispersion relation of an interacting photon with polarization $\lambda$ in Lorenz gauge and an axion is given by (see, for example, Ref.~\cite{Hardy:2016kme})
\be
\label{eq:dispoffdiag}
    \begin{pmatrix} 
 \w^2-k^2-\Pi^\lambda_{AA} & -\Pi^\lambda_{aA}  
\\ -\Pi^\lambda_{Aa}    & \w^2-k^2-m_a^2- \Pi_{aa}
\end{pmatrix}\begin{pmatrix} 
  A_\lambda  
\\ a
\end{pmatrix}=0~,
\ee
where $\Pi_{aa}$ is the axion self-energy, $\Pi_{aA}^\lambda = -e_\mu^\lambda \,  \Pi_{aA}^\mu$ is the mixed axion-photon self-energy projected onto the photon polarization vector $e^\lambda_\mu$, and $\Pi_{AA}^\lambda = - e^\lambda_\mu \,  \Pi_{AA}^{\mu \nu} \, e^\lambda_{\nu}$ is the photon self-energy. In the absence of the axion, the photon self-energy simply maps onto the normal dispersion relation for the photon, i.e., $n_\lambda^2 \, \w^2 = k^2$ where $n_\lambda$ is the refractive index for the $\lambda$ polarization, giving $\Pi_{AA}^\lambda = \w^2 \, (1-n_\lambda^2)$. The fields $a$ and $A$ refer to the free axion and photon states, respectively (i.e., the states defined in the absence of interactions between the axion and the photon). 

To understand what an ``axion" or ``photon" looks like inside a medium, we must diagonalize the mixed dispersion relations in \Eq{dispoffdiag}, which determines the axion-like and photon-like propagation eigenstates (sometimes also referred to as mass eigenstates). Due to the small axion coupling, the propagation states mostly correspond to the free axion and photon eigenstates, albeit with small admixtures of the opposing interaction states. At lowest nontrivial order (quadratic) in the coupling, we find the dispersion relations
\be
\w_\lambda^2 \simeq k^2 + \begin{cases} \Pi^\lambda_{AA}-\dfrac{\Pi_{a A}^\lambda \, \Pi_{Aa}^\lambda}{ m_a^2 - \Pi_{AA}^\lambda} & \text{(photon-like)} \vspace{2mm} \\ 
m_a^2+\Pi_{aa}+\displaystyle\sum_\lambda\dfrac{\Pi_{a A}^\lambda \, \Pi_{Aa}^\lambda}{ m_a^2 - \Pi_{AA}^\lambda} & \text{(axion-like)} 
~.
\end{cases}
\ee

Analogous to the classical computation of the previous section, we must evaluate the imaginary component of the frequency for the propagating axion-like state in order to find the energy absorbed from the axion. Such a procedure gives, for any axion interaction and target electronic structure, the absorption rate
\begin{align}
    R \simeq - \frac{\rhodm}{\rhoT \, m_a^2} \, \Im \bigg( \Pi_{aa} + \sum_\lambda \frac{\Pi_{a A}^\lambda \, \Pi_{Aa}^\lambda}{ m_a^2 - \Pi_{AA}^\lambda} \bigg) ~.
    \label{eq:SE_abs_per_axion_1}
\end{align}
At the onset, \Eq{SE_abs_per_axion_1} can be simplified if we assume that the target is isotropic. In this limit, the photon self-energy is independent of polarization, $\Pi_{AA}^\lambda = \Pi_{AA}$, and the sum over the photon states can be removed using the completeness relation, $\sum_\lambda e_\mu^\lambda \, e_\nu^\lambda = - g^{\mu \nu}$. This reduces \Eq{SE_abs_per_axion_1} to
\begin{align}
    R \simeq - \frac{\rhodm}{\rhoT m_a^2} \Im \left( \Pi_{aa} + \frac{\Pi_{a A}^i \Pi_{Aa}^i}{ m_a^2 - \Pi_{AA}} \right) ~ .
    \label{eq:SE_abs_per_axion_2}
\end{align}

These self-energies can then be computed diagrammatically using the relevant interactions present in the nonrelativistic Lagrangian~\cite{Mitridate:2021ctr} (ignoring subdominant terms dependent on the axion momentum),
\begin{align}
    \Lag_\text{NR} & \supset -i \frac{e}{m_e} \mathbf{A} \cdot \psi^\dagger \, \nabla \, \psi + \frac{e^2}{2 m_e} \mathbf{A}^2 \psi^\dagger \psi - i g_{ae} \, \frac{\dot{a}}{m_e} \, \psi^\dagger \,\sigmav \cdot \nabla \, \psi + g_{ae} \frac{e \dot{a}}{m_e} \mathbf{A} \cdot \psi^\dagger \, \sigmav \, \psi \, ,
    \label{eq:NR_lag}
\end{align}
where the first two terms are contributions from nonrelativistic QED, and the last two are from the axioelectric interaction, $\Lag \supset g_{ae} \, \dot{a} \psi^\dagger \, \Piv \cdot \sigmav \, \psi / m_e$. The self-energies in \Eq{SE_abs_per_axion_2} are given diagrammatically by
\vspace{1em}
\begin{align}
    \begin{fmffile}{feyn_1}
        \begin{fmfgraph*}(80,30)
            \fmfleft{i} \fmfright{f}
            \fmf{dashes,label=$\overset{k^\mu}{\longrightarrow}$,label.dist=-0.2w}{i,m1}
            \fmf{plain,right,tension=.4}{m1,m2}
            \fmf{plain,right,tension=.4}{m2,m1}
            \fmf{dashes}{m2,f}
            \fmflabel{$\Pi_{aa}: \quad a$}{i}
            \fmffreeze
            \fmfforce{vloc(__m1) shifted (0,-5.35mm)}{b1}
            \fmfforce{vloc(__m2) shifted (0,-5.35mm)}{b2}
            \fmfforce{vloc(__m1) shifted (0,5.35mm)}{t1}
            \fmfforce{vloc(__m2) shifted (0,5.35mm)}{t2}
            \fmfset{arrow_len}{2.5mm}
            \fmf{phantom_arrow}{b1,b2}
            \fmf{phantom_arrow}{t2,t1}
        \end{fmfgraph*}
    \end{fmffile}
\end{align}
\begin{align}
    \begin{fmffile}{feyn_3}
        \begin{fmfgraph*}(80,30)
            \fmfleft{i} \fmfright{f}
            \fmf{dashes,label=$\overset{k^\mu}{\longrightarrow}$,label.dist=-0.2w}{i,m1}
            \fmf{plain,right,tension=.4}{m1,m2}
            \fmf{plain,right,tension=.4}{m2,m1}
            \fmf{photon}{m2,f}
            \fmflabel{$\Pi_{aA}: \quad a$}{i}
            \fmflabel{$A \quad +$}{f}
            \fmffreeze
            \fmfforce{vloc(__m1) shifted (0,-5.35mm)}{b1}
            \fmfforce{vloc(__m2) shifted (0,-5.35mm)}{b2}
            \fmfforce{vloc(__m1) shifted (0,5.35mm)}{t1}
            \fmfforce{vloc(__m2) shifted (0,5.35mm)}{t2}
            \fmfset{arrow_len}{2.5mm}
            \fmf{phantom_arrow}{b1,b2}
            \fmf{phantom_arrow}{t2,t1}
        \end{fmfgraph*}
    \end{fmffile}
    \quad \quad\quad\quad\quad
    \begin{fmffile}{feyn_4}
        \begin{fmfgraph*}(80,30)
            \fmfleft{i} \fmfright{f}
            \fmf{dashes,label=$\overset{k^\mu}{\longrightarrow}$,label.dist=-0.2w}{i,m1}
            \fmf{plain}{m1,m1} 
            \fmf{photon}{m1,f}
            \fmflabel{$a$}{i}
            \fmflabel{$A$}{f}
            \fmffreeze
            \fmfset{arrow_len}{2.5mm}
            \fmfforce{vloc(__m1) shifted (-3.5mm,9.25mm)}{tl}
            \fmfforce{vloc(__m1) shifted (3.25mm,9.25mm)}{tr}
            \fmf{phantom_arrow}{tl,tr}
        \end{fmfgraph*}
    \end{fmffile}
\end{align}
\begin{align}
    \begin{fmffile}{feyn_5}
        \begin{fmfgraph*}(80,30)
            \fmfleft{i} \fmfright{f}
            \fmf{photon,label=$\overset{k^\mu}{\longrightarrow}$,label.dist=-0.2w}{i,m1}
            \fmf{plain,right,tension=.4}{m1,m2}
            \fmf{plain,right,tension=.4}{m2,m1}
            \fmf{photon}{m2,f}
            \fmflabel{$\Pi_{AA}: \quad A$}{i}
            \fmflabel{$A \quad +$}{f}
            \fmffreeze
            \fmfforce{vloc(__m1) shifted (0,-5.35mm)}{b1}
            \fmfforce{vloc(__m2) shifted (0,-5.35mm)}{b2}
            \fmfforce{vloc(__m1) shifted (0,5.35mm)}{t1}
            \fmfforce{vloc(__m2) shifted (0,5.35mm)}{t2}
            \fmfset{arrow_len}{2.5mm}
            \fmf{phantom_arrow}{b1,b2}
            \fmf{phantom_arrow}{t2,t1}
        \end{fmfgraph*}
    \end{fmffile}
    \quad \quad\quad\quad\quad
    \begin{fmffile}{feyn_6}
        \begin{fmfgraph*}(80,30)
            \fmfleft{i} \fmfright{f}
            \fmf{photon,label=$\overset{k^\mu}{\longrightarrow}$,label.dist=-0.2w}{i,m1}
            \fmf{plain}{m1,m1} 
            \fmf{photon}{m1,f}
            \fmflabel{$A$}{i}
            \fmflabel{$A~ .$}{f}
            \fmffreeze
            \fmfset{arrow_len}{2.5mm}
            \fmfforce{vloc(__m1) shifted (-3.5mm,9.25mm)}{tl}
            \fmfforce{vloc(__m1) shifted (3.25mm,9.25mm)}{tr}
            \fmf{phantom_arrow}{tl,tr}
        \end{fmfgraph*}
    \end{fmffile}
\end{align}
Following the formalism in Ref.~\cite{Mitridate:2021ctr} these diagrams evaluate to
\begin{align}
    \Pi_{aa} & = g_{ae}^2 \, \frac{\w^2}{m_e^2} \, \bar{\Pi}_{\sigmav \cdot \mathbf{p}, \sigmav \cdot \mathbf{p}} \label{eq:SE_1} \\ 
    \Pi_{aA}^i & = - i e \, g_{ae} \, \frac{\w}{m_e} \, \Big( \bar{\Pi}_{\sigma^i} - \frac{1}{m_e} \, \bar{\Pi}_{\sigmav \cdot \mathbf{p}, p^i} \Big) \label{eq:SE_2}\\ 
    \Pi_{AA} & = - \frac{e^2}{m_e} \, \Big( \bar{\Pi}_1 - \frac{1}{3 m_e} \bar{\Pi}_{p^i, p^i}\Big) ~ , \label{eq:SE_3}
\end{align}
using the Feynman rules determined by the nonrelativistic Lagrangian in Eq.~\eqref{eq:NR_lag}, and noting that $\Pi_{Aa}^i = - \Pi_{aA}^i$. The quantities $\bar{\Pi}$ contain all the information about the target electronic structure and for operators $\mathcal{O}$ (such as momentum ``$p^i$" or identity ``1" operators) are defined as
\begin{align}
    \bar{\Pi}_{\mathcal{O}_1, \mathcal{O}_2} & = - \frac{1}{V} \, \sum_{I \, I'} \, \frac{f_I - f_{I'}}{\w - \Delta \w_{II'} + i \, \delta_{II'}} ~ \langle I | \mathcal{O}_1 | I' \rangle \, \langle I' | \mathcal{O}_2 | I \rangle \label{eq:self_energy_general} \\
    \bar{\Pi}_{\mathcal{O}} & = - \frac{1}{V} \, \sum_{I} \, f_I \, \langle I | \mathcal{O} | I \rangle ~,
\end{align}
where $I$ and $I'$ index the electronic states $|I \rangle$ with energy $\w_I$, $\Delta \w_{II'} = \w_{I'} - \w_I$, $\delta_{II'} = \delta ~ \text{sign}(\w_{I'} - \w_I)$ with $\delta$ the width of the electronic states, and $V$ is the target volume. The filling fraction satisfies $f_I = 1$ if a state is occupied and $f_I = 0$ otherwise.

We will focus on targets where the states can be indexed by a band number $b$ and spin quantum number $s \in \{ \uparrow, \downarrow \}$, such that a state label is given by  $I = \{ b, s \}$. This is possible when spin is a good quantum number and allows the state to be split into spatial and spin degrees of freedom, i.e., $| I \rangle = | b \rangle \otimes | s \rangle$. We will find it useful to introduce self-energies with a spin index, such that \Eq{self_energy_general} generalizes to
\begin{align}
    \bar{\Pi}^{ss'}_{\mathcal{O}_1, \mathcal{O}_2} & = - \frac{1}{V} \, \sum_{b \, b'} \, \frac{f_{b\,s} - f_{b'\, s'}}{\w - \Delta \w_{bs\,b's'} + i \delta_{bs\,b's'}} ~ \langle b, s | \mathcal{O}_1 | b', s' \rangle \, \langle b', s' | \mathcal{O}_2 | b, s \rangle\\ 
    \bar{\Pi}^s_{\mathcal{O}} & = - \frac{1}{V} \,  \sum_{b} \, f_{b, s} \, \langle b, s | \mathcal{O} | b, s \rangle ~ .
\end{align}
If the target is spatially isotropic, i.e., $\bar{\Pi}_{p^i, p^j}^{ss'} =  \bar{\Pi}_{p^i, p^i}^{ss'} \, \delta^{ij} / 3$, the self-energies in \Eq{SE_3} can be simplified further, to
\begin{align}
    \Pi_{aa} & = g_{ae}^2 \, \frac{\w^2}{m_e^2} \, \bigg( \, \frac{1}{3} \, \Big(\bar{\Pi}_{ p^i, p^i}^{\uparrow \uparrow} + \bar{\Pi}_{ p^i, p^i}^{\downarrow \downarrow} \Big) + \frac{2}{3} \, \Big( \bar{\Pi}_{ p^i, p^i}^{\downarrow \uparrow} + \bar{\Pi}_{ p^i, p^i}^{\uparrow \downarrow} \Big) \bigg) \label{eq:pi_aa_general}\\ 
    \Pi_{aA}^i & =  - i e \, g_{ae} \,  \frac{\w}{m_e} \, \hat{s}^i \, \bigg( \bar{\Pi}^\uparrow_1 - \bar{\Pi}^\downarrow_1 - \frac{1}{3m_e} \, \Big( \bar{\Pi}^{\uparrow \uparrow}_{p^i, p^i} - \bar{\Pi}^{\downarrow \downarrow}_{p^i, p^i} \Big) \bigg)\label{eq:pi_aA_general}\\ 
    \Pi_{AA} & = -\frac{e^2}{m_e} \, \bigg(\bar{\Pi}^\uparrow_1 + \bar{\Pi}^\downarrow_1 - \frac{1}{3m_e} \, \Big(  \bar{\Pi}^{\uparrow \uparrow}_{p^i, p^i} + \bar{\Pi}^{\downarrow \downarrow}_{p^i, p^i} \Big) \bigg) ~, \label{eq:pi_AA_general}
\end{align}
where $\hat{s}^i = \langle s | \, \sigma^i \, | s \rangle$ is the expectation value of spin (as used in, e.g., \Eq{Eeff}). Note that since $\bar{\Pi}_1^s$ is simply the expectation value of the identity operator, it is straightforwardly evaluated as $\bar \Pi_1^s = - N_e^s / V$, where $N_e^s$ is the number of electrons in spin state $s$.

Before computing the absorption rate in a spin-polarized target, we reproduce the standard result for a target with no spin polarization. In the absence of a net spin polarization, $f_{bs} = f_{b}$ (i.e., if a spin-up state is filled, then so is the corresponding spin-down state). Furthermore, we assume that the energy levels are spin independent, $\w_{bs} \simeq \w_{b}$. For a target satisfying these assumptions, we have $\bar{\Pi}^{ss'}_{\mathcal{O}_1, \mathcal{O}_2} = \bar{\Pi}_{\mathcal{O}_1, \mathcal{O}_2}$, and therefore Eqs.~\ref{eq:pi_aa_general}$-$\ref{eq:pi_AA_general} simplify to
\begin{align}
    \Pi_{aa} & = 2 g_{ae}^2 \, \frac{\w^2}{m_e^2} ~ \bar{\Pi}_{ p^i, p^i} \label{eq:pi_aa_standard} \\ 
    \Pi_{aA}^i & = 0 \label{eq:pi_aA_standard} \\ 
    \Pi_{AA} & = -\frac{2 e^2}{m_e} \, \Big(\bar{\Pi}_1 - \frac{1}{3m_e} \,  \bar{\Pi}_{p^i, p^i} \Big) ~ . \label{eq:pi_AA_standard}
\end{align}
Notice that since $\Pi_{aA}^i = 0$, there will be no mixing or screening effects; the photon and axion do not directly interact, which leaves the propagation eigenstates simply as the free states (this reflects the absence of any electromagnetic backreaction that screens the signal in unpolarized targets, as discussed in the classical derivation above). Substituting these self-energies into \Eq{SE_abs_per_axion_2} gives
\begin{align}
    R & = - \frac{\rhodm}{\rhoT} \, \frac{2 g_{ae}^2}{m_e^2} \, \Im \big( \bar{\Pi}_{p^i, p^i} \big) = - \frac{\rhodm}{\rhoT} \, \frac{3 g_{ae}^2}{ e^2} \, \Im \big( \Pi_{AA} \big) = \frac{\rhodm}{\rhoT} \, \frac{3 g_{ae}^2 m_a^2}{e^2} \, \Im \big( \, \eps(m_a) \, \big) ~,
\end{align}
where in the last step we used $\Pi_{AA}(\w) \simeq \w^2 \, \big( 1 - \eps(\w) \big)$ for $k \ll \w$, in agreement with the first case of \Eq{Rabs}. 

A similar calculation can be performed to determine the absorption rate for a target that is ``super-polarized," e.g., one where all the electrons are polarized in the  spin-up $\uparrow$ state. In this case, $f_{b,\downarrow} = 0$, for all $b$, and $f_{b, \uparrow} = 1$ when $\w_{b \uparrow} < 0$ (corresponding to the electrons below the Fermi surface). In this limit, some of the self-energies simplify, such as  $\bar{\Pi}^\downarrow_\mathcal{O} = 0$ and $\bar{\Pi}^{\downarrow\downarrow}_{\mathcal{O}_1, \mathcal{O}_2} = 0$. Additionally, the contribution from $\bar{\Pi}^{\downarrow\uparrow}_{p^i, p^i}$ to $\Im \left( \Pi_{aa} \right)$ is negligible in the small width limit, since there are no allowed $\downarrow$ to $\uparrow$ transitions in a spin-up super-polarized target. Therefore, in this case, the self-energies in Eqs.~\ref{eq:pi_aa_general}$-$\ref{eq:pi_AA_general} reduce to
\begin{align}
    \Pi_{aa} & = g_{ae}^2 \, \frac{\w^2}{3 m_e^2} \, \Big( \bar{\Pi}_{ p^i, p^i}^{\uparrow \uparrow} + 2 \, \bar{\Pi}_{p^i, p^i}^{\uparrow\downarrow} \Big) \label{eq:pi_aa_super}\\ 
    \Pi_{aA}^i & =  - i e \, g_{ae} \, \frac{\w}{m_e} \, \hat{s}^i \, \Big( \bar{\Pi}^\uparrow_1 - \frac{1}{3m_e} \, \bar{\Pi}^{\uparrow \uparrow}_{p^i, p^i} \Big)\label{eq:pi_aA_super}\\ 
    \Pi_{AA} & = -\frac{e^2}{m_e} \, \Big(\bar{\Pi}^\uparrow_1 - \frac{1}{3m_e} \, \bar{\Pi}^{\uparrow \uparrow}_{p^i, p^i} \Big) ~ . \label{eq:pi_AA_super}
\end{align}
Unlike the unpolarized case, the polarized spins do induce a mixing between the axion and the photon (consistent with the expectation of screening generated by an electrical backreaction, as discussed in the classical derivation above). These self-energies can be simplified further by noticing that $\Pi_{aA}^i = - i g_{ae} \, m_a\, \hat{s}^i\, \Pi_{AA} / e$, which can be used in the second term of \Eq{SE_abs_per_axion_1}. Additionally, note that $\Im\left( \Pi_{aa} \right)$ has two contributions in \Eq{pi_aa_super}. The first term corresponds to where the spin is fixed as $\uparrow$ which incorporates the classical force described in \Sec{HeisEOM}. The second term instead corresponds to a spin-flip transition, which does not have a classical analogue. The first term can be simply related to the self-energy of the photon $\Pi_{AA}$ which is consistent with the picture of the axioelectric effect as a spin-coupled effective electric field (i.e., when all spins are polarized one can simply relate charges and spins and so electrons should have a similar effect on both the axion and photon). However, this statement does not hold for the spin-flip term. The absorption rate is then given by
\begin{align}
\label{eq:Rgen}
    R \simeq g_{ae}^2 \, \frac{\rhodm}{\rhoT} \, \bigg( \frac{m_a^2}{e^2} \, \Im\bigg( \frac{\tsw{}{-} 1}{\eps(m_a)} \bigg) + \frac{2}{3 m_e^2}\, \Im\bigg(- \bar{\Pi}_{p^i, p^i}^{\uparrow \downarrow} \bigg) \bigg) ~,
\end{align}
where the first term matches the result in \Sec{abs_classical}, as expected. Since the second term includes spin-flip transitions, it is natural that it could not be incorporated in the classical derivation. Furthermore, in the limit of large spin splitting, when the spin-flip transitions are kinematically unavailable, $\Im\left(- \bar{\Pi}_{p^i, p^i}^{\uparrow \downarrow} \right) \rightarrow 0$, and the absorption rate reduces to \Eq{Rabs}. Note that the dependence on $\Im(-1/\eps)$ comes about due to the mixing of the axion with the photon and gives the same functional form as a direct electromagnetically-coupled axion. Thus, while the couplings are different in either case, the resulting physics is similar in the spin-polarized limit. In fact, we can also rederive the form of the axion-induced $E$-field by considering the basis that diagonalizes the dispersion relation of \Eq{dispoffdiag} to leading order, which we will label as $(\tilde A,\tilde a)$. To convert from propagation basis to the interaction $(A,a)$ basis, we can use the rotation matrix
\be
  \begin{pmatrix} 
   A  
\\  a
\end{pmatrix}\simeq  \begin{pmatrix} 
 1 & \frac{\Pi_{aA}}{\Pi_{AA}-m_a^2}  
\\ \frac{-\Pi_{aA}}{\Pi_{AA}-m_a^2}    & 1
\end{pmatrix}\begin{pmatrix} 
  \tilde A  
\\ \tilde a
\end{pmatrix}\,.
\ee
From this matrix, and noting that at lowest nontrivial order the magnitudes of $a$ and $\tilde a$ are the same, we see that axion-photon mixing induces an $E$-field in an infinite medium of the form
\be
    E^i \simeq i\w A^i \simeq i\w \, \frac{\Pi^i_{aA}}{\Pi_{AA}-m_a^2} \, \tilde a \simeq -i \, \frac{\Pi_{aA}^i}{\eps} \, \frac{a}{\w}~.\label{eq:mixingE}
\ee
An alternative definition of the spin-polarized contribution to the permittivity $\eps_\sigma$ can then be obtained by equating \Eqs{mixingE}{Einfinite},
\be
\label{eq:epsaPi}
    \epsa \equiv 1+\frac{e^2}{\w^2\, m_e} \, \bigg( \bar{\Pi}^\uparrow_1 - \bar{\Pi}^\downarrow_1 - \frac{1}{3m_e} \, \Big( \bar{\Pi}^{\uparrow \uparrow}_{p^i, p^i} - \bar{\Pi}^{\downarrow \downarrow}_{p^i, p^i} \Big) \bigg)\,.
\ee
For ``super-polarized" spins, we can use $\Pi_{aA}^i = - i \, g_{ae} \,m_a\, \hat{s}^i\, \Pi_{AA} / e$ to explicitly show that $\eps_\sigma = \eps$; conversely, if the spins are unpolarized, then $\eps_\sigma=1$, as mentioned previously. Note that this form for $\epsa$ is analogous to the classical sum over spin weighted susceptibilities in \Eq{epsa} where the contributions to spin-up and spin-down states are added electron by electron.

While the derivation incorporates the entirety of electronic states, further simplifications can be made when the core electrons are much more tightly bound than the valence electrons, as in typical solids. In this case, the core electrons' contributions to the self-energies are negligible, due to their suppressed electron propagator in \Eq{self_energy_general}. This can be understood physically from the dielectric function, which encodes the target electronic response to an impinging electric field. If the core electrons are very tightly bound, then they respond less efficiently to an electric field. If the electronic response can be approximated by only including the least bound electrons, then only these electrons need to be spin-polarized for the ``super-polarized" approximation discussed previously to hold. This is a much weaker, and more realistic, condition than requiring every electron to be spin-polarized.

\section{Dipole Moments and Energy Shifts}
\label{sec:EDMandDeltaE}

In this section, we discuss two cases in which simple parametric estimates can yield misleading results. In \Sec{EDM}, we show how one can use field redefinitions to work in a Hamiltonian lacking manifest axion shift symmetry. This leads to apparent electric dipole moment effects, which are actually spurious. In \Sec{energyshifts}, we show that the axioelectric atomic energy level shift is smaller than the naive expectation.

\subsection{Spurious Electric Dipole Moments}
\label{sec:EDM}

The axion-fermion coupling in \Eq{IntroLag} has an approximately equivalent nonderivative form, which has been the subject of some recent confusion. Let us first review how the nonderivative coupling is derived. In the original Lagrangian,
\be \label{eq:deriv_lag}
\Lag \supset \Psibar (i \slashed{\del} - \mf) \Psi + g_{af} \, (\del_\mu a) \, \Psibar \g^\mu \g^5 \Psi
~,
\ee
one can perform a chiral redefinition of the fermion field, $\Psi \to e^{-i g_{af} a\g^5} \Psi$. This yields 
\be \label{eq:nonderiv_lag}
\Lag \supset \Psibar (i \slashed{\del} - e^{2 i g_{af} a \g^5} \mf) \Psi - \frac{q_f^2}{8 \pi^2}  \, g_{af} \, a \, F_{\mu \nu} \, \tilde{F}^{\mu \nu} 
~,
\ee
where the derivative coupling has been rotated away. The last term in \Eq{nonderiv_lag}, which arises due to the chiral anomaly, shifts the axion-photon coupling; it affects computations of loop-induced processes but will be irrelevant to the physical signatures considered in this work. Since these two Lagrangians are related by a field definition, they must yield precisely the same $S$-matrix elements and thus the same physical predictions~\cite{Chisholm:1961tha,Kamefuchi:1961sb}. While it may naively appear that \Eq{nonderiv_lag} implies the presence of physical effects even for time- and space-indendent axion fields, $a(x, t) = a_0$, this cannot be the case, as the derivative coupling in \Eq{deriv_lag} would simply vanish exactly. 

Truncating the exponential in \Eq{nonderiv_lag} at $\order{g_{af}}$ and dropping the anomaly term yields a nonderivative interaction
\be 
\Lag \supset - 2 \, \mf \, g_{af} \, a \, \Psibar i \g^5 \Psi ~. \label{eq:IntroLagAlt}
\ee
As shown in \App{nonderiv}, it results in an alternative low-energy Hamiltonian for the fermion of the form
\be 
\label{eq:HamAlt}
H_{\text{alt}} \simeq \frac{\Piv^2}{2 \mf} + q_f \, \phi  - \frac{q_f}{2 \mf} \, \B \cdot \sigmav 
- g_{af} \, (\grad a) \cdot \sigmav - \frac{g_{af}}{4 \mf} \, \{\dot{a}, \Piv \cdot \sigmav\} + \frac{q_f \, g_{af}}{2 \mf} \,  a \, \E \cdot \sigmav 
~.
\ee
Compared to the original Hamiltonian of \Eq{HamDeriv}, \Eq{HamAlt} has an axioelectric term with coefficient smaller by a factor of two, as well as a ``nonrelativistic EDM'' term proportional to $a$ that seems to violate the axion's shift symmetry. 

The resolution of this paradox is that the two Hamiltonians describe exactly the same physics, but differ in their labeling of position states. To see this explicitly, note that the chiral field redefinition of $\Psi$ modifies the nonrelativistic field $\psi$, defined in \Eq{4to2}, by a momentum-dependent phase factor,
\be
\psi \to \psi - i g_{af} \, a \, \psibar \simeq \exp\left(- \frac{i g_{af} \,  a \, \sigmav \cdot \Piv }{2 m_f} \right) \psi,
\ee
where we used \Eq{psibarEOM}. This implies a shift of the particle's coordinate position by $\Delta \v{x} = g_{af} \, a \, \sigmav / 2 \mf$. While the original Hamiltonian in \Eq{HamDeriv} describes the fermion using a position operator that coincides with its center of charge, in the alternative Hamiltonian of \Eq{HamAlt} it is displaced from the center of charge by $\Delta \v{x}$, thereby leading to an EDM of $\v{d} = - q \, \Delta \v{x}$. In other words, the apparent dependence on $a$ is purely an artifact of the description. As we discuss in detail in \App{nonderiv}, the physical equivalence of these Hamiltonians is completely general and occurs because they can be related by a unitary transformation generated by the Hermitian operator $S \propto \{a, \sigmav \cdot \Piv\}$. While calculations of experimental observables using the alternative Hamiltonian may contain $a$-dependent intermediate quantities, they must drop out of the final result. 

In particular, electron EDM experiments measure the coefficient $d$ of the full relativistic EDM operator $(d/2) \, \Psibar \g^5 \sigma^{\mu\nu} \Psi \, F_{\mu\nu}$, where $\sigma^{\mu \nu} \equiv \left[\g^\mu, \g^\nu \right]/2$, which reduces to $- d \, \sigmav \cdot \E$ in the nonrelativistic limit~\cite{khriplovich2012cp}. Several recent works~\cite{Alexander:2017zoh,Chu:2019iry,Wang:2021dfj} have claimed extremely strong limits on $g_{ae}$ by implicitly assuming that these EDM experiments constrain the coefficient of $\sigmav \cdot \E$, which carries an $a$-dependent contribution in \Eq{HamAlt}. This is incorrect, because the operator $- d \, \sigmav \cdot \E$ produces no $\order{d}$ energy level shift by Schiff's theorem~\cite{Schiff:1963zz}. Instead, the signal from a true EDM arises entirely from its relativistic effects, which are not shared by the apparent EDM induced by an axion field. As discussed in detail in \App{nonderiv}, for a constant axion field, the apparent axion-induced EDM is spurious, and thereby does not shift energy levels or give rise to any other physical effects. These cancellations, which seem mysterious at the Hamiltonian level, arise from the axion's non-manifest shift symmetry. 

On the other hand, a time-varying axion field can lead to physical effects in EDM experiments, but they are suppressed at low $m_a$ by the small ratio of the axion mass to typical atomic energy scales. These effects are most straightforward to calculate with the derivative coupling, since the nonrelativistic EDM term lacks an explicit axion time derivative. Ref.~\cite{Stadnik:2013raa} found that for nonrelativistic electrons bound in an atom, the leading energy shift linear in the external electric field is suppressed by $\order{m_a^2/\text{Ry}^2}$, where the Rydberg constant $\text{Ry}=\alpha^2 m_e/2$ gives the scale of electronic energy levels. As a result, the sensitivity of EDM searches to the axion-electron coupling is much weaker than existing astrophysical bounds. By contrast, two recent works~\cite{Smith:2023htu,DiLuzio:2023ifp} missed this suppression factor, thereby overestimating the sensitivity of EDM experiments by many orders of magnitude. As explained further in \App{nonderiv}, such overestimates can be avoided by taking care to compute the relevant physical observables for each experiment.

\subsection{Suppressed Shifts of Electronic Energy Levels}
\label{sec:energyshifts}

The axion-electron coupling also shifts electronic energy levels in the absence of external fields. The leading-order contributions to these shifts are from the axion wind, $H_\text{wind} = - g_{ae} \,  (\nabla a) \cdot \sigmav$, and axioelectric, $H_\text{ae} = - (g_{ae}/m_e) \, \dot{a} \, \Piv \cdot \sigmav$, terms in the nonrelativistic Hamiltonian of \Eq{HamDeriv}. Thus, to first order in perturbation theory, a naive estimate for the energy level shifts to an electronic state is $\Delta E_{\text{wind}} \simeq \bra{\psi} H_{\text{wind}} \ket{\psi} \sim g_{ae} \, |\nabla a|$ and $\Delta E_{\text{ae}} \simeq \bra{\psi} H_{\text{ae}} \ket{\psi} \sim g_{ae} \, \dot{a} \, v_e$, where $v_e \sim Z_\text{eff} \, \alpha$ is the characteristic electron velocity bound to an ion with effective charge $Z_\text{eff}$. 

The above estimate for the axion wind energy level shift is parametrically correct and is used to infer the reach of atomic clock experiments. However, we illustrate here that the effect of the axioelectric term is parametrically overestimated. To understand why, consider the leading-order terms in the nonrelativistic electron Hamiltonian, $H_0 = \Piv^2 / 2 m_e + V(\xv)$, where the spin-independent potential $V(\xv)$ includes, e.g., the electrostatic potential energy. For electronic states governed by this Hamiltonian, $\Piv = i m_e \left[ H_0, \mathbf{x} \right]$ and thus
\begin{align}
\label{eq:EnergyShift}
\Delta E_{\text{ae}} &= \langle \psi | H_\text{ae} | \psi \rangle = - \frac{g_{ae}}{m_e} \, \dot{a} \, \langle \psi | \, \Piv \cdot \sigmav \, | \psi \rangle =  i g_{ae} \,  \dot{a} \, \langle \psi | \, \xv \cdot \left[ H_0, \sigmav \right]  \, | \psi \rangle = 0
~,
\end{align}
such that the naive leading-order energy level shift from the axioelectric term vanishes, where in the last equality we used that $H_0$ commutes with the spin operator, by definition. $\Delta E_{\text{ae}}$ can be intuitively understood as the work done by the axion effective electric field. To see this, note that in the Heisenberg picture and using the Heisenberg equation of motion for the spin operator, $\Delta E_{\text{ae}}$ can be rewritten as $\Delta E_{\text{ae}} = - e \, \langle \xv_H \cdot \hat{\E}_\text{eff} \rangle$, where $\hat{\E}_\eff \equiv - (g_{ae} / e) \, \dot{a} \, \dot{\sigmav}_H$, in analogy with \Eq{Eeff}. Hence, if the spin orientation is fixed, then \Eq{EnergyShift} holds. 

Instead, any energy level shift from the axioelectric term must arise from higher-order contributions to $H_0$ that break the derivation in \Eq{EnergyShift}. For example, the relativistic kinetic energy correction shifts $[H_0, \xv]$ so that it is no longer proportional to $\Piv$, while the electrons' spin-orbit or spin-spin interactions contribute to $[H_0, \sigmav]$. All of these terms are suppressed by $v_e^2$ relative to $H_0$, leading to the parametric estimate  $\Delta E_{\text{ae}} \sim g_{ae} \, \dot{a} \, v_e^3$, which is smaller than the naive estimate by a factor of $\alpha^2 \sim 10^{-4}$ for electrons with $Z_{\text{eff}} \sim 1$. For light elements, this suppresses the projected reach in Ref.~\cite{Arvanitaki:2021wjk}, which used the naive estimate discussed previously; however, it can be alleviated for very heavy atoms, with $Z \gg 1$. Future calculations of this energy shift, such as those relevant for Ref.~\cite{Roising:2021lpv}, must be careful to account for these suppressions, as well as including any contributions from the vector potential $e \, \A$ when computing $\langle \psi | \Piv \cdot \sigmav | \psi \rangle$. 

\section{Outlook}
\label{sec:conclusion}

The direct coupling of axion dark matter to Standard Model fermions leads to new experimental signatures. We have provided a firm theoretical foundation in \Sec{the_NR_limit} that clarifies the nature of these signatures, thereby resolving some existing disagreements in the literature in \Sec{EDMandDeltaE}. We have also identified several new experimental strategies, focusing predominantly on the axion-electron coupling. Among these, the magnetized multilayer experiment discussed in \Sec{MADMAX} appears to be the most promising. Like previous ferromagnetic haloscope concepts, it relies on detecting the radiation emitted from the magnetization current induced by the axion wind torque on electron spins. The enhanced sensitivity of our setup comes from a combination of various factors: the use of inexpensive polycrystalline ferrite materials enables a much larger detector volume and a multilayer geometry enhances the effective coupling to the axion field. As a result, orders of magnitude of new parameter space can be explored with existing technology, for axion masses ranging from $\mu\eV$ to $\meV$.

If nearly background-free single photon detectors are developed, they would enable magnetized multilayers to fully explore the DFSZ QCD axion model space when the abundance is set by the well-motivated post-inflationary misalignment mechanism~\cite{Kawasaki:2014sqa,Vaquero:2018tib,Klaer:2017ond,Buschmann:2019icd,Gorghetto:2018myk,Kawasaki:2018bzv,OHare:2021zrq,Buschmann:2021sdq}. Single photon detection for frequencies in the $(1-100) \ \mu\eV$ range has recently seen promising advances using, e.g., Rydberg atoms and superconducting qubits, though both technologies are relatively new and not background free~\cite{Dixit:2020ymh,Graham:2023sow}. For higher frequencies, near a $\meV$, transition edge sensors and kinetic inductance detectors can be used but currently suffer from large dark counts~\cite{BREAD:2021tpx}. 

In future work, we will explore further refinements in determining the sensitivity of a magnetized multilayer haloscope. This includes a detailed numerical analysis that explores optimal tuning and effects from the axion wind's $\order{1}$ daily modulation. More generally, enhancing the axion's effective coupling to collective excitations in a multilayer geometry by operating slightly off resonance may also prove useful for other experiments, such as TOORAD~\cite{Schutte-Engel:2021bqm}.

We also plan on further exploring the findings of \Sec{magnon}, which showed that the axion absorption rate into spin excitations can be inferred directly from measurements of the magnetic energy loss function $\Im(-1/\mu)$, circumventing the need for a detailed microscopic model of magnetic materials. In particular, it would be interesting to expand this result to more general spin-coupled absorption and scattering of particles, which would simplify the formalism needed for dark matter searches, by connecting the signal rate directly to experimentally measurable parameters.

Although we have focused almost exclusively on the axion-electron coupling throughout most of this work, many of our results also translate to the axion-nucleon coupling. In particular, the mechanical resonators mentioned in \Sec{Mechsig} would be sensitive to comparable values of the axion-nucleon coupling, albeit only if $\order{1}$ nuclear spin-polarization can be achieved. Furthermore, the magnetization current discussed in \Sec{MADMAX} also exists for nucleon couplings, but is suppressed by the small magnitude of nuclear magnetic dipole moments. However, both of these issues may be alleviated in materials with strong hyperfine interactions between nuclear and electronic spins~\cite{Chigusa:2023hmz}. 

There are also a number of experimental concepts we have not considered in detail. In \Sec{Mechsig}, we discussed how the axion wind may excite toroidal modes in mechanical resonators, which could lead to competitive sensitivity in the $\text{kHz} - \text{MHz}$ frequency range not covered by existing torsion pendulums or electromagnetic experiments. Developing this idea further would require considering the quality factors of such modes, the form factors for coupling to them, and mechanisms to read out their excitations. Finally, we note that if electron spins are made to precess at an angular frequency $\w_{\text{spin}} \gg m_a$, then both the mechanical and magnetization current signatures can be upconverted to a higher frequency $\w_{\text{sig}} \simeq m_a + \w_{\text{spin}}$. Such a ``heterodyne'' approach has been applied to cavity experiments for photon-coupled axions~\cite{Berlin:2019ahk,Lasenby:2019prg,Thomson:2019aht,Thomson:2023moc}, enabling  sensitivity to axion masses parametrically smaller than the cavity's mode frequencies, but at the cost of introducing additional noise sources, which must be carefully analyzed.

The axion-fermion coupling leads to a rich variety of experimental signatures, which have previously been underexplored. Going forward, we hope that this work serves as a firm foundation for new ideas and future efforts. 

\acknowledgments

We thank Mina Arvanitaki, Ryan Janish, Ciaran O'Hare, Christopher Smith, and Natalia Toro for helpful discussions, and Maxim Pospelov for very valuable insight regarding axioelectric energy level shifts. This research was supported by the U.S.~Department of Energy's Office of Science under contract DE–AC02–76SF00515, as well as by Fermilab's Superconducting Quantum Materials and Systems Center (SQMS) under contract number DE-AC02-07CH11359. Fermilab is operated by the Fermi Research Alliance, LLC under Contract DE-AC02-07CH11359 with the U.S.~Department of Energy. This work was completed in part at the Perimeter Institute. Research at Perimeter Institute is supported in part by the Government of Canada through the Department of Innovation, Science and Economic Development Canada and by the Province of Ontario through the Ministry of Colleges and Universities.

\appendix
\addappheadtotoc
\renewcommand\thefigure{\thesection.\arabic{figure}}
\setcounter{figure}{0}

\newpage

\section{Reducing the Axial Current}
\label{app:bilinears}

Here, we show two ways to take the nonrelativistic limit of the axial vector current $\Psibar \g^\mu \g^5 \Psi$. The fastest is to treat $\Psi$ as a relativistic wavefunction and decompose it into two-component spinors, as in \Eq{4to2}, which gives
\be
\label{eq:axial1}
\Psibar \g^\mu \g^5 \Psi = \big( \, \psi^\dagger \psibar + \psibar^\dagger \psi ~,~ \psi^\dagger \sigmav \psi + \psibar^\dagger \sigmav \psibar \, \big)^\mu
~.
\ee
Now, note that \Eqs{psibarEOM}{vEOM} imply $\psibar \simeq (1/2) \, \vv \cdot \sigmav \, \psi$ to leading order in $g_{af}$. Using this in \Eq{axial1} then yields
\be
\label{eq:axial2}
\Psibar \g^\mu \g^5 \Psi \simeq \psi^\dagger \, \big(  \vv \cdot \sigmav ~, \,   \sigmav \big)^\mu \, \psi + \order{v^2} + \order{g_{af}}
~.
\ee
Integrating \Eq{axial2} over space then yields \Eq{classicalaxial}, the axial current for a single particle. 

Another method, which works for multi-particle states, is to treat $\Psi$ as a quantized field with free mode expansion
\be 
\label{eq:mode_exp}
\Psi(\xv) = \int \frac{d^3 \p}{(2 \pi)^3} \, \sum_s \Big( a^s_{\p} \, u_s(\p) \, e^{i \p \cdot \xv} + b^s_{\p} \, v_s(\p) \, e^{-i \p \cdot \xv} \Big)
~,
\ee
where states with momentum $\p$ and spin $s$ are described by $\ket{\p, s} = {a_{\p}^s}^\dagger \ket{0}$ with $\braket{\p, r | \v{q},s} = (2\pi)^3 \,  \delta^{(3)}(\p - \v{q}) \, \delta^{rs}$. In \Eq{mode_exp}, the coefficient of the particle annihilation operator $a^s_{\p}$ is a positive-frequency mode-function which can be obtained by solving the free Dirac equation, yielding
\be 
\label{eq:pos_spinor_defs}
u_s(\p) = \sqrt{\frac{E+\mf}{2E}} ~ \begin{pmatrix} \xi_s \vspace{2mm} \\ \dfrac{\p \cdot \sigmav}{E+\mf} \, \xi_s \end{pmatrix}
~,
\ee
where $E$ is the energy, $\xi_s$ is a two-component spinor, and we used the nonrelativistic normalization $u^\dagger_s(\p) u_s(\p) = 1$. 

We restrict to single particle states in the presence of an axion field with approximately constant uniform $\del_\mu a$. In this case, we can take the classical limit by computing diagonal matrix elements of the axial vector current. As a warm up, let's first evaluate the matrix element for the vector current $\Psibar \g^\mu \Psi$. Its spatial components have expectation values  
\be
\bra{\p, s} \Psibar \g^i \Psi \ket{\p, s} = \overline{u}_s(\p) \g^i u_s(\p) = \frac{1}{E} \, \xi_s^\dagger \, \left( \sigma^i \, \p \cdot \sigmav \right) \, \xi_s = v^i
~,
\ee
where $\vv$ is the velocity and we used the form of the mode-functions in \Eq{pos_spinor_defs}. The expectation value of the temporal component is simply $1$, by our normalization convention. By superposing plane waves, one can construct one-particle wavepackets of spatial spread $r$ momentarily centered around a location $\xv_0$, provided that $r \gg 1/m$. In such a normalized state, the expectation value of the vector current is thus approximately 
\be \label{eq:particle_current}
\langle \Psibar \g^\mu \Psi \rangle \simeq (1, \vv)^\mu ~ \delta^{(3)}(\xv - \xv_0)
\ee
for small $r$, which is a classical particle's number current, as expected. Similarly, for the axial vector current, 
\be
\bra{\p, s} \Psibar \g^0 \g^5 \Psi \ket{\p, s} = \overline{u}_s(\p) \g^0 \g^5 u_s(\p) = \frac{1}{E} \, \xi_s^\dagger \, \p \cdot \sigmav \, \xi_s = \vv \cdot \shat
~,
\ee
where $\shat$ is the unit normalized spin vector, and 
\be
\bra{\p, s} \Psibar \g^i \g^5 \Psi \ket{\p, s} = \overline{u}_s(\p) \g^i \g^5 u_s(\p) = \sqrt{1-v^2} ~ s_i + \frac{v_i \, \vv \cdot \shat}{1 + \sqrt{1-v^2}}
~.
\ee
Thus, in a normalized one-particle wavepacket state, the axial vector current is 
\be 
\label{eq:av_current}
\langle \Psibar \g^\mu \g^5 \Psi \rangle = \big(\vv \cdot \shat \, , \, \shat + \order{v^2} \big)^\mu ~ \delta^{(3)}(\xv - \xv_0)
~,
\ee
which recovers \Eq{classicalaxial}. Note that this derivation implicitly treats $\partial_\mu a$ as constant because it only considers diagonal momentum-space matrix elements; thus, it does not yield the $\grad \dot{a}$ term in the full Hamiltonian \Eq{HamDeriv}. Finally, note that the four-vectors in \Eqs{particle_current}{av_current} are not Lorentz covariant, because they multiply a noncovariant delta function; however, they are related by a factor of $\gamma$ to the familiar four-velocity $u^\mu$ and spin four-vector $s^\mu$~\cite{jackson1999classical,misner1973gravitation}.

\section{Radiation From Slabs and Multilayers}
\label{app:appendixlayer}

In this appendix, we derive the results quoted in \Sec{EMsig} for radiation emitted from single slabs. We then discuss how these results are extended to calculations for multilayer setups. 

\begin{center}
\textit{Polarization Currents}
\end{center}

First, we derive the amplitude of the radiation field outside of a single infinite slab of thickness $d$, with boundaries located at $z = \pm d/2$. As discussed in \Sec{LAMPOST}, at the high frequencies relevant for the axioelectric polarization current, we can neglect the tensorial nature of $\mu$. For concreteness, suppose the slab's spin polarization lies along $\xhat$. Solving \Eq{Efield1} for the electric field within each region gives
\be
E_x = e^{- i m_a t} \times \begin{cases} E_\sig \, e^{- i m_a z} & z < -d/2 \\ E_{\text{in}} \cos(n m_a z) + J_a^P / (i m_a \eps) & -d/2 < z < d/2 \\ E_\sig \, e^{i m_a z} & z > d/2 ~, \end{cases} 
\ee
where $E_\sig$ and $E_{\text{in}}$ are complex amplitudes and we used the fact that the polarization current, and hence the field, is symmetric about $d = 0$. Imposing continuity of $E_x$ at $z = d/2$, we have
\be
E_{\text{in}} \, \cos(n m_a d / 2) + \frac{J_a^P}{i m_a \, \eps} = E_\sig \, e^{i m_a d / 2}
~.
\ee
Furthermore, since there is no axion-induced surface current in this case, $B_y/\mu$ is also continuous at the boundary, such that Faraday's law yields
\be
\frac{i n}{\mu} \, E_{\text{in}} \, \sin(n m_a d / 2) = E_\sig \, e^{i m_a d / 2}
~.
\ee
Solving this system of equations, we arrive at \Eq{ElecSlab}. 

In a situation involving many interfaces, such as a multilayer dielectric haloscope, it is not practical to directly solve for $E_x$ in each region by simultaneously imposing all of the boundary conditions. Instead, the solution can be built up by considering the outgoing radiation generated at each interface. Each of these waves then propagates through the rest of the stack,  as in ordinary electrodynamics. Thus, let us continue by considering an interface at $z = 0$ between two mediums with scalar permittivities $\eps_{1, 2}$ and permeabilities $\mu_{1, 2}$, assuming uniform spin polarization along $\xhat$. For the axioelectric polarization current, the resulting electric field is 
\be
E_x = e^{- i m_a t} \times \begin{cases} E_1^\g \, e^{- i n_1 m_a z} + E_1^a & z < 0 \\ E_2^\g \, e^{i n_2 m_a z} + E_2^a & z > 0 ~, \end{cases}
\ee
where $E_i^a = J_{a,i}^P / (i m_a \eps_i)$. Imposing continuity of $\E_\parallel$ and $\B_\parallel/\mu$, we have
\begin{align}
E_1^\g + E_1^a &= E_2^\g + E_2^a \,,\\
(-n_1 / \mu_1) \, E_1^\g &= (n_2 / \mu_2) \, E_2^\g
~,
\end{align}
whose solution is
\begin{align} 
\label{eq:E_outgoing}
E_1^\g = - E_2^\g = \left(E_2^a-E_1^a\right)\, \frac{\eps_2n_1}{\eps_1n_2+\eps_2n_1}
~. 
\end{align}
If both media are maximally spin-polarized, with $\eps_{\sigma, i} = \eps_i$, then the difference of $E_i^a$ above simplifies to 
\be
E_2^a - E_1^a = \left(\frac{\eps_{\sigma, 1} - 1}{\eps_1} - \frac{\eps_{\sigma, 2} - 1}{\eps_2} \right) E_{\eff} = \left(\frac{1}{\eps_2} - \frac{1}{\eps_1} \right) E_{\eff}
~.
\ee
In this case, \Eq{E_outgoing} exactly matches the analogous result for the axion-photon coupling in, e.g., Sec.~3.1 of Ref.~\cite{Millar:2016cjp}, after making the replacment $g_{a\g \g} \, a \, B_0 \to - E_{\eff}$. Thus, in \Sec{LAMPOST} we may directly recast results from conventional dielectric haloscope calculations, as stated in \Eq{ElecMapping}.

More generally, the layers can have different degrees of spin polarization, which leads to more flexibility than in a conventional dielectric haloscope. If $\eps_2$ is large, then $E_2^a - E_1^a$ is maximized when $\eps_1 = 1$. As such, the most aggressive forecast for the LAMPOST optical dielectric haloscope in Ref.~\cite{Baryakhtar:2018doz} assumed vacuum gaps between layers. These are mechanically challenging at optical wavelengths and thus have been avoided by current LAMPOST and MuDHI prototypes. However, here we can achieve the same enhancement in $E_2^a - E_1^a$ with arbitrary $\eps_1$ by simply not spin-polarizing that medium, such that $\eps_{\sigma, 1} = 1$. This case does not  directly map onto a conventional dielectric haloscope, but we expect it yields a similar signal enhancement without the need for vacuum gaps.

\begin{center}
\textit{Magnetization Currents}
\end{center}

For the axion wind induced magnetization current, we assume that $\grad a$ is uniform and lies in the slab's plane. As discussed in \Sec{MADMAX}, we can treat the circular polarizations of the generated electric field separately. These polarizations have amplitudes
\be
E^\pm = e^{- i m_a t} \times \begin{cases} -E_\sig^\pm \, e^{- i m_a z} & z < -d/2, \\ E_{\text{in}}^\pm \, \sin(n m_a z) & -d/2 < z < d/2, \\ E_\sig^\pm \, e^{i m_a z} & z > d/2 ~, \end{cases} 
\ee
where we used the fact that the magnetization current, and hence the field, is antisymmetric about $d = 0$. At $z = d/2$, $E^\pm$ is continuous while $B^\pm/\mu_\pm$ jumps by $M_a^\pm$, giving
\begin{align}
E_{\text{in}}^\pm \, \sin(n_\pm m_a d / 2) &= E_\sig^\pm \, e^{i m_a d / 2} \\
- \frac{in_\pm}{\mu_\pm} \, E_{\text{in}}^\pm \, \cos(n_\pm m_a d / 2) &= E_\sig^\pm \, e^{i m_a d / 2} + M_a^\pm
~.
\end{align}
Solving these equations and using the definition of $\Mv_a$ in \Eq{mag_def}, along with $B_{\eff}^\pm = B_{\eff}/\sqrt{2}$, then yields \Eq{MagSlab}. 

Once again, to understand a general multilayer, we first consider the radiation emitted from a single interface at $z = 0$. In this case, the radiation is due to a surface current $\v{K}_a = \Mv^\parallel_{a,1} - \Mv^\parallel_{a,2}$ at the boundary. The radiation field is  
\be
E^\pm = e^{- i m_a t} \times \begin{cases} E_1^{\pm} \, e^{- i n_1 m_a z}  & z < 0 \\ E_2^{\pm} \, e^{i n_2 m_a z} & z > 0 ~. \end{cases}
\ee
The relevant boundary conditions,
\begin{align}
E_1^\pm &= E_2^\pm, \\
(-n_1 / \mu_1) E_1^\pm &= (n_2 / \mu_2) E_2^\pm + K_a^\pm
~,
\end{align}
determine $E_{1,2}^\pm$ to be 
\be
E_1^\pm = E_2^\pm = K_a^\pm ~ \frac{\mu_1 \, \mu_2}{\mu_1 \, n_2 + \mu_2 \, n_1}
~.
\ee
Note that this is a completely different structure compared to the analogous result in \Eq{E_outgoing}, and hence has no simple mapping to a conventional dielectric haloscope. 

\section{Power Absorption in Classical Electrodynamics}
\label{app:Ohms}

The power absorbed from the axion field in a spin-polarized medium, through either the axioelectric or axion wind terms, can be computed classically by considering the work done on the electrons. Similar arguments have been used for photon-coupled axions~\cite{Lawson:2019brd,Marsh:2022fmo,ALPHA:2022rxj} and dark photons~\cite{Gelmini:2020kcu}. For electron-coupled axions, one must take care to account for the work done by both the real electromagnetic fields $\E$ and $\B$ and the effective fields $\E_{\eff}$ and $\B_{\eff}$. 

First, consider absorption through the axioelectric term in an infinite medium with $\epsa \simeq \eps$. Both $\E$ and $\E_{\eff}$ do work on the electrons, since the net force is proportional to their sum. Applying \Eq{Einfinite} gives $\E + \E_\eff \simeq \E_{\eff}/\eps$, and combining \Eqs{axioncurrent}{totj} gives a total current density $\jv = (\eps-1) \, \dt \E_\eff / \eps$. The time-averaged power dissipated in the medium per unit volume is then given by a generalized form of Ohm's law, \timec
\be
\label{eq:PowerV}
\frac{P}{V} = \tsw{-}{}
\frac{1}{2} \, \text{Re} \big( (\E + \E_\eff) \cdot \jv^* \big)
\simeq \frac{m_a}{2} \, |E_\eff^2| \, \Im\Big( \frac{\tsw{}{-} 1}{\eps}\Big) 
~,
\ee
where the factor of $1/2$ accounts for the time average.

Next, consider absorption through the axion wind term in an infinite medium with a constant background magnetization $\Mv_0$ and magnetic field $\v{B}$. In this case, the work done is associated with the rotation of magnetic dipoles in the presence of $\B$ and $\B_{\eff}$. Accounting for both of these fields gives 
\be
\frac{P}{V} = \frac{1}{2} \, \text{Re} \big( (\v{H} + \v{H}_{\text{eff}}) \cdot (\del_t \Mv)^* \big) = \frac{1}{2} \, \text{Re} \Big( \big(\B + \B_\eff - \Mv_0 - \Mv_a \big) \cdot \big(\del_t (\Mv_0 + \Mv_a)\big)^* \Big)
~.
\ee
Since $\Mv_0$ is constant and the induced magnetization $\Mv_a$ is orthogonal to $\Mv_0$ and $\B$, the expression above reduces to 
\be 
\label{eq:magPower}
\frac{P}{V} = \frac{1}{2} \, \text{Re} \big( (\B_\eff - \Mv_a) \cdot (\del_t \Mv_a)^* \big) = \frac{m_a}{2} \, |B_{\text{eff}}^2| \, \Im\left( \tsw{}{-} \qhat^T \mu^{-1} \qhat \right) 
~,
\ee
where we applied \Eq{PandM}, used the fact that $\mu$ is Hermitian, and defined $\hat{\v{q}}$ as the unit vector pointing along the axion gradient. The dominant contribution to the power comes from the resonantly amplified ``minus'' circular polarization. Its permeability, given in \Eq{mu_expr}, corresponds to \timec
\be
\Im\bigg( \frac{\tsw{}{-} 1}{\mu_-}\bigg)  = \frac{m_a \, \w_M / 2 Q}{(m_a - \w_B)^2 + (\w_B / 2 Q)^2}
~.
\ee
The power absorption rate is maximized when the axion frequency matches the infinite-medium Kittel frequency, $m_a \simeq \w_B$, and $\hat{\v{q}}$ is perpendicular to the background magnetization. In this case, we have
\be \label{eq:final_magnon_rate}
\frac{P}{V} = \frac12 \, |B_{\text{eff}}^2|\, Q \, \w_M
~.
\ee
Alternatively, if one averages over all directions of $\qhat$, then the power is reduced by a factor of $2/3$.

The results above for the power density $P/V$ match those of \Sec{absorb} after rescaling to a rate per unit target mass, $R = (P / V) / (\rho_T \, m_a)$, and substituting in the definitions of the effective fields. In particular, \Eq{PowerV} corresponds to the second line of \Eq{Rabs} after substituting $|E_{\text{eff}}^2|/2 = (g_{ae} \, m_a/e)^2 \, \rhodm$, and \Eq{magPower} reduces to \Eq{classicalmagnon} after substituting $|B_{\text{eff}}^2|/2 = (g_{ae} \, \vdm/\mu_B)^2 \, \rhodm$.

Though the derivations here assumed infinite media, they also apply to finite media, provided that the characteristic size  of the medium is much larger than the screening length and boundary conditions are unimportant. Moreover, these calculations can be straightforwardly generalized for arbitrary geometries, provided one accounts for how the boundary conditions affect the real electromagnetic fields, as was done in Ref.~\cite{ALPHA:2022rxj}. On the other hand, since in-medium fields and currents are always macroscopically averaged, these arguments cannot be applied to unpolarized media, where those quantities would simply vanish. For the unpolarized case, one must instead use the methods of \Sec{absorb}.

\section{Analysis of Ferrite Multilayers}
\label{app:YIG}

In this appendix, we justify the choices of ferrite material properties listed in Table~\ref{tab:noise}, discuss analytic estimates of the signal power in a multilayer setup, and present numeric results for single crystal YIG.

\begin{center}
\textit{Properties of Ferrite Materials}
\end{center}

Ferrite materials are well-studied~\cite{nicolas1980microwave,fuller1987ferrites,dionne2009magnetic,ozgur2009microwave}, and here we present some of their measured properties. First, for polycrystalline spinel ferrites, we infer the saturation magnetization and microwave permittivity $\eps$ from a commercial datasheet~\cite{datasheet}. This datasheet also gives the microwave permittivity of YIG, which is comparable to that of the spinel ferrites. The saturation magnetization of YIG significantly increases when it is cooled. In particular, at cryogenic temperatures it is $M_S = 0.25 \ \mathrm{T}$~\cite{maier2017temperature}, half that of spinel ferrite. 

Ferrites can be fabricated with small dielectric losses, $\tan \delta_\eps = \Im(\eps) / \text{Re}(\eps) \lesssim 10^{-4}$, which are largely due to the presence of divalent Fe ions or other impurities~\cite{nicolas1980microwave}. Thus, magnetic losses dominate for all materials we consider here. The magnetic quality factor $Q$ is the most difficult parameter to infer; for simplicity, we first focus on $Q$ for single crystal YIG. 

The quality factor of undoped single crystal YIG spheres has been directly measured at room temperature and ranges from $Q \sim (0.7 - 1.8) \times 10^4$ at microwave frequencies~\cite{klingler2017gilbert,krupka2018ferromagnetic,pacewicz2019rigorous}. However, for YIG, $Q$ has a complex dependence on temperature. It is generally inferred from the linewidth of the ferromagnetic resonance, $\Delta H \simeq H/Q$. As the temperature is reduced, the linewidth is observed to initially reduce, but then sharply increase below $T \sim 100 \ \text{K}$~\cite{spencer1959low,spencer1961low,maier2017temperature}. These trends are understood theoretically to be arising from the reduced density of thermal phonons and the increased effect of fast-relaxing rare earth ions, respectively~\cite{dionne2009magnetic}. At even lower temperatures, $T \lesssim 10 \ \text{K}$, the linewidth reduces again as the effects of rare earth ions freezes out. However, recent measurements down to $T = 30 \ \text{mK}$ have found that the linewidth increases again due to coupling to two-level systems~\cite{boventer2018complex,pfirrmann2019magnons}. Remarkably, the linewidth at such low temperatures is comparable to that at room temperature, even though the loss mechanisms are completely different. Given this ambiguity, in our discussion below, we adopt an intermediate value of $Q \sim 10^4$ for single crystal YIG. 

In polycrystalline materials, the resonance linewidth is much larger. For instance, Ref.~\cite{datasheet} found that $\Delta H \lesssim 200 \ \text{Oe}$ for a millimeter wave ferrite. Taking this number at face value and assuming that this material is accurately described by \Eq{LL_equation} then yields $Q \sim 20$. However, in reality the linewidth for polycrystalline materials is mostly due to inhomogeneities in the material's microscopic structure and, thus, cannot be used to infer the damping away from resonance~\cite{vrehen1969absorption,patton1972review,white1978magnetic}. Instead, one must measure the frequency-dependent ``effective linewidth'' $\Delta H_\eff$, which determines the parameter $Q$ appearing in \Eq{mu_expr}. Away from resonance, $\Delta H_\eff$ is much smaller than $\Delta H$ for polycrystalline materials and can even approach the value of $\Delta H$ for single crystals~\cite{mo2007low}. For instance, Ref.~\cite{nicolas1980microwave} quotes a linewidth of $\Delta H_\eff = 4 \ \text{Oe}$ for lithium ferrites, corresponding to $Q \sim 10^3$. However, given the many other uncertainties, we adopt a conservative intermediate value of $Q \sim 10^2$ for polycrystalline spinel ferrite. 

In both cases, a detailed estimate of the experimental sensitivity requires dedicated measurements in appropriate conditions. Since magnetic losses depend on the geometry of the sample and applied field, such measurements should use thin slab samples with an orthogonal applied field. They must also be performed at cryogenic temperatures and low input power, well away from resonance.

\begin{center}
\textit{Analytic Estimates}
\end{center}

As noted in \Sec{MADMAX}, the quality factor of polycrystalline spinel ferrite is too low to derive accurate, simple analytic expressions. However, for single crystal YIG, analytic approximations provide an excellent description of the numeric result. In deriving them below, we assume that $\eps$ is a real $\order{1}$ number and $\Delta \w \ll m_a, \w_M$. We also assume that the setup is operated many linewidths away from the resonance, $\Delta \w \gg m_a / Q$, which increases the screening length and allows the use of many layers. These approximations are consistent provided that $Q \gg 1$ and $Q \, \w_M / m_a \gg 1$, which are satisfied, for both polycrystalline ferrite and YIG, for all masses we consider. 

In this regime, the transparent mode slab thickness is 
\be
d \simeq \frac{\pi}{m_a} \sqrt{\frac{\w_m}{\eps \, \Delta \w}}~, \label{eq:d_eq}
\ee
which is much larger than the vacuum gaps between slabs. The sensitivity bandwidth is 
\be
\Delta \w_s \simeq \frac{\pi}{2} \, \frac{m_a}{Q}~, \label{eq:ws_eq}
\ee
which scales with $Q$ just as in typical resonant experiments. The number of layers grows with $\Delta \w$,
\be
N \simeq \frac{4Q}{\pi} \frac{\Delta \w}{m_a}~, \label{eq:N_eq_2} \\ 
\ee
which is large when $\Delta \w \gg m_a/Q$, as anticipated. Finally, the form factor has nontrivial dependence on $\Delta \w$, even though we have fixed $\text{Re}(n_-) m_a d = \pi$, because the argument of the cotangent in \Eq{MagSlab} has an imaginary part. In particular, the form factor is approximately
\be
\FF_- \simeq \bigg| \frac{1}{\mu - i n \tan(i/2N)} \bigg| \simeq \bigg| \frac{1}{\mu + n/2N} \bigg| \simeq \bigg| \frac{\Delta \w}{\w_M} + \frac{\pi \sqrt{\eps}}{8 Q} \, \frac{m_a}{\sqrt{\w_M \, \Delta \w}} \bigg|^{-1}~, \label{eq:FF_eq} 
\ee
so that increasing $\Delta \w$ decreases the cotangent term, leading to a peak in $\FF_-$ at nonzero $\Delta \w$. 

\begin{figure*}[t]
\includegraphics[width=2.35in]{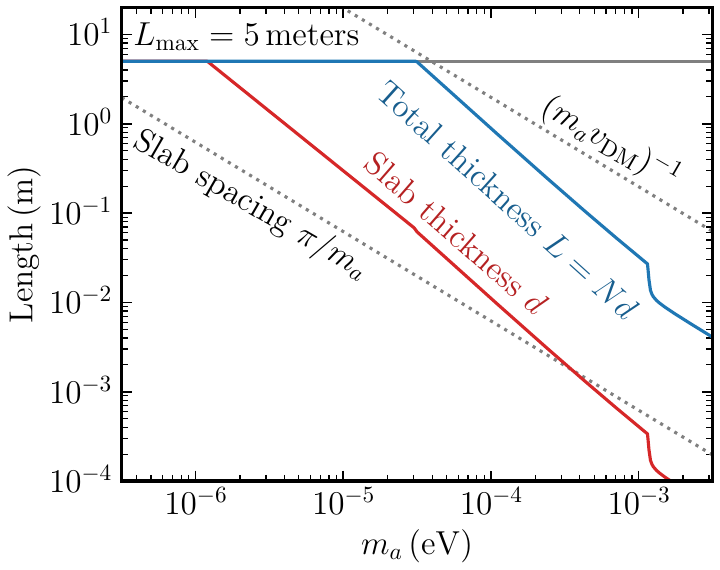} \hspace{-0.1in}
\includegraphics[width=2.35in]{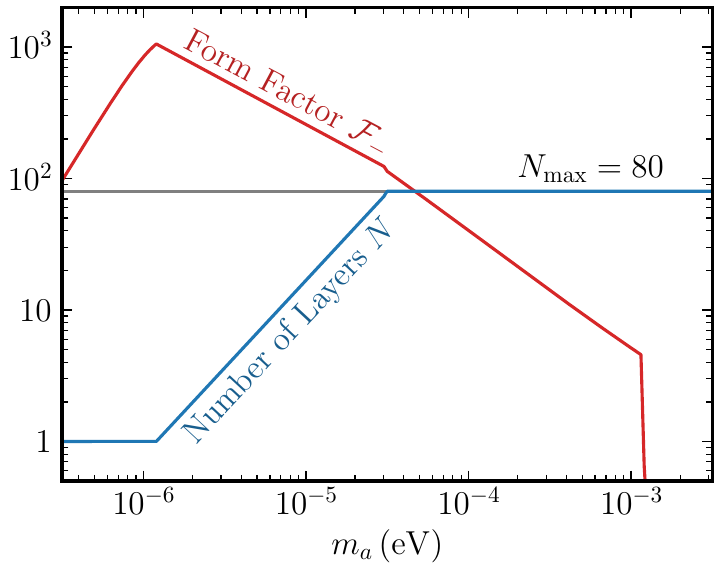} \hspace{-0.1in}
\includegraphics[width=2.35in]{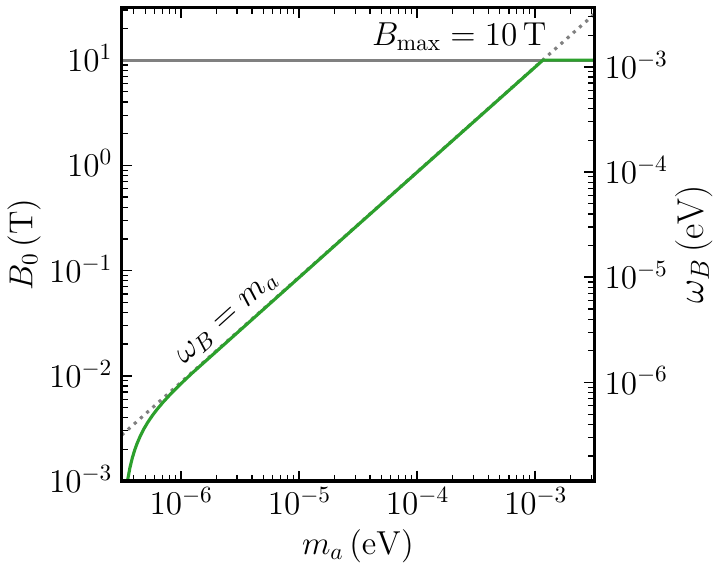} 
\caption{Analogue of \Fig{values}, but for single crystal YIG.}
\label{fig:values_yig} 
\end{figure*}

The signal power is determined by the product $N \, \FF_-$, which monotonically increases with $\Delta \w$ until it saturates at $N \, \FF_- \sim (4/\pi) \, (Q \, \w_M / m_a)$ at frequency separations $\Delta \w \gtrsim \Delta \w_0 \equiv (m_a \, \sqrt{\eps \, \w_M}/Q)^{2/3}$. In this regime, $N$ and $\FF_-$ are inversely proportional, yielding a simple estimate for the signal power,
\be
\label{eq:Psigopt2}
P_\sig \simeq \frac{4}{\pi^2} \bigg(\frac{Q \, \w_M}{m_a} \bigg)^2 \, | B_\eff^2 | \, A
~.
\ee

Now, we can perform some cross-checks. First, note that for a fixed $\Delta \w$, the total volume of the slabs scales as $V = N \, A \, d \propto Q$, so that \Eq{Psigopt2} can be rewritten as $P_\sig \propto Q \, V$ as generically expected. Second, \Eq{Psigopt2} was derived without considering the limits on the number of layers and total slab length, both of which increase with $\Delta \w$. Demanding that these limits do not take effect until the maximum signal power is reached at $\Delta \w \gtrsim \Delta \w_0$ implies
\begin{align}
N_{\text{max}} &\gtrsim \left( \frac{Q \, \w_M}{m_a} \right)^{1/3} \label{eq:N_cons} \\
m_a L_{\text{max}} &\gtrsim \left( \frac{Q \, \w_M}{m_a} \right)^{2/3} \label{eq:L_cons}
~,
\end{align}
where we have dropped numeric factors. For both materials, \Eq{N_cons} is true for all masses we consider, and \Eq{L_cons} is true only in the upper half of the mass range. For lower masses, the signal power is limited by the constraint on $L$. 

Finally, for self-consistency we must ensure that the optimal $\Delta \w$ is not too large for our approximations to break down. Since it is favorable to increase $\Delta \w$ whenever our approximations hold, its value is dictated by either $N_{\text{max}}$ or $L_{\text{max}}$. When the limit comes from the former, $N = N_{\text{max}}$, demanding that $\Delta \w \ll \w_M, m_a$ yields the condition
\begin{align}
Q \gg N_{\text{max}}, (m_a / \w_M) N_{\text{max}}~.
\end{align}
For single crystal YIG, this is satisfied for the entire mass range we consider, but for polycrystalline ferrite it is not satisfied at all; in the latter case a signal power higher than \Eq{Psigopt2} is possible. 

\begin{center}
\textit{Sensitivity with Single Crystal YIG}
\end{center}

\begin{figure*}
\includegraphics[width=12cm]{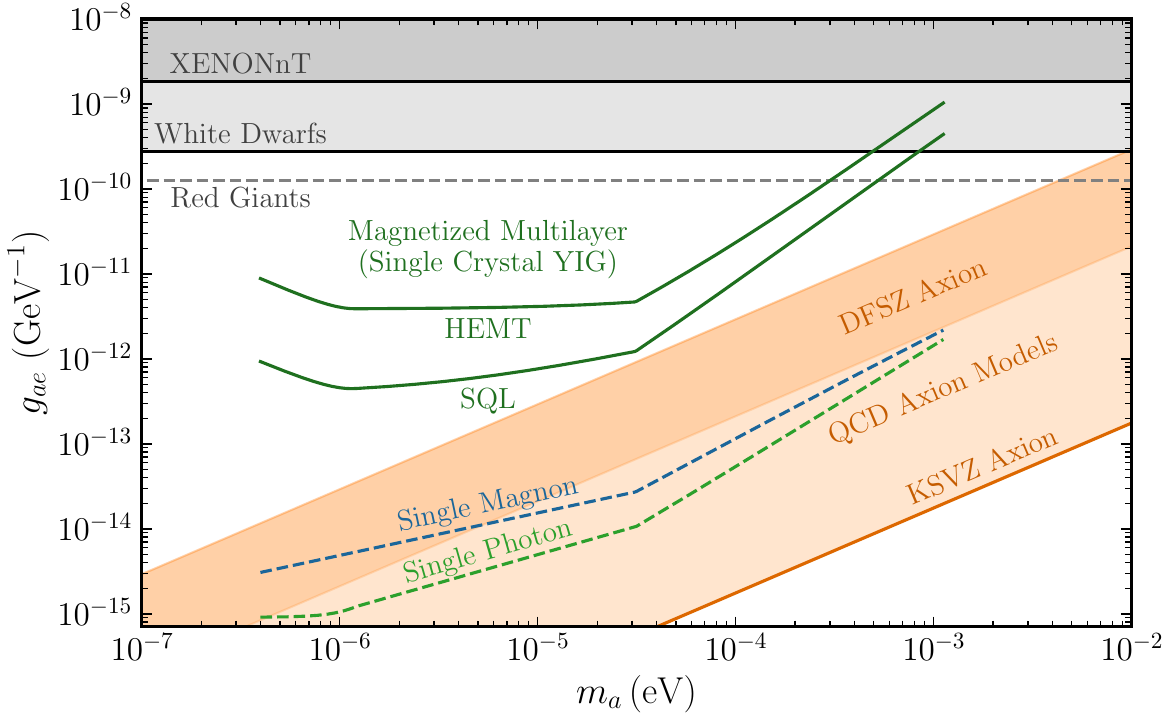} 
\caption{Analogue of \Fig{projection_ferrite}, but for single crystal YIG. We assume material parameters $M_S = 0.25 \ \mathrm{T}$, $Q = 10^4$, and $\eps = 15$, while all other parameters are as in Table~\ref{tab:noise}.}
\label{fig:projection_YIG}
\end{figure*}

For the sake of comparision, we compute the sensitivity for single crystal YIG, corresponding to $M_S = 0.25 \ \mathrm{T}$, $Q = 10^4$, and $\eps = 15$, using the same procedure as in \Sec{MADMAX}. Some intermediate quantities from the numeric calculation are shown in \Fig{values_yig}. They are qualitatively similar to the corresponding results for polycrystaline ferrite, but they match the analytic results quite closely for $m_a \gtrsim 3 \times 10^{-5} \, \eV$, where the multilayer is not limited by $L_{\text{max}}$. 

The resulting sensitivity for a YIG multilayer setup is shown in \Fig{projection_YIG}. Numerically, it is only slightly greater than the result in \Fig{projection_ferrite} for polycrystalline ferrite. Though the signal power is enhanced by $Q^2$, this improvement is partially cancelled by the decrease in $\w_M$ and in the sensitivity bandwidth. In addition, the signal power for polycrystalline ferrite is somewhat greater than the approximate analytic result in \Eq{Psigopt2}. Overall, for a magnetized multilayer setup, using single crystal YIG provides only a small sensitivity enhancement at a greatly increased cost. 

\section{Equivalence of the Nonderivative Coupling}
\label{app:nonderiv}

The axion-fermion coupling has equivalent derivative and nonderivative forms, which we denote with subscripts $d$ and $n$, respectively. For brevity, we will drop all other subscripts in this appendix, replacing $q_f$, $m_f$, and $g_{af}$ with $q$, $m$, and $g$, respectively. In this work, we have focused mostly on the derivative form of the coupling,
\be
\label{eq:app_ld}
\Lag_d = g \, (\del_\mu a) \, \Psibar \g^\mu \g^5 \Psi
~.
\ee
In \Sec{EDM}, we discussed how this coupling is equivalent at $\order{g}$ via a field redefinition to a nonderivative form
\be
\label{eq:app_ln}
\Lag_n = - 2 m \, g a \, \Psibar i \g^5 \Psi ~,
\ee
where we neglected the axion-photon coupling generated by the chiral anomaly. In this appendix, we explain how these couplings are equivalent at the level of their nonrelativistic Hamiltonians. In \Sec{LowHam}, we showed that, up to higher-order terms in $1/m$, the derivative coupling $\Lag_d$ corresponds to
\be 
\label{eq:app_hd}
H_d = H_0 - g \, (\grad a) \cdot \sigmav - \frac{g}{2 m} \, \{ \dot{a} , \Piv \cdot \sigmav\} 
~,
\ee
where $H_0$ is the usual Pauli Hamiltonian. In \App{nr_ham}, we will show that $\Lag_n$ instead corresponds to 
\be 
\label{eq:app_hn}
H_n = H_0 - g \, (\grad a) \cdot \sigmav - \frac{g}{4 m} \, \{\dot{a}, \Piv \cdot \sigmav\} + \frac{q \, g}{2 m} \,  a \, \E \cdot \sigmav 
~,
\ee
to the same order in $1/m$. In \App{ham_equiv}, we show that $H_d$ and $H_n$ are related by a unitary transformation and thereby physically equivalent. These results are consistent with those in recent works. 

The two Hamiltonians superficially do not appear to be physically equivalent, which has led to substantial recent confusion. First, $H_d$ and $H_n$ do not have the same coefficient for the axioelectric term. In \App{app_axioelectric}, we show that the axioelectric term encodes measurable relative acceleration effects. As a result, when one considers a multi-particle Hamiltonian, $H_n$ necessarily contains complicated additional terms which precisely compensate for the smaller coefficient of its axioelectric term. Second, for constant $a$, $H_n$ contains a term of the form $d \, \sigmav \cdot \E$ with $d \propto q \, a$, which appears to violate the axion's underlying shift symmetry. In \App{nredm}, we explain why this ``nonrelativistic EDM'' term has no $\order{a}$ physical effects in the nonrelativistic limit. The essential reason is that such a term is just an artifact of choosing to describe an ordinary charged particle with a shifted position operator. Of course, as we discuss in \App{true_EDM}, true EDMs do have physical effects, but those effects arise solely through their higher-order and relativistic corrections. By contrast, this axion-induced spurious EDM does not share these corrections and therefore does not produce a signal proportional to $a$ in any EDM experiment, reflecting the axion's underlying shift symmetry.

\stoptocentries
\subsection{Deriving The Nonderivative Hamiltonian}
\starttocentries
\label{app:nr_ham}

To derive $H_n$, we use the Pauli elimination method, as in \Sec{LowHam}. For $\Lag_n$, the fermion's equation of motion is
\be \label{eq:nonderiv_eom}
(i \slashed{\del} - m - q \, \slashed{A} - 2 i m \, g a \, \g^5) \, \Psi = 0 ~,
\ee 
which in terms of two-component fields is
\begin{align}
(i \del_t - q \phi) \, \psi &= ( \Piv \cdot \sigmav + 2 i m \, g a) \, \psibar  \label{eq:psi_eom_2} \\
(i \del_t + 2 m - q \phi) \, \psibar &= ( \Piv \cdot \sigmav - 2 i m \, g a) \, \psi
~. \label{eq:chi_eom_2}
\end{align}
In these equations, the axion coupling always appears with the fermion mass $m$, so we must work to higher order in the nonrelativistic expansion to capture the desired $\order{g/m}$ terms in the Hamiltonian. To do so, we write $\psibar$ as
\begin{align}
\psibar &= \frac{1}{2m} (\Piv \cdot \sigmav - 2 i m \, g a) \, \psi - \frac{i \del_t - q \phi}{2m} \, \psibar \label{eq:chi_expr_1} \\
&= \frac{1}{2m} \left( 1 - \frac{i \del_t - q \phi}{2m} \right) (\Piv \cdot \sigmav - 2 i m \, g a) \, \psi + \order{1/m^3} ~, \label{eq:chi_expr_2}
\end{align}
where we iterated \Eq{chi_expr_1} to reach \Eq{chi_expr_2}. Substituting this back into \Eq{psi_eom_2} yields
\be \label{eq:psi_intermediate}
(i \del_t - q \phi)\,  \psi \simeq \frac{1}{2m} \, (\Piv \cdot \sigmav + 2 i m \, g a) \left( 1 - \frac{i \del_t - q\phi}{2m} \right) (\Piv \cdot \sigmav - 2 i m \, g a) \, \psi
~,
\ee
which is accurate up to $\order{1/m^3}$ terms. Note that this power counting argument treats the momentum as $\order{1}$, so that an expansion in $1/m$ is also an expansion in velocity. 

At this point, we can read off a fiducial Hamiltonian $\bar{H}_n$ by defining $i \del_t \psi = \bar{H}_n \psi$. As shown in Refs.~\cite{strange1998relativistic,berestetskii1982quantum}, the $g$-independent terms at $\order{1/m^2}$ will contain the usual fine-structure corrections to the Pauli Hamiltonian. We are instead interested in the $\order{g/m}$ terms, which can arise from either the first or last factor in \Eq{psi_intermediate}. The contribution from the first factor of \Eq{psi_intermediate} is
\be
\bar{H}_n \supset iga \left( 1 - \frac{i \del_t - q\phi}{2m} \right) \Piv \cdot \sigmav \simeq iga \, \Piv \cdot \sigmav - \frac{ga}{2m} \, q \dot{\A} \cdot \sigmav ~, \label{eq:partial_h_1}
\ee
where we used the fact that $(i \del_t - q \phi) \, \psi = \order{1/m}$ and can thus be neglected. The last factor of \Eq{psi_intermediate} gives
\begin{align}
\bar{H}_n &\supset \Piv \cdot \sigmav \left( 1 - \frac{i \del_t - q\phi}{2m} \right) (-iga) \\
&\simeq -iga \, \Piv \cdot \sigmav - g \,  (\grad a) \cdot \sigmav - \frac{ga}{2m} \, q (\grad \phi) \cdot \sigmav + \frac{ig}{2m} \, (\grad \dot{a}) \cdot \sigmav - \frac{g\dot{a}}{2m} \, \Piv \cdot \sigmav ~ . \label{eq:partial_h_2}
\end{align}
Combining the two yields the $\order{g/m}$ part of the Hamiltonian,
\be
\label{eq:Hbaralt}
\bar{H}_n \supset - g \, (\grad a) \cdot \sigmav + \frac{ga}{2m} \, q \E \cdot \sigmav + \frac{ig}{2m} \, (\grad \dot{a}) \cdot \sigmav - \frac{g\dot{a}}{2m} \, \Piv \cdot \sigmav ~ .
\ee

However, this Hamiltonian is not physically appropriate, as it is not Hermitian. The reason is that the equations of motion preserve the norm of the full four-component wavefunction $\Psi$, which is
\be
\int d^3 \xv \, \Psi^\dagger \Psi = \int d^3 \xv \, \psi^\dagger \psi + \psibar^\dagger \psibar \simeq \int d^3 \xv \, \psi^\dagger \left(1 + \frac{ (\Piv \cdot \sigmav)^2}{4m^2} - \frac{g \, (\grad a) \cdot \sigmav}{2m} \right) \psi
~,
\ee
where we used \Eq{chi_expr_2} and again dropped $\order{1/m^3}$ terms. We thus define a renormalized two-component wavefunction whose norm is preserved,
\be \label{eq:def_rescaled}
\psi_{\text{nr}} \simeq M \, \psi \equiv \left(1 + \frac{ (\Piv \cdot \sigmav)^2}{8m^2} - \frac{g \, (\grad a) \cdot \sigmav}{4m} \right) \psi
~,
\ee
which defines the rescaling coefficient $M$. This subtlety was irrelevant for the derivation of $H_d$ in \Sec{LowHam}, since there the contributions to $M$ started at $\order{g/m^2}$. The physical Hamiltonian is then identified using $i \del_t \psi_{\text{nr}} = H_n \psi_{\text{nr}}$, giving 
\be
H_n \simeq \bar{H}_n + [M, \bar{H}_n] + i \del_t M
~.
\ee
As discussed in Refs.~\cite{strange1998relativistic,berestetskii1982quantum}, these $M$-dependent correction terms are essential for reproducing fine-structure corrections; to give a more recent application, such a field renormalization is also required to derive correct results in heavy quark effective theory~\cite{Balk:1993ev,Gardestig:2007mk}. For our purposes, we are interested in the $g$-dependent correction term, $i \del_t M \supset - i g \, (\grad \dot{a}) \cdot \sigmav / 4m$. Including this term, we arrive at the claimed result \Eq{app_hn}, which matches the one found in Ref.~\cite{Smith:2023htu} using the methods of Foldy--Wouthuysen transformations. The Pauli elimination method was also used in Refs.~\cite{Alexander:2017zoh,Wang:2021dfj,DiLuzio:2023ifp}, yielding the same EDM term, though only Ref.~\cite{DiLuzio:2023ifp} kept track of the full Hamiltonian. 

\stoptocentries
\subsection{Equivalence of the Hamiltonians}
\starttocentries
\label{app:ham_equiv}

\begin{center}
\textit{Truncating the Interaction}
\end{center}

Before we explain why $H_d$ and $H_n$ are equivalent, let us first dispatch a red herring. It is tempting to conclude that the mismatch arises from truncating the exponential at $\order{g}$ in the full nonderivative form of the interaction in \Eq{nonderiv_lag}, resulting in $\Lag_n$. Historically, this was the resolution to a related puzzle which arose in the study of pion-nucleon interactions, which have ``pseudovector'' and ``pseudoscalar'' forms closely analogous to $\Lag_d$ and $\Lag_n$, respectively. In terms of our axion-based language, Dyson claimed that the couplings were equivalent on the basis of the axion wind terms being the same~\cite{dyson1948interactions}. However, the two couplings give different amplitudes at $\order{g^2}$~\cite{slotnick1949charge}, and it was quickly realized that the couplings were only equivalent up to additional interaction terms of $\order{g^2}$~\cite{case1949equivalence,berger1952equivalence,drell1952pseudoscalar}. To give a more recent example, for axions produced in supernova, both axion-nucleon and pion-nucleon interactions are relevant. Both interactions can be rotated to the nonderivative form, but in that case the cross-term $(g \, m / f_\pi) \, \bar{N} a \, \pi N$ must be kept to compute a correct axion bremsstrahlung rate~\cite{Raffelt:1987yt,Carena:1988kr,Choi:1988xt}. 

By contrast, in our case there is nothing wrong with truncating at $\order{g}$. We are focusing on axion dark matter searches, where all relevant amplitudes are $\order{g}$, since $\order{g^2}$ and higher effects are negligible. Furthermore, the mismatch between the Hamiltonians starts at $\order{g}$, so it cannot be resolved by considering higher-order terms. 

Incidentally, we note that $\Lag_n$ is often derived by applying integration by parts to $\Lag_d$, and then ``simplifying'' with the free Dirac equation. This is not exact, but it yields the correct result because applying the equations of motion preserves $S$-matrix elements up to the addition of higher-order terms that are negligible here~\cite{Arzt:1993gz}.

\begin{center}
\textit{Unitary Equivalence}
\end{center}

To relate the two Hamiltonians, we note that any quantum theory can be reparametrized by applying a unitary transformation to the states. In particular, for any Hermitian operator $S$, we can define ``primed" states by $\ket{\psi'} = e^{iS} \ket{\psi}$. If the original states evolve as $i \del_t \ket{\psi} = H \ket{\psi}$, the primed states evolve as $i \del_t \ket{\psi'} = H' \ket{\psi'}$, where 
\be
\label{eq:ham_pr}
H' = e^{i S}\,  (H - i \del_t) \, e^{- i S} = H + i [S, H] - \del_t S + \order{S^2}
~.
\ee
In particular, suppose we start with the derivative Hamiltonian, $H = H_d$, and choose
\be
\label{eq:S_expr}
S = - \beta  \, \frac{g}{4m} \, \{a, \Piv \cdot \sigmav \}
~,
\ee
where $\beta$ is a dimensionless parameter. Working to $\order{g/m}$, the primed Hamiltonian is then
\be \label{eq:h_pr_def}
H' \simeq H_d + \beta  \, \frac{g}{4m} \, \{\dot{a}, \Piv \cdot \sigmav \} + \beta  \, \frac{q \, g}{2m} \, a \, \sigmav \cdot \E
~.
\ee
For $\beta = 1$, we recover the nonderivative Hamiltonian $H_n$. Evidently, this transformation is the nonrelativistic Hamiltonian analogue of the field redefinition used to convert $\Lag_d$ to $\Lag_n$. Of course, the fact that the Hamiltonian can be transformed in appearance does not imply that experimental results are ambiguous. If an experiment measures an observable $A$ in the original picture, then in the primed picture it measures a primed observable $A'$ with the same matrix elements, $\bra{\psi} A \ket{\psi} = \bra{\psi'} A' \ket{\psi'}$, which implies that operators transform correspondingly as
\be 
\label{eq:operator_transformation}
A' = e^{i S} \, A \, e^{- i S} = A + i [S, A] + \order{S^2} ~.
\ee
Since matrix elements match by construction, experimental results are exactly the same in either picture.

This insight was crucial to resolve an old controversy in the theory of pion-nucleon interactions. In our language, Refs.~\cite{cheon1968foldy,barnhill1969ambiguity} pointed out that the coefficient of the axioelectric term could be modified by a unitary transformation. This led to confusion, with some claiming that pion absorption amplitudes were ambiguous. However, in Ref.~\cite{Friar:1974dk}, it was pointed out that such a transformation also generates an additional term involving the nuclear potential (the analogue of the EDM term above), and that when this is properly accounted for, all physical results are unchanged. 

A similar lesson can be drawn from the relativistic quantum mechanics of spin-$1/2$ particles, where a unitary transformation can be used to map the Dirac representation to the Foldy--Wouthuysen representation~\cite{Foldy:1949wa}. In the latter representation, the so-called ``Zitterbewegung" (or jittering) of the fermion's position and spin is eliminated, but its coupling to electromagnetic fields becomes nonlocal~\cite{DeVries:1970pbg,Costella:1995gt}. Despite these radical differences, the two representations are perfectly physically equivalent. Similarly, we will see that all values of $\beta$ give equivalent physical predictions, but $\beta = 0$ is by far the easiest choice to work with. For any $\beta \neq 0$, proper calculations of physical observables will display ``mysterious'' cancellations that reflect the non-manifest axion shift symmetry. 

\stoptocentries
\subsection{The Physical Axioelectric Term}
\starttocentries
\label{app:app_axioelectric}

In this section, we set the axion gradient $\nabla a$ to zero for simplicity. We also consider only neutral particles, $q = 0$, so that we can discuss the axioelectric term in isolation. In this case, the primed Hamiltonian in \Eq{h_pr_def} is
\be 
\label{eq:simpler_unitary}
H' \simeq \frac{p^2}{2 m} - \Big( 1 - \frac{\beta}{2} \Big) \, \frac{g \dot{a}}{m} \, \p \cdot \sigmav
~.
\ee
Thus, for neutral particles it appears that one can shift the coefficient of the axioelectric term to an arbitrary value, without any other consequences. On this basis, Ref.~\cite{Smith:2023htu} concluded that the axioelectric term is unphysical for neutral particles, and should be eliminated by choosing $\beta = 2$. 

The problem with this reasoning is that for a single particle, \textit{any} force, whether real or fictitious, can be removed by performing a unitary transformation which maps the laboratory frame to the particle's frame. The difference between the two cases is that real forces can produce relative accelerations between distinct colocated particles, while fictitious forces do not. By this standard, the axioelectric force is real and its effects can be measured experimentally. 

To see this explicitly, we must generalize \Eq{simpler_unitary} to a Hamiltonian with multiple interacting particles. One simple way to do this is to consider a Lagrangian with two neutral fermion fields $\Psi_1$ and $\Psi_2$ of mass $m$, with interactions
\be \label{eq:ld_12}
\Lag_d^{(12)} = g \, (\del_\mu a) \, \left( \Psibar_1 \g^\mu \g^5 \Psi_1 + \Psibar_2 \g^\mu \g^5 \Psi_2 \right) + c \, \Psibar_1 \Psi_1 \Psibar_2 \Psi_2
~,
\ee
where $c$ is a coupling constant for the contact interaction. Following the same procedure as in \Sec{LowHam} yields 
\be
i \dt \, \psi_1 \simeq \Big( \frac{p_1^2}{2m} - \frac{g \dot{a}}{m} \, \p_1 \cdot \sigmav_1 - c \, \Psibar_2 \Psi_2 \Big) \, \psi_1
~,
\ee
where a similar equation holds for $\psi_2$. In the limit where there is a single nonrelativistic well-localized particle of each type, we can simplify the bilinears as in \App{bilinears}, giving a Hamiltonian
\be \label{eq:original_contact}
H \simeq \frac{p_1^2}{2 m} + \frac{p_2^2}{2m} - c \, \delta^{(3)}(\xv_1 - \xv_2) - \frac{g \dot{a}}{m} \, (\p_1 \cdot \sigmav_1 + \p_2 \cdot \sigmav_2) 
~.
\ee
Finally, applying the unitary transformation generated by \Eq{S_expr} to each particle yields 
\be 
\label{eq:contact_hamiltonian}
H' \simeq \frac{p_1^2}{2 m} + \frac{p_2^2}{2m} - c \, \delta^{(3)}(\xv_1 - \xv_2) - \left(1 - \frac{\beta}{2} \right) \frac{g \dot{a}}{m} \, (\p_1 \cdot \sigmav_1 + \p_2 \cdot \sigmav_2) + \frac{\beta}{2} \, \frac{ga \, c}{m} \, (\sigmav_1 - \sigmav_2) \cdot \grad_1 \delta^{(3)}(\xv_1 - \xv_2)
~.
\ee
For the unprimed Hamiltonian in \Eq{original_contact}, the axioelectric term produces a measurable relative acceleration effect. For example, if the particles begin in their lowest bound state, with different spin directions, then a time-varying axion field can induce a transition to an excited state. On the other hand, for the primed Hamiltonian in \Eq{contact_hamiltonian} with $\beta = 2$, the axioelectric term is indeed rotated away, but it is replaced with a new term which induces precisely the same transitions. Similar reasoning would hold for any interaction between a particle and a measurement apparatus. One is free to work with $H'$, but for $\beta \neq 0$ it will always contain complicated additional terms which encode the physical effect of the axioelectric force. 

One might argue that the final term in \Eq{contact_hamiltonian} is only present because we chose to start with the derivative form of the interaction. Indeed, if we had started with
\be \label{eq:lag_n12}
\Lag_n^{(12)} = - 2 m \, g a \, \left( \Psibar_1 i \g^5 \Psi_1 + \Psibar_2 i \g^5 \Psi_2 \right) + c \, \Psibar_1 \Psi_1 \Psibar_2 \Psi_2
~,
\ee
then the corresponding nonrelativistic Hamiltonian would be \Eq{contact_hamiltonian} with $\beta = 1$ but without the final term. However, \Eq{lag_n12} is not an appropriate choice of Lagrangian because it violates the shift symmetry of the axion. The correct way to derive $\Lag_n^{(12)}$ is to start from \Eq{ld_12} and apply a chiral field redefinition, which instead yields 
\be \label{eq:lag_n12_better}
\Lag_n^{(12)} = - 2 m \, g a \, \left( \Psibar_1 i \g^5 \Psi_1 + \Psibar_2 i \g^5 \Psi_2 \right) + c \, \Psibar_1 e^{2 i g a \g^5} \Psi_1 \, \Psibar_2 e^{2 i g a \g^5} \Psi_2
~.
\ee
It is straightforward to check that the nonrelativistic Hamiltonian corresponding to this Lagrangian is indeed given by \Eq{contact_hamiltonian} with $\beta = 1$. In other words, regardless of what Lagrangian we start from, the axioelectric force is most directly described by taking $\beta = 0$.

\stoptocentries
\subsection{The Unphysical Nonrelativistic EDM Term}
\starttocentries
\label{app:nredm}

In this section, we instead set the axion gradient $\nabla a$ and time derivative $\dot{a}$ to zero for simplicity, so that we can discuss the nonrelativistic EDM term in isolation. In this case, the primed Hamiltonian in \Eq{h_pr_def} is
\be 
\label{eq:EDM_hams_2}
H'(\v{x}, \p) \simeq \frac{(\p - q \, \A)^2}{2 m} + q \, \phi - \frac{q}{2 m} \, \B \cdot \sigmav - d \, \E \cdot \sigmav
~,
\ee
where $d \equiv - \beta q \, g a / (2m)$, and all of the external fields above are evaluated at $\v{x}$. The final term in \Eq{EDM_hams_2} is indeed the nonrelativistic limit of the contribution of a true EDM to the Hamiltonian density, $\mathcal{H}_{\text{EDM}} = (d/2) \, \Psibar \g^5 \sigma^{\mu\nu} \Psi \, F_{\mu\nu}$. However, perhaps surprisingly, the term $- d \, \E \cdot \sigmav$ in isolation has no $\order{d}$ physical effects in the nonrelativistic limit, so that \Eq{EDM_hams_2} does not directly imply physical effects proportional to the axion field. 

\begin{center}
\textit{Energy Levels of Bound Electrons}
\end{center}

The simplest and most general way to see this is to note that $H^\prime$ is unitarily equivalent to $H_d$, which has no nonrelativistic EDM term. Crucially, unlike the axioelectric term in \App{app_axioelectric}, the Hamiltonian without the nonrelativistic EDM term does not contain any other $\order{d}$ terms, so its effect can truly be removed without consequence. This remains true even when interparticle Coulomb interactions are included. 

In particular, EDMs are often measured through the energy level shifts they induce in atoms placed in an external, constant electric field. If the atoms are described nonrelativistically, with only pairwise Coulomb interactions, then a nonrelativistic EDM term can be eliminated at $\order{d}$ by a unitary transformation generated by $S \propto \Piv \cdot \sigmav$. Since this $S$ is time-independent, \Eq{ham_pr} implies that it preserves energy levels. We thus conclude that nonrelativistic EDMs do not have any $\order{d}$ effects on atomic energy levels, which is simply the famous statement of Schiff's theorem~\cite{Schiff:1963zz}. 

\begin{center}
\textit{Shifting the Position Operator}
\end{center}

While the physical equivalence of $H_n$ and $H_d$ decisively rules out an $\mathcal{O}(d)$ EDM, it is unintuitive that the $- d \, \E \cdot \sigmav$ term is unphysical for charged particles. This is because such a term seems to suggest the generation of electric dipole radiation and spin precession in an electric field. Indeed, the neutron EDM \textit{can} be measured using the latter effect~\cite{Smith:1957ht}. However, the situation for charged particles is fundamentally different. The reason is that a particle of charge $q$ can always be artificially described by a position operator shifted relative to the usual one by $\Delta \v{x}$, which introduces an electric dipole moment $\v{d} = q \, \Delta \v{x}$. This is the fundamental reason an EDM term appears in \Eq{EDM_hams_2}, and it implies that the same physical effects can be described without it. 

To make this statement more precise, consider the position operator $\v{x}$ in the unprimed Hamiltonian, which has no EDM term. As usual, the acceleration of the particle is proportional to $\E(\v{x})$, its static Coulomb field is centered at $\v{x}$, and when the particle accelerates, the electric dipole radiation it produces is governed by $d^2 \v{x}/dt^2$. In light of these facts, we say that $\v{x}$ is the particle's ``center of charge'' $\v{x}_q$. However, after a unitary transformation to $H'$, \Eq{operator_transformation} implies that the center of charge becomes
\be \label{eq:centers_of_charge}
\xv_q' = \xv + (d/q) \, \sigmav + \order{d^2}
~,
\ee
which is shifted from $\v{x}$ precisely as anticipated in the previous paragraph. By construction, the center of charge evolves the same way in both pictures, so that any effect of the apparent EDM can be equivalently explained without it. 

For example, suppose a free electron is placed in a uniform magnetic field $\B_0$, so that the spin precesses at the Larmor frequency $\w_L = q B_0 / m$. If the particle is at rest under the unprimed Hamiltonian, $d \v{x}/dt = 0$, then $d \v{x}_q/dt = 0$, which implies that no electric dipole radiation is produced. When we consider the same situation under the primed Hamiltonian, the particle has a precessing EDM, but simultaneously orbits in a circle of radius $d/q$ in the opposite direction at the cyclotron frequency $\w_c = q B_0 / m$. These two effects compensate each other, so that $d \v{x}_q'/dt = 0$ and, again, no electric dipole radiation is produced. Conversely, if the particle were at rest under the primed Hamiltonian, electric dipole radiation would be produced, but it would be equivalently described under the unprimed Hamiltonian as a consequence of the particle's circular motion. 

\begin{center}
\textit{EDM-Induced Spin Precession}
\end{center}

Similarly, it naively seems possible to unambiguously identify the EDM through its effect on spin precession, but this is also impossible. To illustrate this point we will consider two more thought experiments. 

First, Ref.~\cite{Alexander:2017zoh} proposed an approach which is analogous to certain searches for the neutron EDM. Its authors claimed that if an electron was placed in a uniform electric field, the EDM would cause its spin to precess, resulting in an observable transverse magnetic field from its magnetic dipole moment. This led Ref.~\cite{Chu:2019iry} to project very strong experimental sensitivity for atomic magnetometers, shown in \Fig{future}. However, this idea has a fundamental problem. If the electron was free, it would immediately accelerate out of the experimental apparatus. If it was instead bound in a nonrelativistic atom, it would experience zero average electric field and hence undergo no spin precession~\cite{garwin1959electric}. 

Therefore, the experiment proposed in Ref.~\cite{Alexander:2017zoh} cannot work as stated. For a free electron, we could instead allow the electron to fly away, but subsequently measure its spin dynamically. Concretely, suppose a free electron is prepared at rest with vertical spin, and then experiences a uniform horizontal electric field. After passing through this field it encounters a Stern--Gerlach apparatus, i.e., a vertical but nonuniform magnetic field, where the field gradient deflects the electron according to its spin. This appears, in principle, to be a way to measure the EDM-induced spin precession without requiring it to be bound. 

The problem with this idea is that the unprimed Hamiltonian, with no EDM, yields the same deflection. In this picture, the electron's location is shifted by $(d/q) \, \sigmav$, so it flies through the Stern--Gerlach apparatus off center, yielding an additional $q \v{v} \times \B$ Lorentz force. It is straightforward to show that the center of charge behaves the same way to $\order{d}$ in both pictures. Once again, the apparent EDM effect can be equivalently explained without an EDM. 

\stoptocentries
\subsection{True EDMs and Spurious EDMs}
\starttocentries
\label{app:true_EDM}

There are many experiments, reviewed in Refs.~\cite{Chupp:2017rkp,Alarcon:2022ero}, which are sensitive to constant electron EDMs. Unfortunately, the shift symmetry of the axion implies that a constant axion field cannot produce a signal in any of them. Here, we explain how a number of these experiments work and elaborate on why they cannot measure an axion signal.

\begin{center}
\textit{Scattering and Spin Precession of Free Electrons}
\end{center}

A true electron EDM affects the relativistic scattering of electrons on nuclei~\cite{burleson1960experimental,goldemberg1963upper,rand1965determination}. True EDMs can also cause spin precession when a particle moves at relativistic speeds through a \textit{magnetic} field; this was used in the first searches for the electron and muon EDM~\cite{berley1958electric,nelson1959search,berley1960search,charpak1961new} and today is relevant in storage ring experiments. Crucially, both of these signatures do not follow from the nonrelativistic EDM term, but rather come from relativistic corrections present in the true EDM's full Hamiltonian density, $\mathcal{H}_{\text{EDM}} = (d/2) \, \Psibar \g^5 \sigma^{\mu\nu} \Psi \, F_{\mu\nu}$. 

Since the axion contributes a nonrelativistic EDM term of $\order{ga/m}$, such corrections would first appear in the Hamiltonian at $\order{ga/m^2}$. However, Ref.~\cite{Smith:2023htu} computed the Hamiltonian out to this order and showed that the $\order{ga/m^2}$ terms actually exactly vanish for a constant uniform axion field.\footnote{Note that Ref.~\cite{Smith:2023htu} uses a different convention for the coupling; in their variables the nonrelativistic EDM term is $\order{1/m^2}$ and the leading relativistic corrections to it are $\order{1/m^3}$.} In other words, while a true EDM and a constant axion field modify the Hamiltonian in the same way at leading order in the nonrelativistic expansion, the axion does not induce the relativistic corrections that make true EDMs observable for free particles.

\begin{center}
\textit{Energy Levels of Bound Electrons}
\end{center}

The most sensitive modern electron EDM experiments measure shifts of atomic or molecular energy levels. As we have discussed in \App{nredm}, Schiff's theorem states that $\order{d}$ shifts of atomic energy levels vanish in the nonrelativistic limit. Therefore, to find a nonzero effect one must either work to $\order{d^2}$~\cite{feinberg1958effects,salpeter1958some} or account for relativistic effects such as length contraction, in whose presence there are $\order{dv^2}$ energy level shifts~\cite{sandars1965electric,sandars1968electric,commins2007electric}.

Recalling that our expansion in $1/m$ is equivalent to an expansion in $v$, these facts naively suggest that a constant axion field shifts energy levels at $\order{g^2 a^2 / m^2}$ and at $\order{ga/m^3}$, respectively. The problem with this reasoning is, again, that a constant axion field does not enter the Hamiltonian in the same way as a true EDM. The two have the same nonrelativistic limit, but differ in their relativistic and higher-order corrections. In order to compute these purported axion-induced energy level shifts consistently, one must expand the Hamiltonian to $\order{g^2 a^2 / m^2}$ and $\order{ga/m^3}$, respectively. This produces additional terms which completely cancel off any energy level shift. Although such a lengthy computation has not been explicitly demonstrated in the literature, such a cancellation \textit{must} occur to all orders due to the underlying shift symmetry of the axion. These ``mysterious" cancellations are the hallmark of a non-manifest symmetry and are one of the reasons it is often easier to work with manifest symmetries. \\

\begin{center}
\textit{Recasting EDM Experiments}
\end{center}

In an EDM experiment of measurement time $\tau$, Ref.~\cite{Wang:2021dfj} asserted that the axion-induced nonrelativistic EDM term acts like a constant true EDM when $m_a \tau \ll 1$, thereby claiming a bound on the axion-electron coupling many orders of magnitude stronger than existing bounds. However, this is incorrect; as we have just discussed, the signal from a true EDM arises from relativistic corrections which are not shared by the spurious axion EDM. 

On the other hand, when the axion field has nontrivial time dependence, a term proportional to $a(t) \, \sigmav \cdot \E$ can produce observable effects, such as shifts in electronic energy levels. However, such shifts are generally suppressed by powers of $m_a / \text{Ry}$, where $\text{Ry} = \alpha^2 m_e/2$ is the scale of electronic energy levels. For example, Ref.~\cite{Stadnik:2013raa} showed that the energy level shift for bound electrons, linear in the external electric field, is suppressed by $(m_a / \text{Ry})^2$. 

Recently, Refs.~\cite{Smith:2023htu,DiLuzio:2023ifp} noted that for \textit{free} electrons, axion-induced effects are suppressed by powers of $m_a \tau$. For example, Ref.~\cite{DiLuzio:2023ifp} computed the time evolution operator for a free particle, from $t = 0$ to $t = \tau$, and noted that it depended on the quantity
\be
\frac{1}{\tau} \int_0^\tau dt ~ d_{\eff}(t) \equiv \frac{1}{\tau} \int_0^\tau dt ~ (d(t) - d(\tau)) \label{eq:deff}
~,
\ee
where $d_{\text{eff}}$ is an ``effective'' EDM. The left-hand side has a simple physical interpretation given by \Eq{centers_of_charge}; for a free particle at rest in the primed picture, it is simply $q$ times the difference of the center of charge's final location and its average location. \Eq{deff} is indeed suppressed for $m_a \tau \ll 1$, and unsuppressed for $m_a \tau \gtrsim 1$. However, Refs.~\cite{Smith:2023htu,DiLuzio:2023ifp} assumed without justification that the signal in a \textit{bound} electron EDM experiment is not suppressed when $m_a \tau \gtrsim 1$, even though it is governed by completely different observables. Thus, for $m_a \tau \sim 1$ these works overestimate the signal strength for bound electrons by powers of $\text{Ry} \cdot \tau \gg 1$. 

\begin{center}
\textit{Color Charged Fermions}
\end{center}

Throughout this work, we have exclusively considered the case where $\Psi$ is a color neutral fermion. The story is somewhat different when $\Psi$ is a quark field, since the chiral field redefinition leading to the nonderivative coupling $\Lag_n$ also produces a term proportional to $a \, G_{\mu\nu} \, \tilde{G}^{\mu\nu}$. Both this term and $\Lag_n$ can produce true hadronic EDMs for a constant axion field $a$. If one starts with only a derivative coupling, then their contributions will cancel, in accordance with the fact that the derivative coupling $\Lag_d$ itself vanishes for constant $a$. But in a generic QCD axion model, there will be true hadronic EDMs induced, which can be probed by both existing EDM experiments and in dedicated experiments which resonantly amplify the effect at nonzero $m_a$~\cite{Graham:2013gfa}. For a recent review of such efforts, see Ref.~\cite{Berlin:2022mia}.

\bibliographystyle{utphys3}
\bibliography{AxionFermion}

\end{document}